\documentclass{aa} 

\usepackage{graphicx}
\usepackage{multirow}
\usepackage{color, colortbl}
\definecolor{Gray}{gray}{0.9}
\usepackage{xcolor}

\usepackage{hyperref}
\hypersetup{
  colorlinks,
  filecolor=blue,
  citecolor=blue,
  linkcolor=blue,
  urlcolor=blue}
  
\usepackage{txfonts}
\begin{document}

   \title{The massive relic galaxy NGC~1277 is dark matter deficient\thanks{The reduced spectra and the information necessary to reconstruct the kinematic maps presented in this paper are available in electronic form at the CDS via anonymous ftp to \url{cdsarc.cds.unistra.fr} (130.79.128.5) or via \url{https://cdsarc.cds.unistra.fr/cgi-bin/qcat?J/A+A/}}}
   \subtitle{From dynamical models of integral-field stellar kinematics out to five effective radii}

   \author{Sébastien~Comerón\inst{1,2}, Ignacio~Trujillo\inst{2,1}, Michele~Cappellari\inst{3}, Fernando~Buitrago\inst{4,5}, Luis~E.~Garduño\inst{6}, Javier~Zaragoza-Cardiel\inst{6,7}, Igor~A.~Zinchenko\inst{8,9}, Maritza~A.~Lara-López\inst{10,11}, Anna~Ferré-Mateu\inst{2,1}, and Sami~Dib\inst{12}}

   \institute{Departamento de Astrof\'isica, Universidad de La Laguna, E-38200, La Laguna, Tenerife, Spain\\ \email{lsebasti@ull.edu.es}
              \and Instituto de Astrof\'isica de Canarias E-38205, La Laguna, Tenerife, Spain
              \and Sub-Department of Astrophysics, Department of Physics, University of Oxford, Denys Wilkinson Building, Keble Road, Oxford OX1 3RH, UK
              \and Departamento de Física Teórica, Atómica y Óptica, Universidad de Valladolid, E-47011 Valladolid, Spain
              \and Instituto de Astrof\'{i}sica e Ci\^{e}ncias do Espa\c{c}o, Universidade de Lisboa, OAL, Tapada da Ajuda, PT1349-018 Lisbon, Portugal
              \and Instituto Nacional de Astrofísica, Óptica y Electrónica (INAOE), Luis Enrique Erro No.~1, Tonantzintla, Puebla, CP 72840, México
              \and Consejo Nacional de Ciencia y Tecnolog\'ia, Av.~Insurgentes Sur 1582, 03940 M\'exico City, México
              \and Faculty of Physics, Ludwig-Maximilians-Universit{\"a}t, Scheinerstr.~1, 81679 Munich, Germany
              \and Main Astronomical Observatory, National Academy of Sciences of Ukraine, 27 Akademika Zabolotnoho St, 03680, Kyiv, Ukraine
              \and Departamento de Física de la Tierra y Astrofísica, Universidad Complutense de Madrid, E-28040 Madrid, Spain
              \and Instituto de Física de Partículas y del Cosmos IPARCOS, Fac.~de Ciencias Físicas, Universidad Complutense de Madrid, E-28040 Madrid, Spain
              \and Max Planck Institute for Astronomy, K\"onigstuhl 17, 69117, Heidelberg, Germany}

% \abstract{}{}{}{}{}
% 5 {} token are mandatory
 
  \abstract{According to the Lambda cold dark matter ($\Lambda$CDM) cosmology, present-day galaxies with stellar masses $M_\star>10^{11}\,{\rm M}_\odot$ should contain a sizable fraction of dark matter within their stellar body. Models indicate that in massive early-type galaxies (ETGs) with $M_\star\approx1.5\times10^{11}\,{\rm M}_\odot$, dark matter should account for $\sim15\%$ of the dynamical mass within one effective radius ($1\,R_{\rm e}$) and for $\sim60\%$ within $5\,R_{\rm e}$. Most massive ETGs have been shaped through a two-phase process: the rapid growth of a compact core was followed by the accretion of an extended envelope through mergers. The exceedingly rare galaxies that have avoided the second phase, the so-called relic galaxies, are thought to be the frozen remains of the massive ETG population at $z\gtrsim2$. The best relic galaxy candidate discovered to date is NGC~1277, in the Perseus cluster. We used deep integral field George and Cynthia Mitchel Spectrograph (GCMS) data to revisit NGC~1277 out to an unprecedented radius of 6\,kpc (corresponding to $5\,R_{\rm e}$). By using Jeans anisotropic modelling, we find a negligible dark matter fraction within $5\,R_{\rm e}$ ($f_{\rm DM}(5\,R_{\rm e})<0.05$; two-sigma confidence level), which is in tension with the $\Lambda$CDM expectation. Since the lack of an extended envelope would reduce dynamical friction and prevent the accretion of an envelope, we propose that NGC~1277 lost its dark matter very early or that it was dark matter deficient ab initio. We discuss our discovery in the framework of recent proposals, suggesting that some relic galaxies may result from dark matter stripping as they fell in and interacted within galaxy clusters. Alternatively, NGC~1277 might have been born in a high-velocity collision of gas-rich proto-galactic fragments, where dark matter left behind a disc of dissipative baryons. We speculate that the relative velocities of $\approx2000\,{\rm km\,s^{-1}}$ required for the latter process to happen were possible in the progenitors of the present-day rich galaxy clusters. 
  
  } 
   \authorrunning{S.~Comerón et al.}

   \maketitle

   %
%________________________________________________________________

\section{Introduction}

The first proposed formation scenarios for massive early-type galaxies (ETGs) suggested a monolithic collapse \citep{Larson1969} similar to what had been posited for the oldest stellar populations in the Milky Way \citep{Eggen1962}. These models fell out of fashion with the advent of hierarchical formation models within a cold dark matter cosmology \citep{White1978, Klypin1983}. However, recent computational studies assuming a Lambda cold dark matter ($\Lambda$CDM) cosmology indicate that nature borrows from both scenarios and that the actual massive ETG formation process can be broadly divided into two phases \citep{Naab2009, Oser2010, Johansson2012, Lackner2012, Zolotov2015, RodriguezGomez2016}. At first, the massive ETG progenitors suffer a dissipative collapse that results in a compact rotating object (known as a blue nugget) whose star formation is rapidly quenched \citep{Huang2013, Dekel2014, Zolotov2015, MartinNavarro2019, Lapiner2023}, perhaps due to active galactic nucleus (AGN) feedback \citep[e.g.][]{Choi2018}. During the second phase, the passive evolution of the stellar population turns the blue nugget into a red one, and most of the growth of the stellar mass of the object comes from the accretion of an extended pressure-supported envelope through dry accretion.

This simplified two-phase formation narrative is supported by observations indicating that passively evolving galaxies at high redshift were much more compact than local massive ETGs \citep{Daddi2005, Zirm2007, Cimatti2008, Toft2017, Kubo2018}, and they have grown by a factor of roughly four in effective radius since redshift $z\sim2$ \citep{Trujillo2006, Trujillo2007, Buitrago2008, Wel2011, Szomoru2012}. Further credence to this framework is given by the fact that stellar envelopes have been shown to grow with cosmic time \citep{Damjanov2011, Ryan2012, Wel2014, Bai2014, Buitrago2017}. Also, the recently identified compact, star-forming, and presumably rotating ultra-red flattened objects \citep{Nelson2023} could be dust-shrouded blue nuggets, which would give additional evidence for the two-phase formation scenario.

Both models and observations indicate that red nuggets are the evolved remnants of the infant massive galaxies, offering a sneak peek at the epoch at which the first giant galaxies assembled, at $z\gtrsim2$. Since red nuggets are expected to survive within at least a fraction of massive galaxies \citep{Pulsoni2021}, they can be examined and resolved in nearby galaxies. However, this often comes at the expense of suffering contamination from the envelope. Local red nuggets embedded within elliptical or even disc galaxies have been studied through structural decompositions \citep{Huang2013, Graham2015, Rosa2016, Costantin2021} and resolved stellar populations \citep{Yildirim2017, MartinNavarro2018, FerreMateu2019, Barbosa2021}. Also, some red nuggets could be identified with the kinematically decoupled components (KDCs) often observed in the central regions of slow rotator ETGs \citep[see review by][]{Cappellari2016}. Cleaner nuggets can be studied at intermediate and high $z$ \citep[e.g.][]{Damjanov2009, Damjanov2013, Damjanov2014, Dokkum2009, Auger2011, Stockton2014, Hsu2014, Saulder2015, Oldham2017, Charbonnier2017, Buitrago2018, Scognamiglio2020, Spiniello2021, Spiniello2021a, Lisiecki2023, DAgo2023}, but the price to pay is to suffer cosmological dimming and poor angular resolution.

Local red nuggets that have somehow avoided accreting an envelope are therefore key to accessing good-quality uncontaminated data from the progenitors of massive ETGs. They constitute snapshots of galaxies that have evolved passively, and whose growth stopped at a freezing redshift $z_{\rm f}>1$. Such galaxies are exceedingly rare \citep[see the number density estimates in][]{Quilis2013} and have been named `relic galaxies' since the study by \citet{Trujillo2009}. Ironically, no relic galaxy was found in the study defining them. Later searches were more successful, and nowadays a few tens of candidates and spectroscopically confirmed relics are known \citep{Trujillo2014, FerreMateu2015, FerreMateu2017, Spiniello2021, Spiniello2021a}.

The detailed study of NGC~1277 has provided unique insight into the extreme conditions under which nuggets grow. \citet{MartinNavarro2015} found that the initial mass function (IMF) of NGC~1277 is bottom-heavy throughout the galaxy, which implies a large stellar mass-to-light ratio $\Upsilon_\star$, as confirmed by \citet{Yildirim2015}. Similar although slightly less extreme properties have also been confirmed for two more relic galaxies \citep{FerreMateu2017}.  Also, dynamical modelling has unveiled that virtually no dark matter is necessary to explain the stellar kinematics in the innermost $13\arcsec$ of NGC~1277 \citep{Yildirim2015}.

NGC~1277 is a fast rotator, a property that is common among compact ETGs \citep{Yildirim2017}. It is found in a cluster environment, which is in agreement with the expectations for relic galaxies from some cosmological simulations \citep{Stringer2015, PeraltadeArriba2016} and observational findings \citep{Poggianti2013, Buitrago2018, Baldry2021}. The reason for this environmental preference has been posited to be that the high relative velocities between cluster galaxies are efficient at preventing prolonged interactions. However, other evolutionary pathways are also possible \citep[see][for more details]{Buitrago2018, Tortora2020}.

NGC 1277 was originally reported to have an unexpectedly massive black hole with $M_{\rm BH}=17\times10^9\,{\rm M}_\odot$ from Schwarzschild dynamical modelling based on rather low spatial resolution long-slit data \citep[][assumed distance of $d=73\,{\rm Mpc}$]{Bosch2012}. However, a reanalysis of the same data by \citet[][$d=73\,{\rm Mpc}$]{Emsellem2013} using an $N$-body particle realisation based on the Jeans Anisotropic Modelling \citep[\texttt{JAM};][]{Cappellari2008} formalism suggested that the black hole was likely overestimated and proposed $M_{\rm BH}\approx5\times10^9\,{\rm M}_\odot$. This lower value of $M_{\rm BH}$ was later confirmed with higher spatial resolution integral field data from both Schwarzschild \citep[][$d=71\,{\rm Mpc}$]{Walsh2016} and \texttt{JAM} models \citep{Krajnovic2018}. Even the lower black hole mass estimates place NGC~1277 more than one order of magnitude above the scaling relations linking the mass of the black hole with the luminosity of the spheroid \citep[the $M_{\rm BH}-L$ relation;][]{Gueltekin2009, Sani2011}, but would be consistent with the $M_{\rm BH}-\sigma$ relation \citep{Ferrarese2000, Gebhardt2000} for a stellar velocity dispersion of $400\,{\rm km\,s^{-1}}$ at the centre of the galaxy \citep{Bosch2012}. A possible explanation for the $M_{\rm BH}-L$ anomaly in NGC~1277 is that the black hole assembly occurs chiefly during the early collapse phase, with little to no mass increase during the envelope growth period \citep{Trujillo2014, FerreMateu2015, FerreMateu2017}. Since the scaling relations are established using mostly ETGs that have undergone an envelope growth, NGC~1277 would naturally have a lower mass than regular massive ETGs hosting black holes with a similar mass. Another possibility is that the central black hole in NGC~1277 was captured. Indeed, NGC~1275, which is the central dominant elliptical in the Perseus cluster, has an under-massive black hole. It has been postulated that a former black hole sitting at the centre of NGC~1275 might have been lost due to recoil during a black hole fusion and ended up being captured by NGC~1277 \citep{Shields2013}. Alternatively, the mass of the central black hole in NGC~1277 might have been overestimated and be as low as $M_{\rm BH}=1.2\times10^{9}\,{\rm M}_\odot$ \citep{Graham2016} due to the degeneracy between the black hole mass and $\Upsilon_\star$ in the central parts of the galaxy \citep[see also][]{Emsellem2013}.

We revisit NGC~1277 with new deep spectroscopic data obtained with the George and Cynthia Mitchel Spectrograph (GCMS), an integral field spectrograph at the 2.7\,m Harlan J.~Smith telescope at the McDonald Observatory \citep{Hill2008, Tufts2008}. These data are a significant improvement over single-slit observations \citep{Trujillo2014, MartinNavarro2015}. Our observations also improve upon previous integral field observations with a narrower radial coverage than ours \citep[$13\arcsec$ versus $18\arcsec$, roughly an increase of 50\% in the area coverage;][]{Yildirim2015, Yildirim2017}. This is critical to explore the outskirts of the galaxy, where dark matter is expected to contribute the most to the mass budget, and test the claims that a dark matter halo is not required to describe the stellar kinematics. An additional advantage of our dataset is that the field of view simultaneously covers NGC~1278, a regular massive ETG in the Perseus cluster with a mass similar to that of NGC~1277. Thus, we can study a compact relic galaxy alongside more extended object that has an envelope that has probably been accreted (as indicated by the age, metallicity, and abundance gradients obtained from an analysis of the stellar populations; companion paper Ferré-Mateu et al.~in prep.) with exactly the same instrumental setup, enabling a straightforward comparison of their properties. In order to mitigate the coarse GCMS $\sim4^{\prime\prime}$ angular resolution and better constrain the kinematics in the vicinity of the central black hole, we are complementing our data with kinematics of the inner $1\farcs6\times1\farcs6$ of NGC~1277  obtained by \citet{Walsh2016} using the adaptive-optics-assisted  Near-infrared Integral Field Spectrometer \citep[NIFS;][]{McGregor2003} at the Gemini North telescope.

In this paper, we examine the stellar kinematics of NGC~1277 and NGC~1278 and produce detailed dynamical Jeans models with \texttt{JAM}\footnote{We used v6.3.2 of the \texttt{Python} package \texttt{JamPy}, available from \url{https://pypi.org/project/jampy/}.}  \citep{Cappellari2008, Cappellari2020}. We assume NGC~1277 to be isolated and non-interacting. This is based on the observation by \citet{Trujillo2014} that the object has no signs of tidal features down to a level of $\mu_r=26.8\,{\rm mag\,arsec^{-2}}$. In order to be consistent with the dark matter halo parametrisation from \citet[][see Sect.~\ref{dmhs}]{Child2018}, we assume a flat Universe with a Hubble-Lema\^itre constant of $H_0=71\,{\rm km\,s^{-1}\,Mpc^{-1}}$ and a matter density $\Omega_{\rm m0}=0.265$. Given a redshift of $z=0.017195$ for the Perseus Cluster \citep{Bilton2018}, we determine that the comoving distance to the galaxies is $d=72.4\,{\rm Mpc}$. The luminosity distance is $d_{\rm L}=73.6\,{\rm Mpc}$ and the angular diameter distance is $d_{\rm A}=71.1\,{\rm Mpc}$, which means that $1\arcsec$ corresponds to $345\,{\rm pc}$.

The goal of this article is to describe the dark matter distribution of NGC~1277 out to an unprecedented radius of five effective radii ($5\,R_{\rm e}$, corresponding to 6\,kpc). Because of the relic nature of the target, this is equivalent to studying the dark matter halo properties of a massive ETG at $z\gtrsim2$. The results of the experiment provide tight constraints to test the accuracy of the dark matter distributions that are predicted by the $\Lambda$CDM model.  

This paper is organised as follows: in Sect.~\ref{data} we describe the GCMS data and their reduction process. In Sect.~\ref{technique} we detail how the stellar kinematics and the multi-Gaussian expansion (\texttt{MGE}) models of the {\it Hubble Space Telescope} ({\it HST}) images and mass maps were obtained. In Sect.~\ref{jam} we describe the Jeans dynamical modelling of NGC~1277 and NGC~1278. We discuss our results in Sect.~\ref{discussion} and present our summary in Sect.~\ref{conclusion}. We also include Appendix~\ref{appendix}, where we discuss the accuracy of our black hole mass determination if we ignore the high-resolution {\it HST} and NIFS data. This is crucial to assess the feasibility of the black hole mass determination for the progenitors of relic galaxies at high redshift using adaptive optics-assisted or {\it James Webb Space Telescope} ({\it JWST}) data.

\section{Data and data reduction}

\label{data}

\begin{figure*}
\begin{center}
  \includegraphics[scale=0.48]{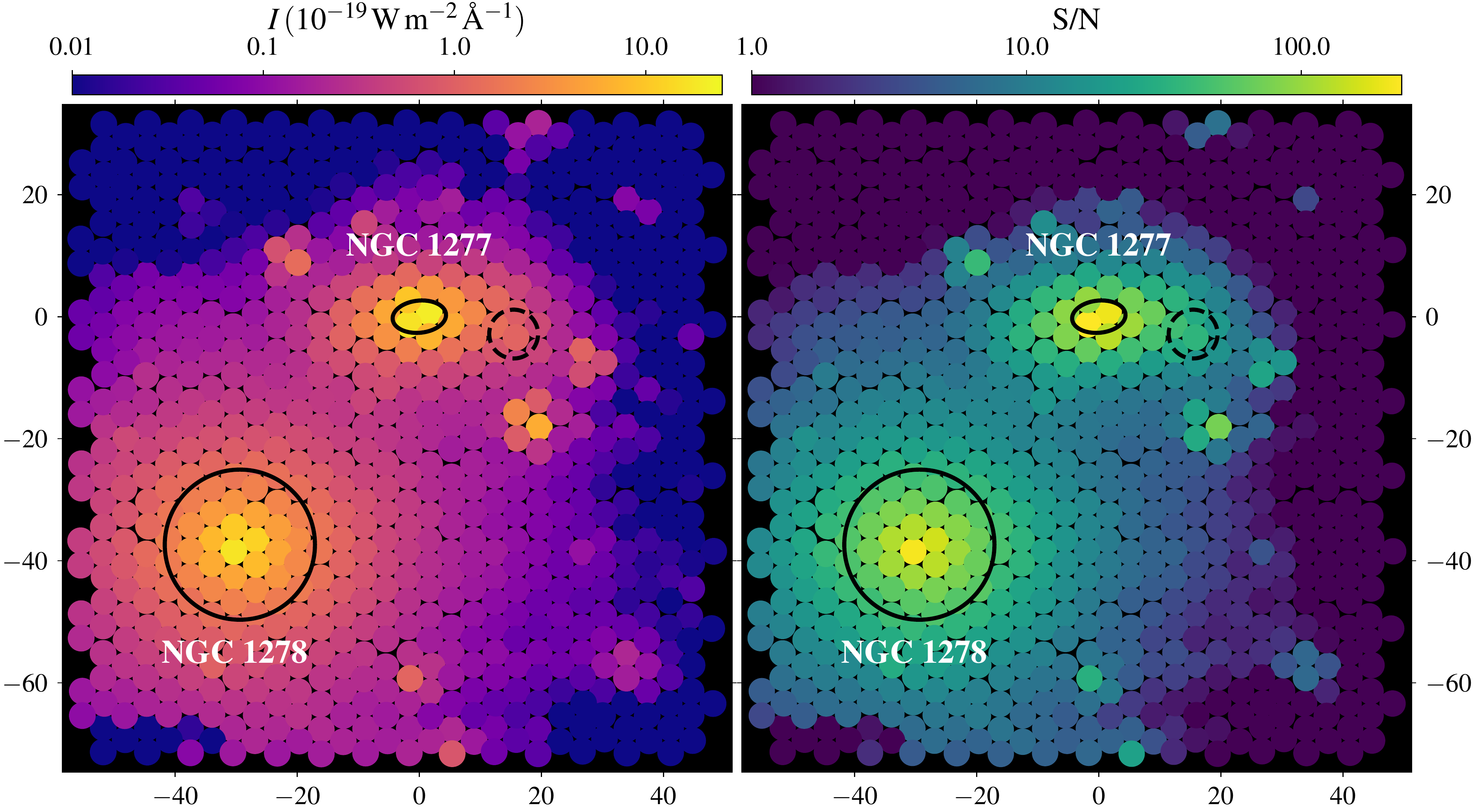}\\
  \end{center}
  \caption{\label{map_snr} Map of the median intensity ({\it left} panel) and S/N per spectral pixel of 1.1\,\AA\ ({\it right} panel) in the wavelength range $4400\,{\rm \AA}-6650\,{\rm \AA}$. The coordinates are centred in NGC~1277 and are in arcseconds. The black continuous shapes denote the effective radii of the galaxies as measured in Sect.~\ref{mges}. The effective radius of NGC~1277 has been de-circularised assuming an axial ratio $q^{\prime}_\star=0.6$, as determined from an ellipse fit to the outer regions of the galaxy (see Sect.~\ref{mges}). The dashed circle indicates the location of the galaxy that is in the line of sight of NGC~1277 (Sect.~\ref{kinematics}).}
\end{figure*}

The George Mitchell and Cynthia Spectrograph (GCMS; formerly known as \mbox{VIRUS-P}), is hosted at the 2.7\,m Harlan J.~Smith telescope at the McDonald Observatory \citep{Tufts2008, Hill2008}. The integral field unit comprises 246 $4\farcs16$ in diameter fibres that are distributed in a fixed pattern with a filling factor of roughly one-third. The total field of view of the instrument is of $100\arcsec\times102\arcsec$. At the distance of NGC~1277, the fibre diameter corresponds to 1.4\,kpc and the field of view is $\sim35\,{\rm kpc}$ per side.

The data were taken with the `red setup' on 29--31 December 2019 as part of the Metal-THINGS programme \citep{LaraLopez2021, LaraLopez2023}. The observations were made with the VP1 grism and covered the $4400\,{\rm \AA}-6800\,{\rm \AA}$ spectral range with a spectral resolution of 5.3\,\AA. In order to ensure a high filling factor, a three-pointing dither pattern was followed, resulting in a 738-fibre coverage of the field of view. The dithering pattern was designed so as to avoid the position of the fibres to overlap within each three-pointing dataset.

Ten three-point dither datasets were taken, each with an exposure time of 900\,s per dither for a total of 9000\,s (2.5\,hours) per position. Throughout the acquisition, the full width at half maximum (FWHM) of the seeing varied between 1\farcs7 and 3\farcs5. Only one of the datasets had a seeing above $2\farcs8$. We checked that ignoring the latter dataset did not improve the quality of the recovered kinematics.

The basic data reduction (bias subtraction, flat frame correction, and wavelength calibration) was performed using \texttt{P3D}\footnote{The software can be obtained from \url{https:/p3d.sourceforge.io}.} \citep{Sandin2010}. The sky emission of the ten datasets was estimated by computing the median spectrum of a few hand-picked sky fibres. After the sky subtraction, the datasets were flux-calibrated using \texttt{IRAF} \citep{Tody1986} and combined to produce a final science-ready data product with a single spectrum for each of the 738 fibre locations (from now on and for the sake of brevity, we simply call them fibres) in the field of view. We cropped the reddest portion of the spectrum, above the rest frame 6650\,\AA, to avoid some calibration problems on that portion of the spectrum. Nine fibres in the south-eastern corner of the field of view have problematic spectra and were excluded from further analysis. Figure~\ref{map_snr} shows maps of the median spectral intensity and the signal-to-noise ratio (S/N) per spectral pixel of 1.1\,{\AA}. The S/N was obtained by estimating the noise with the \texttt{DER\_SNR} algorithm by \citet{Stoehr2008} and comparing it to the median intensity.

\section{Galaxy resolved kinematics and \texttt{MGE} modelling}

\label{technique}

\subsection{The resolved kinematics of NGC~1277 and NGC~1278}

\label{kinematics}

The kinematics of the field were obtained with \texttt{pPXF}\footnote{We used v7.4.5 of the \texttt{pPXF} package available from \url{https://pypi.org/project/ppxf/}.} \citep{Cappellari2004, Cappellari2022}. This code fits the spectra with a linear combination of spectral energy distribution templates convolved with a line-of-sight velocity distribution (LOSVD). The templates were drawn from the E-MILES single stellar population models\footnote{The E-MILES spectral energy distribution libraries can be downloaded from \url{http://research.iac.es/proyecto/miles/pages/spectral-energy-distributions-seds/e-miles.php}.} \citep{Vazdekis2015}, which uses the MILES \citep{SanchezBlazquez2006, FalconBarroso2011} and CaT \citep{Cenarro2007} stellar templates. We used the models with a Kroupa initial mass function \citep{Kroupa2001} and BaSTI isochrones \citep{Pietrinferni2004, Pietrinferni2006, Pietrinferni2009, Pietrinferni2013, Cordier2007}, but the choice of a specific library plays little effect in the precise determination of the kinematics. To speed up the processing, we excluded the templates that, a priori, would not describe a massive ETG, that is those with low metallicity ($[{\rm Z}/{\rm H}]<-0.35$) and young ages ($<1\,{\rm Gyr}$). We have tested that producing the fits with the complete MILES model library would not introduce significant changes in the derived kinematics.

We only fitted the first two momenta of the LOSVD, that is the mean recession velocity and the velocity dispersion $[v,\sigma]$. The characterisation of higher-order momenta requires high-S/N data, which would go against our goal of covering as much of NGC~1277 as possible. In some lines of sight, two objects appear in projection, so their spectra should be fitted with two sets of templates with distinct kinematics. This is why each spectrum was at first fitted with single sets of templates with initial values $[v=5066\,{\rm km\,s^{-1}},\sigma=250\,{\rm km\,s^{-1}}]$ \citep[corresponding to the redshift $z=0.016898$ of NGC~1277;][]{Falco1999}, $[v=6066\,{\rm km\,s^{-1}},\sigma=250\,{\rm km\,s^{-1}}]$ (roughly the velocity of NGC~1278), $[v=4066\,{\rm km\,s^{-1}},\sigma=50\,{\rm km\,s^{-1}}]$ (the velocity of a small galaxy overlapping with part of the western side of NGC~1277; see Fig.~\ref{map_snr}), and $[v=-34\,{\rm km\,s^{-1}},\sigma=50\,{\rm km\,s^{-1}}]$ (this is a velocity compatible with that of Galactic objects and it was chosen so to describe foreground stars), and with pairs of sets of templates combining the above initial values. The best fit was chosen based on the chi-squared values after penalising by a factor 1.1 fits with two sets of templates. The latter factor was manually selected after extensively testing which value was required not to use two templates in fitting spectra where a visual inspection clearly indicated the presence of a single object.

A necessary input of the Jeans modelling (Sect.~\ref{jam}) is a $V_{\rm rms}\equiv\sqrt{V^2+\sigma^2}$ map, where $V$ is the velocity corrected for the mean recession velocity of the galaxy $v_{\rm rec}$ ($V\equiv v-v_{\rm rec}$). The recession velocity was obtained by assuming a symmetric rotation curve and by finding the average of the peak velocities of the receding and the approaching sides. We obtained a recession velocity of $v_{\rm rec}=5097\,{\rm km\,s^{-1}}$ for NGC~1277 and $v_{\rm rec}=6072\,{\rm km\,s^{-1}}$ for NGC~1278.

\begin{figure*}
\begin{center}
  \includegraphics[scale=0.48]{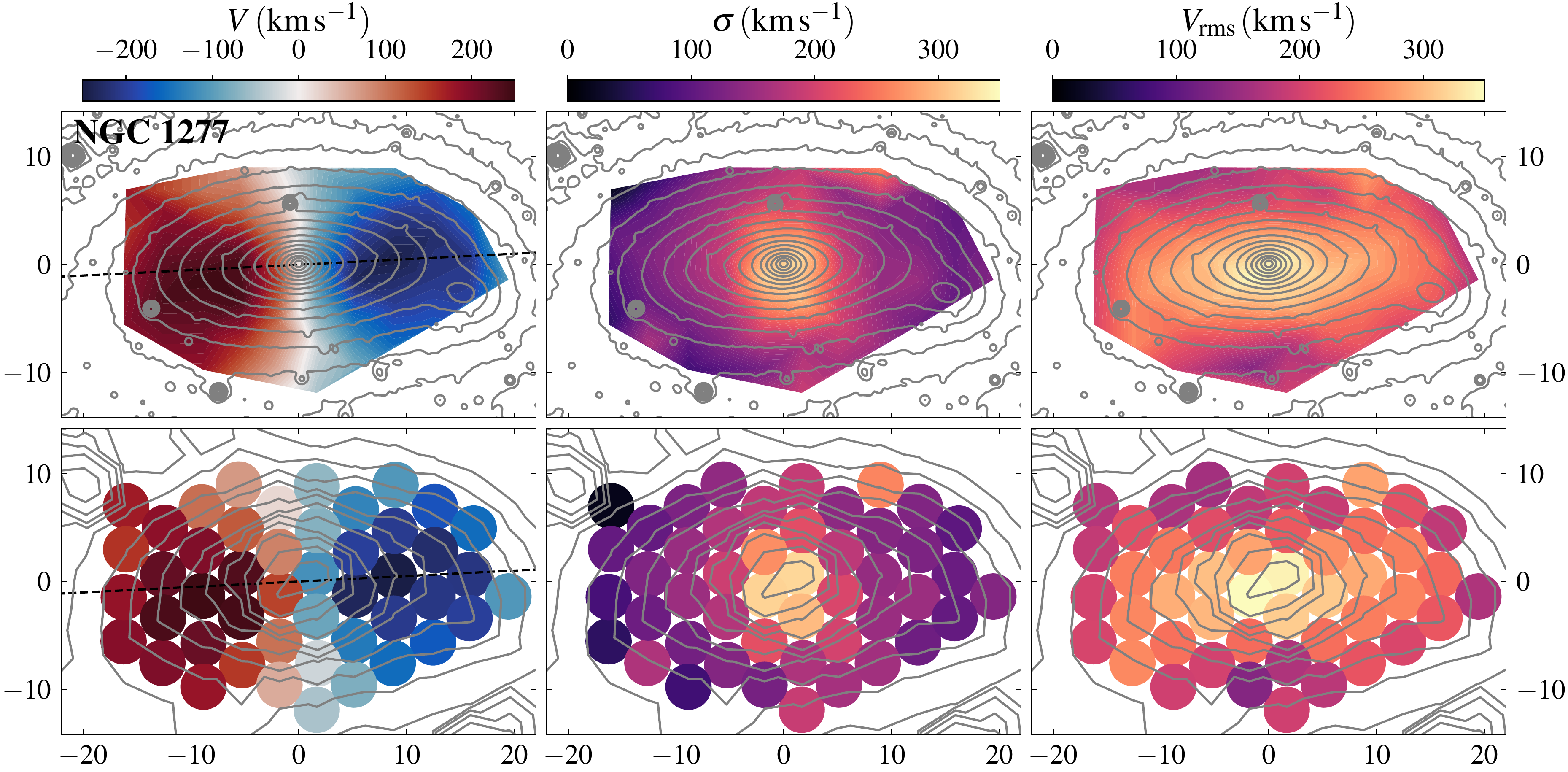}\\
  \end{center}
  \caption{\label{NGC1277_kinmap} Kinematic maps of NGC~1277. From {\it left} to {\it right}, $V$, $\sigma$, and $V_{\rm rms}$ maps of NGC~1277. The dashed line in the {\it left} panel indicates the kinematic major axis determined with \texttt{kinemetry}. The {\it top} row shows an interpolated version of the maps, whereas the {\it lower} one shows the individual fibre values. The grey contours denote isophotes separated by $0.5\,{\rm mag\,arcsec^{-2}}$ intervals obtained from an {\it HST} F625W image  ({\it top} row) and from the integral field data integrated between 6150\,\AA\ and 6650\,\AA\ ({\it lower} row). The axes are in units of arcseconds.}
\end{figure*}

\begin{figure*}
\begin{center}
  \includegraphics[scale=0.48]{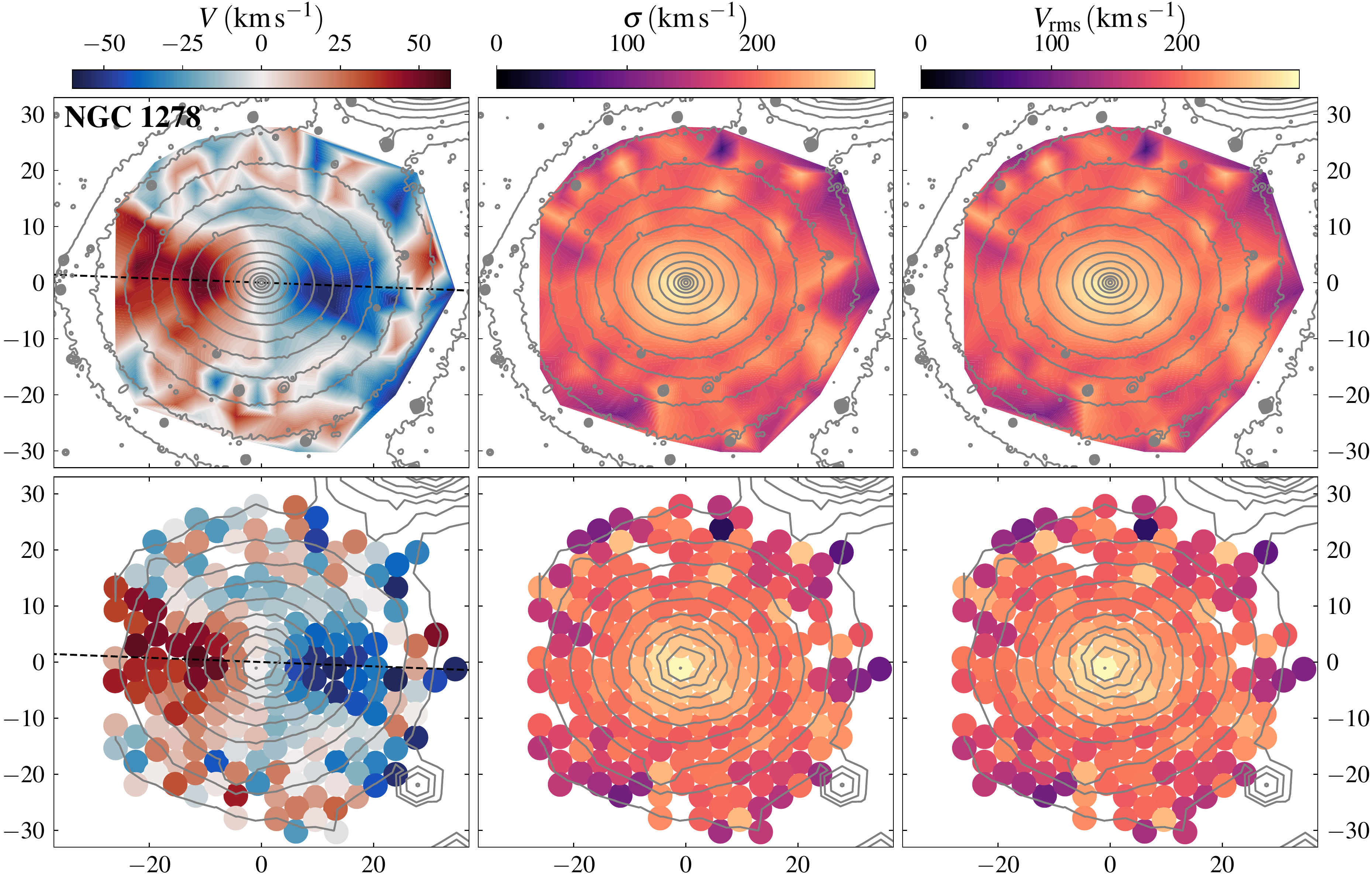}\\
  \end{center}
  \caption{\label{NGC1278_kinmap} Same as Fig.~\ref{NGC1277_kinmap}, but for NGC~1278. In order to highlight the kinematic structures, the colour scales differ from those in Fig.~\ref{NGC1277_kinmap}. In this case, the {\it top}-row isophotes are based on an {\it HST} F850LP image, which is the one used for the \texttt{MGE} modelling detailed in Sect.~\ref{mges}. Since for this galaxy $V\ll\sigma$, the $V_{\rm rms}$ map is virtually identical to the $\sigma$ map.}
\end{figure*}

Figures~\ref{NGC1277_kinmap} and \ref{NGC1278_kinmap} show the $V$, $\sigma$, and $V_{\rm rms}$ maps for NGC~1277 and NGC~1278, respectively. We are only plotting the fibres that were used for the Jeans modelling, that is those with high enough S/N (${\rm S/N}>7$ per spectral pixel), low kinematic errors ($\epsilon\left(V_{\rm rms}\right)<65\,{\rm km\,s^{-1}}$; formal errors as derived by \texttt{pPXF}), and for which \texttt{pPXF} has converged to reasonable $\sigma$ values ($10\,{\rm km\,s^{-1}}<\sigma <350\,{\rm km\,s^{-1}}$). Furthermore we only selected those with $V$ within $400\,{\rm km\,s^{-1}}$ of $v_{\rm rec}$ for NGC~1277 and within $60\,{\rm km\,s^{-1}}$ for NGC~1278. Those values are selected to be slightly larger than the amplitude of the rotation curve of the galaxies. Overall, the maps include 47 fibres for NGC~1277 and 183 for NGC~1278.  The coverage of NGC~1277 is roughly elliptical in shape, with a semi-major axis of $\sim18^{\prime\prime}$ (6\,kpc at the distance of NGC~1277, corresponding to five effective radii as measured in Sect.~\ref{mges}) and a semi-minor axis of $\sim10^{\prime\prime}$ (3.5\,kpc). The coverage of NGC~1278 is roughly circular, with a radius of $\sim30^{\prime\prime}$ (10\,kpc, or more than two effective radii).

Because of the large angular size of the fibres compared to the size of NGC~1277, we had to devise a method to pinpoint the centre of the galaxy in the CGMS data. Initially, we aligned the GCMS data of both galaxies simultaneously with an {\it HST} Advanced Camera for Surveys \citep[ACS;][]{Ford1998} F850LP image \citep[programme ID GO 14215;][]{Trujillo2015}. This was done by comparing the flux of the fibres selected for the Jeans modelling with that within circular apertures in the image with the same radius as the fibres. The {\it HST} image had previously been convolved with a Gaussian kernel with a sigma of one arcsecond so to match the seeing of the GCMS observations. The CGMS fluxes were calculated in the $6150\,{\rm \AA}-6650\,{\rm \AA}$ spectral range. We excluded bluer wavelengths in order to avoid a too large wavelength difference between the integral field and the {\it HST} data. Ideally, an {\it HST} image in a wavelength matching that of the CGMS data should have been used, but unfortunately the available ACS F625W image \citep[programme ID GO 10546;][]{Fabian2005} does not fully cover NGC~1278. The determination of the exact alignment was obtained by minimising
\begin{equation}
\label{centering}
 L=\sum_i\left(\frac{F_{\rm GCMS,i}-F_{HST,i}}{F_{\rm GCMS,i}}\right)^2,
\end{equation}
where $F_{\rm GCMS,i}$ and $F_{\it HST,i}$ denote the flux for the $i^{\rm th}$ bin in the CGMS fibre and the corresponding {\it HST} area, respectively. To deal with the differences in wavelength range between the two instruments and with possible flux calibration issues, we normalised $F_{\rm GCMS,i}$ by setting $\sum_iF_{\rm GCMS,i}=\sum_iF_{HST,i}$ before applying Eq.~\ref{centering}. The minimisation was performed with the adaptive \citet{Metropolis1953} for Bayesian analysis code \citep[\texttt{adamet}\footnote{We used v2.0.7 of the \texttt{Python} package \texttt{adamet}, available from \url{https://pypi.org/project/adamet/}.};][]{Cappellari2013}, which is an implementation of the algorithm by \citet{Haario2001}.

\begin{figure}
\begin{center}
  \includegraphics[scale=0.48]{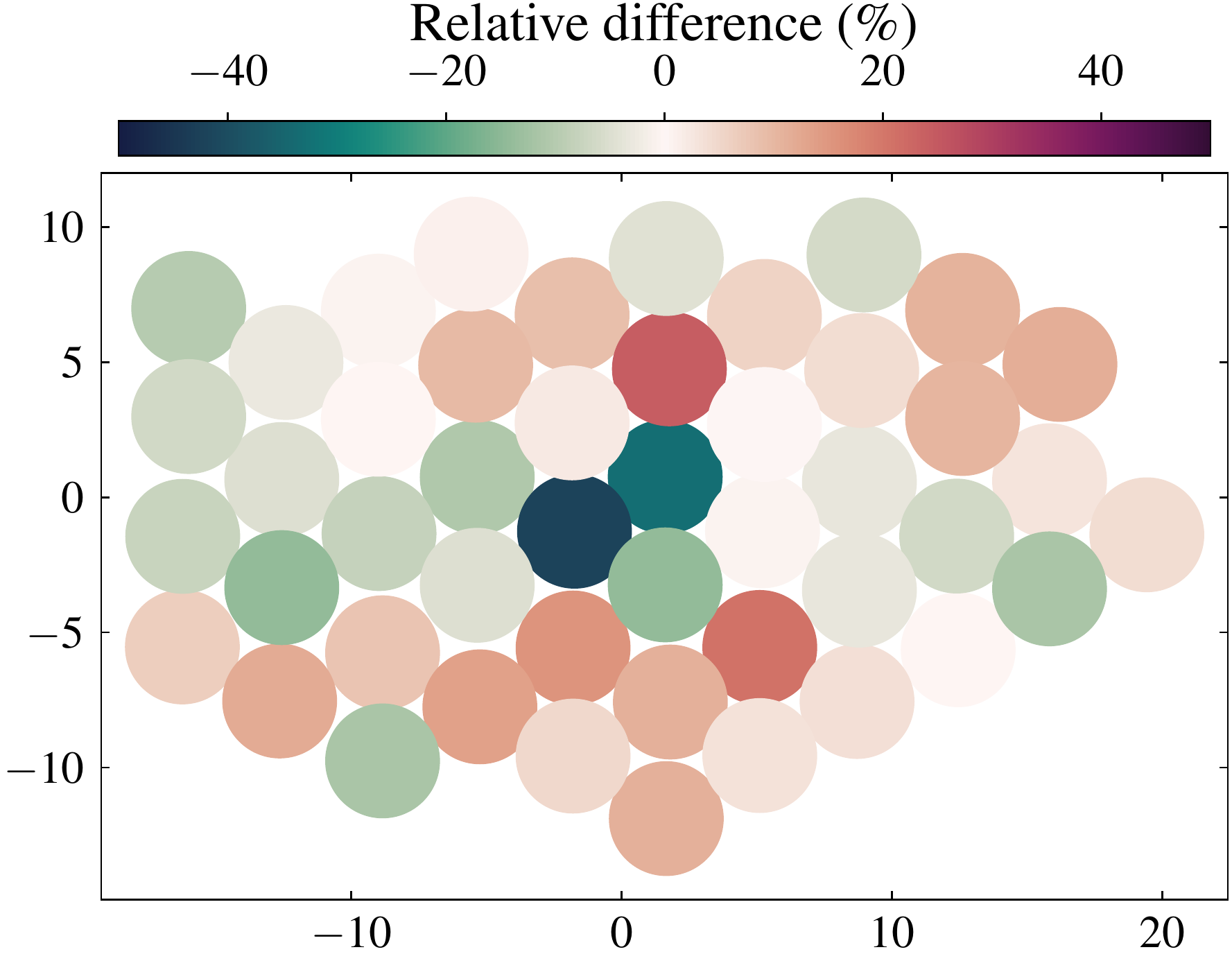}\\
  \end{center}
  \caption{\label{regNGC1277} Relative difference between the $6150\,\AA-6650\,\AA$ normalised flux of the GCMS fibres and the flux from matched apertures of an {\it HST} F625W image smoothed to mimic the seeing of the ground-based observations.}
\end{figure}

\begin{figure}
\begin{center}
  \includegraphics[scale=0.48]{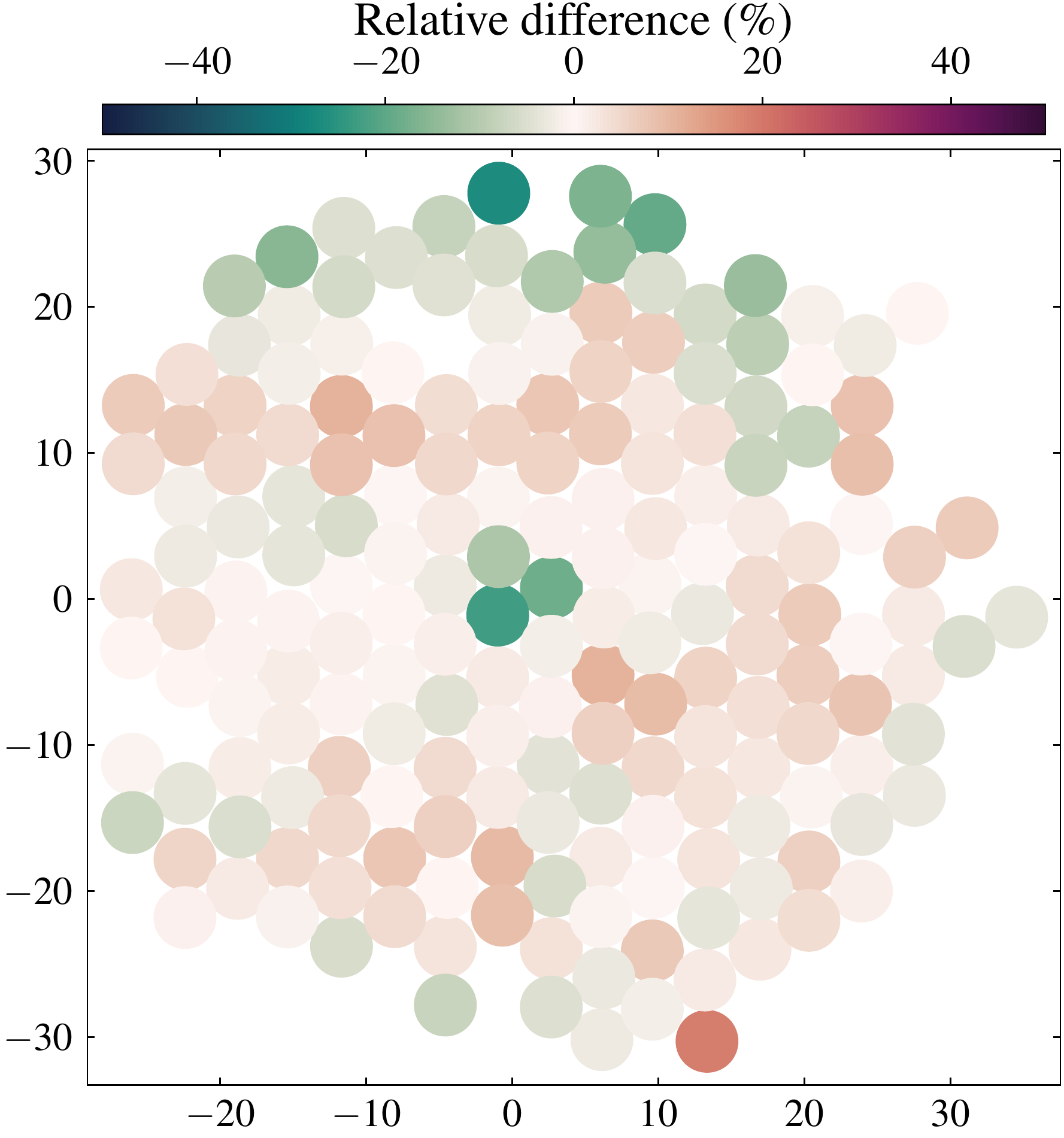}\\
  \end{center}
  \caption{\label{regNGC1278} Same as Fig.~\ref{regNGC1277}, but for NGC~1278 and a F850LP image.}
\end{figure}

The simultaneous registration of the two galaxies led to small systematics in the ratio $(F_{\rm GCMS,i}-F_{HST,i})/F_{\rm GCMS,i}$, indicating small inaccuracies in the coordinates of either the GCMS fibres or the {\it HST} pixels. As a consequence, we redid the registration by processing the two galaxies separately (using the F625W image for NGC~1277). Because a few fibres have a very different photometry from that of the matched {\it HST} apertures, we introduced a two sigma clipping to remove the outliers in $(F_{\rm GCMS,i}-F_{HST,i})/F_{\rm GCMS,i}$. The separate registration of the two galaxies resulted in centres that are $\approx0\farcs5$ and $\approx1\farcs0$ away from those obtained from a joint registration for NGC~1277 and NGC~1278, respectively. In Figs.~\ref{regNGC1277} and \ref{regNGC1278}, we show the relative differences of the GCMS fluxes and those derived from the matched apertures in the {\it HST} images. The residuals are generally at a level of $\sim10\%$ except at the very centre. We checked whether the latter discrepancies could be reduced by fine-tuning the seeing used for blurring the {\it HST} images, but failed to find a value leading to a significant improvement. This might either indicate that the Gaussian shape assumed for the point spread function (PSF) is too simplistic or a non-linearity in the behaviour of the GCMS.

As an alternative method for the GCMS data registration, we aligned four bright stars in the {\it HST} with their GCMS counterparts. The alignment differences between the two methods were of $\approx0\farcs2$ only for NGC~1277. We decided to use the former registration method, but we have checked that using the latter alignment does not significantly change any of the results obtained from the \texttt{JAM} fit (Sects.~\ref{modelNGC1277} and \ref{modelNGC1278}). Actually, according to our tests, the fitted parameters of NGC~1277 are robust to changes in the centre of up to at least $0\farcs5$.

\begin{figure*}
\begin{center}
  \includegraphics[scale=0.48]{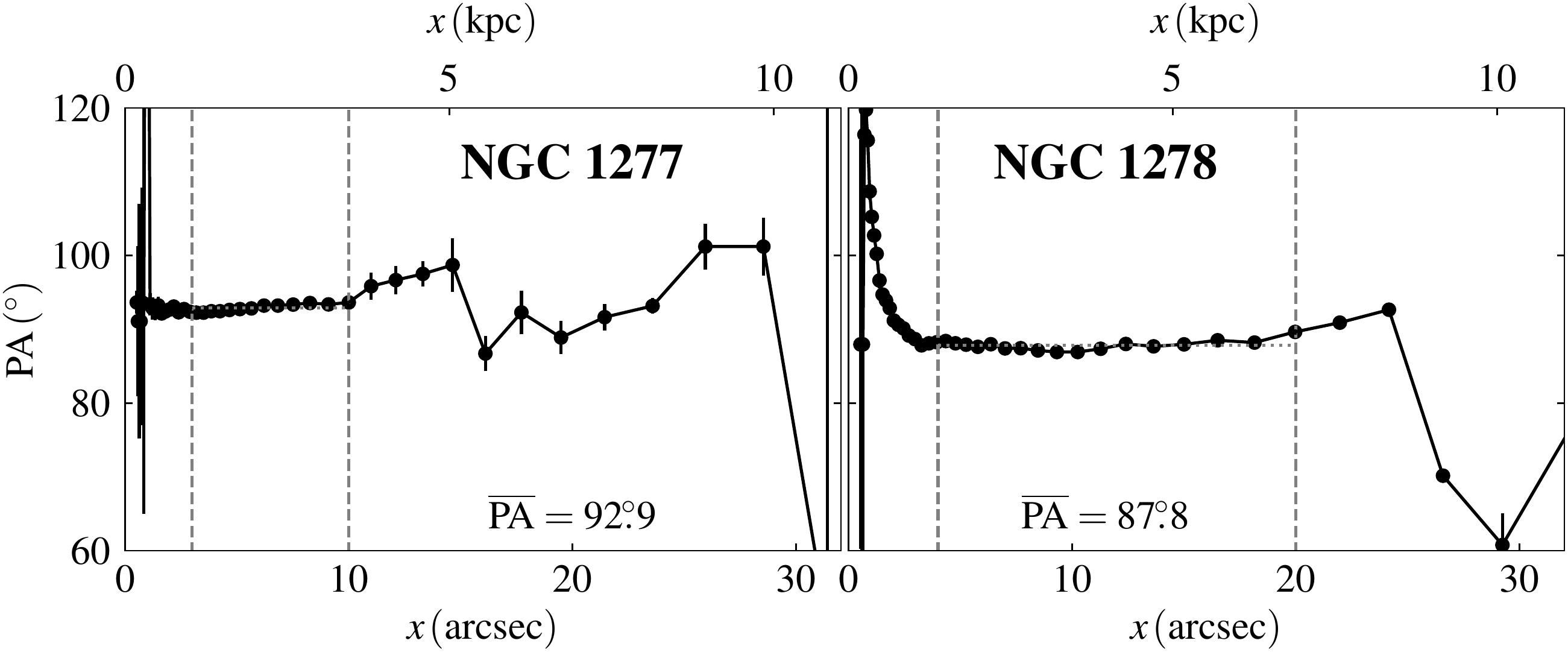}\\
  \end{center}
  \caption{\label{profiles} Position angle as a function of distance along the major axis for the ellipse fit of the {\it HST} imaging of NGC~1277 ({\it left}) and NGC~1278 ({\it right}). The error bars correspond to the one-sigma uncertainties. The vertical dashed lines indicate the radial interval over which the kinematical axis was determined. Its estimated value is indicated by the number in the lower part of the panels and the horizontal dotted lines.}
\end{figure*}

We determined the major axis of the galaxies photometrically by running the \texttt{isophote} routine from the \texttt{Photutils} package that uses the ellipse fitting method from \citet{Jedrzejewski1987} over the {\it HST} F625W image (for NGC~1277) and the F850LP image (for NGC~1278). Point sources, overlapping galaxies were manually masked. The {\it HST} images reveal that both galaxies have a tiny circumnuclear dust ring smaller than $1\arcsec$ in radius which is visually similar to some of those described in \citet{Lauer2005} and \citet{Comeron2010}. The ring in NGC~1277 has already been reported by \citet{Bosch2012} and \citet{Emsellem2013}. The near side of those rings was masked too. The major axis position angles were extracted by averaging the orientation of the tilted rings over radial ranges where it is roughly flat ($3\arcsec\leq R\leq10\arcsec$ for NGC~1277 and $4\arcsec\leq R\leq20\arcsec$ for NGC~1278). We find the position angles $\overline{{\rm PA}}=92\fdg9$ and $\overline{{\rm PA}}=87\fdg8$ for NGC~1277 and NGC~1278, respectively (Fig.~\ref{profiles}). Our determination of the position angle for NGC~1277 is close to the value of ${\rm PA}=92\fdg7$ reported in \citet{Bosch2012} and of ${\rm PA}=96\fdg0$ from HyperLeda\footnote{The HyperLeda database is hosted at \url{http://leda.univ-lyon1.fr/}.} \citep{Makarov2014}.

\begin{figure*}
\begin{center}
  \includegraphics[scale=0.48]{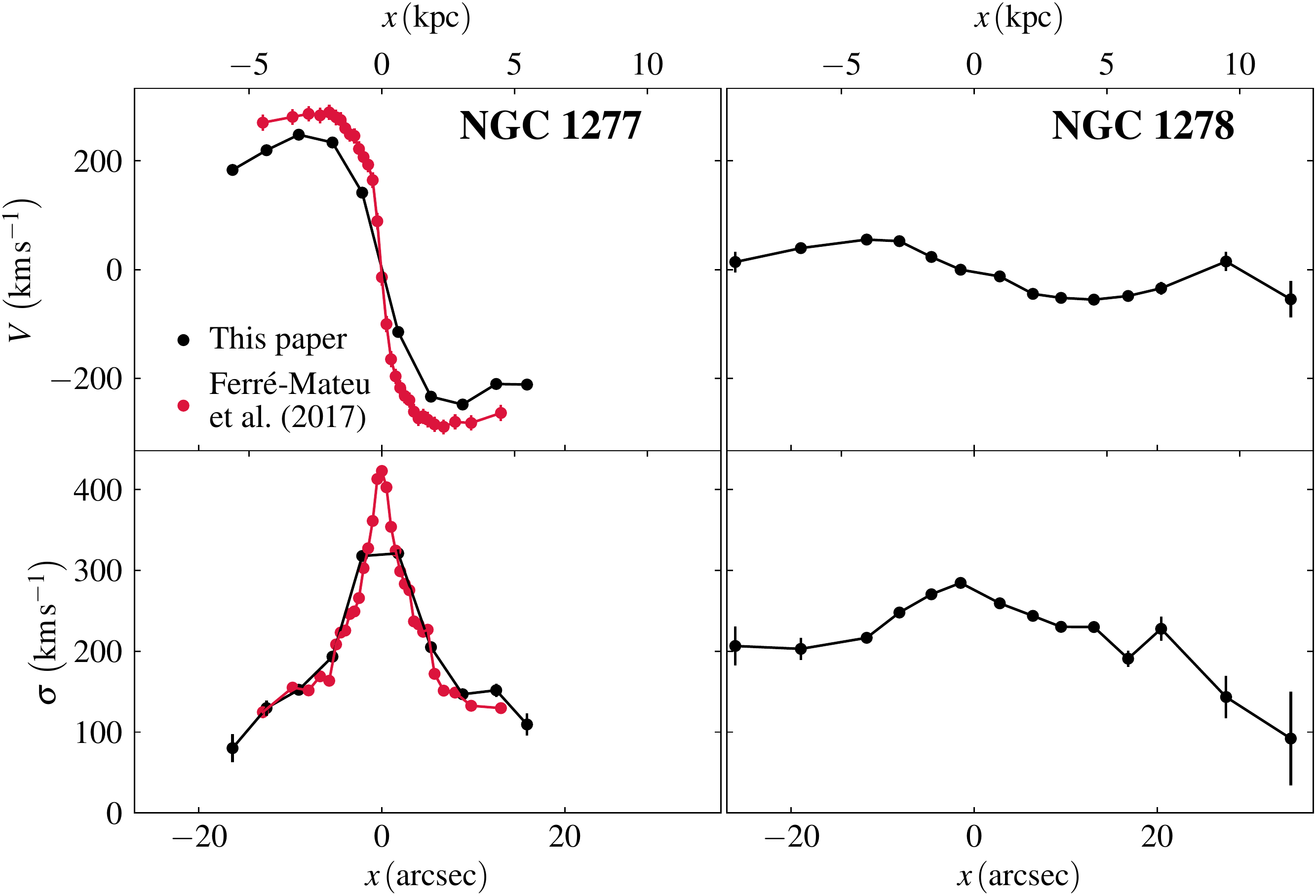}\\
  \end{center}
  \caption{\label{rotcurve} Velocity ({\it top}-row panels) and $\sigma$ ({\it lower}-row panels) profiles for NGC~1277 ({\it left}-column panels) and NGC~1278 ({\it right}-column panels) as measured along the kinematic major axes. The black data points denote our measurements obtained from fibres whose centre is within the fibre radius ($2\farcs08$) of the kinematic major axis. The data in crimson come from \citet{FerreMateu2017}.}
\end{figure*}

We constructed velocity and velocity dispersion profiles using the information from the fibres that are crossed by the kinematic major axis. Figure~\ref{rotcurve} displays the rotation and the velocity dispersion curves for NGC~1277 in black (left panels). The profiles for NGC~1278 are also shown for reference. As a sanity check to assess the quality of our kinematic determination we also plot the previous long-slit measurements of NGC~1277 from \citet{FerreMateu2017} in crimson (right panels). We find a smaller amplitude of the velocity curve than them. This is most likely because, while they explore a relatively narrow slit along the major axis of the galaxy \citep[$1\farcs0$ in width][]{MartinNavarro2015}, our resolution element is $4\farcs16$ in radius, so it gathers light from regions with a significant tangential motion of the stars. Our $\sigma$ profile agrees with that in \citet{FerreMateu2017} except for the central peak which we are missing, most likely due to beam smearing. In principle, a small increase in the sigma of the disc is to be expected in our data because within each fibre we are sampling separated regions with different velocities. However, because these differences in velocity are of the order of at most $\sim50\,{\rm km\,s^{-1}}$, they constitute an almost insignificant increase of up to $\sim10\,{\rm km\,s^{-1}}$ over a typical velocity dispersion of $150\,{\rm km\,s^{-1}}$.

The fact that {in Fig.~\ref{rotcurve}} both galaxies are shown with the same velocity and angular scales highlights the differences between the two. NGC~1277 is a compact fast rotator with a central $\sigma$ cusp, while NGC~1278 is an extended barely rotating object with a much more uniform $\sigma$.

\subsection{\texttt{MGE} models of {\it HST} images and mass maps}

\label{mges}

The density of the luminous tracers whose kinematics dominate the spectral range that we are studying (stars in our case) is a necessary input for \texttt{JAM}. Since we aim to obtain dark matter distributions, we also need to know the baryonic mass distributions to disentangle both kinds of matter.

To trace the luminosity of NGC~1277, we used an {\it HST} ACS F625W image with a Galactic extinction correction $A_{\rm F625W}=0.365\,{\rm mag}$ \citep{Schlafly2011} and also corrected for cosmological dimming. The filter was selected because it is within the spectral range from which we obtained the kinematics ($4400\,{\rm \AA}-6650\,{\rm \AA}$), so it is representative of the stellar population that dominates the stellar kinematics that we are tracing with our instrumental set-up. Unfortunately, NGC~1278 is only partially covered by the exposure. To model this galaxy, we instead used the ACS F850LP image with a correction $A_{\rm F850LP}=0.205\,{\rm mag}$ and corrected for cosmological dimming. This is not ideal, because this band falls outside the CGMS spectral range, and as a consequence does not exactly correspond to the stellar population of the luminous tracers. Because the F850LP image covers both galaxies, it is used as a proxy for the $z$ band in the colour maps (see below).

\begin{table}
\caption{Parameters of the \texttt{MGE} decompositions of the surface brightness maps of NGC~1277 and NGC~1278}
\label{MGElum} 
\centering                                      % used for centering table
\begin{tabular}{c c c}          % centered columns (4 columns)
\hline\hline                        % inserts double horizontal lines
${\rm log}\,I_0\,({\rm log}\,({\rm L}_{\odot}\,{\rm pc^{-2}}))$ & $\sigma_{\star}\,({\rm arcsec})$ & $q^{\prime}_{\star}$ \\    % table heading
\hline                                   
\multicolumn{3}{c}{NGC~1277 (F625W)}\\
\hline
5.119&0.019&0.700\\
4.618&0.114&0.700\\
4.282&0.297&0.700\\
3.771&0.724&0.700\\
3.693&0.732&0.435\\
3.517&1.227&0.700\\
2.913&2.085&0.700\\
2.987&2.836&0.413\\
2.700&5.043&0.400\\
2.345&6.887&0.502\\
1.767&9.974&0.400\\
1.219&14.197&0.400\\
1.631&14.197&0.700\\
\hline
\multicolumn{3}{c}{NGC~1278 (F850LP)}\\
\hline
4.658&0.025&1.000\\
4.184&0.173&1.000\\
3.681&0.479&1.000\\
3.748&0.797&0.834\\
3.378&1.659&0.786\\
3.053&2.757&0.857\\
2.740&3.391&0.750\\
2.673&5.727&0.750\\
2.403&8.461&0.952\\
2.456&9.875&0.750\\
1.331&22.483&0.750\\
2.034&22.483&1.000\\
\hline
\end{tabular}
\tablefoot{$I_0$ stands for the central surface brightnesses of the projected Gaussian components, $\sigma_{\star}$ denotes their dispersion, and $q^{\prime}_{\star}$ their projected axial ratio. An approximate conversion between the F850LP and F625W surface brightnesses can be obtained by dividing the NGC~1278 $I_0$ values by 1.7 (corresponding to $r-z=0.74$, and accounting for the different brightness of the Sun in the two bands). The images were corrected for cosmological dimming before producing the \texttt{MGE} decomposition.}
\end{table}

\begin{table}
\caption{Parameters of the \texttt{MGE} decompositions of the surface mass density maps of NGC~1277 and NGC~1278}
\label{MGEbar} 
\centering                                      % used for centering table
\begin{tabular}{c c c}          % centered columns (4 columns)
\hline\hline                        % inserts double horizontal lines
${\rm log}\,\Sigma_0\,({\rm log}\,({\rm M}_\odot\,{\rm pc^{-2}}))$ & $\sigma_{\star}\,({\rm arcsec})$ & $q^{\prime}_{\star}$ \\    % table heading
\hline                                   
\multicolumn{3}{c}{NGC~1277}\\
\hline
5.513&0.031&0.700\\
5.383&0.100&0.700\\
4.980&0.258&0.700\\
4.774&0.641&0.450\\
4.514&0.759&0.700\\
4.158&1.186&0.700\\
3.649&1.648&0.700\\
3.609&2.552&0.450\\
3.659&3.762&0.450\\
3.259&6.934&0.452\\
2.136&13.951&0.450\\
2.380&13.951&0.700\\
\hline
\multicolumn{3}{c}{NGC~1278}\\
\hline
4.547&0.212&1.000\\
4.345&0.517&1.000\\
4.100&0.923&0.804\\
3.772&1.737&0.815\\
3.675&3.005&0.800\\
3.178&5.477&0.800\\
2.216&8.144&1.000\\
3.131&9.200&0.800\\
1.796&22.529&0.800\\
2.495&22.529&1.000\\
\hline
\end{tabular}
\tablefoot{Same as Table~\ref{MGElum}, but with $\Sigma_0$ indicating the projected central surface mass densities of the Gaussian components. The images in which the mass maps are based were corrected for cosmological dimming.}
\end{table}

The mass maps for both galaxies were constructed by combining the above-mentioned F850LP frame and an extinction- and cosmological dimming-corrected F475W image ($A_{\rm F475W}=0.538$) from the same programme. Equating F475W to the $g$ band and F850LP to the $z$ band \citep{Ryon2022} and assuming a \citet{Chabrier2003} IMF, Table~A1 from \citet{Roediger2015} provides the following relation for the $g$-band mass-to-light ratio:
\begin{equation}
\label{mass1}
 {\rm log}\left(\Upsilon_{\star g}\right)=-1.132+1.116(g-z).
\end{equation}
We did not account for the PSF mismatch because both bands have a Gaussian kernel with a very similar FWHM ($0\farcs086$ and $0\farcs090$ for F475W and F850LP, respectively as estimated with \texttt{IRAF}'s \texttt{imexamine}). We applied the mass-to-light ratio map to the F475W image to obtain the mass surface density map. We considered an absolute solar magnitude of $M_{\odot g}=5.11\,{\rm mag}$ \citep{Willmer2018}. Because of the strong evidence for a bottom-heavy IMF in red nuggets \citep{MartinNavarro2015, FerreMateu2017}, which is consistent with the tendency for the densest galaxies to have the heaviest IMF \citep[Figure~12 in][]{Cappellari2013a}, the mass density map was corrected from a Chabrier to a \citet{Salpeter1955} IMF with the following recipe from \citet{Cimatti2008}
\begin{equation}
\label{mass2}
 {\rm log}\left(M_\star\right)({\rm Salpeter})={\rm log}\left(M_\star\right)({\rm Chabrier})+0.23.
\end{equation}
This recipe does not account for the mass of stellar remnants.

Given that the outer regions of the galaxies contain pixels with no well-defined colours due to negative fluxes introduced by random fluctuations, we fixed the colours beyond a certain radius before producing the mass maps. For NGC~1277 the colours were fixed to those within an elliptical corona oriented along the galaxy major axis. The parameters of the corona were an ellipticity of 0.6 (from our isophote fits in Sect.~\ref{kinematics}), an outer semi-major axis of $9\arcsec$, and a 0.9 times smaller inner semi-major axis. For NGC~1278 we used a circular corona with an outer radius of $15\arcsec$ and an inner radius of 0.9 times that.

Both the surface brightness and the mass density distributions were parametrised using an \texttt{MGE} \citep{Emsellem1994, Cappellari2002}. The fits were made with the \texttt{MgeFit}\footnote{We used v5.0.14 of the \texttt{MgeFit} \texttt{Python} package from \url{https://pypi.org/project/mgefit/}.} software by \citet{Cappellari2002}. This approach describes galaxies as a superposition of two-dimensional Gaussian components which, by knowing the inclination of the galaxy, can be deprojected into three-dimensional oblate axi-symmetric Gaussian components. We fixed the major axis of the \texttt{MGE} components to correspond to those of the kinematic axes measured in Sect.~\ref{kinematics}.

A detailed characterisation of the surface brightness and mass distributions requires knowledge about the PSFs. They were modelled using \texttt{Tiny Tim} \citep{Krist1995, Krist2011} and assuming that the targets sit at the centre of the chip. For the mass map, we adopted the PSF of the reddest filter used to build it, that is F850LP.

We modelled a region significantly larger than that covered by our good-quality integral field data, that is, an ellipse with ${\rm PA}=92\fdg9$ (from the kinematic major axis) and with semi-axes $a=640\,{\rm pixels}$ ($32\arcsec$) and $b=384\,{\rm pixels}$ ($19\farcs2$) for NGC~1277 (the ellipticity was again set to 0.6), and a circle with a radius $a=900\,{\rm pixels}$ ($45\arcsec$) for NGC~1278. We used the same mask as for the isophote fit in Sect.~\ref{kinematics}. As done by \citet{Emsellem2013}, and in order to avoid roundish \texttt{MGE} components for NGC~1277, we imposed a maximum projected axial ratio of 0.7 for the components in this galaxy.

\begin{figure}
\begin{center}
  \includegraphics[scale=0.48]{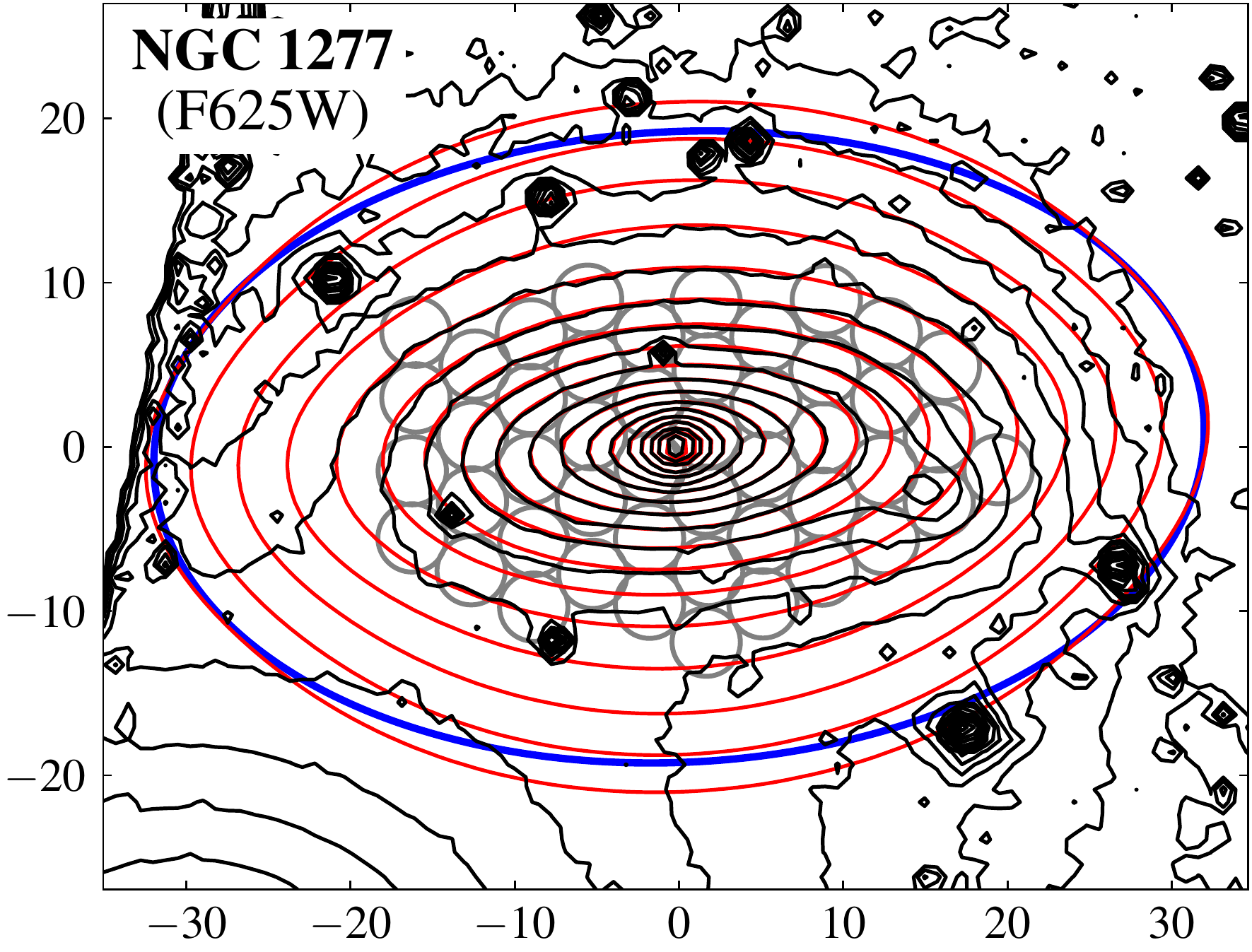}\\
  \end{center}
  \caption{\label{NGC1277_MGE} In black, isophote plot derived from the {\it HST} F625W image of NGC~1277 with $0.5\,{\rm mag\,arcsec^{-2}}$ intervals. The corresponding \texttt{MGE} model isophotes are indicated in red. The blue ellipse denotes the extent of the region that was used to produce the \texttt{MGE} models (an ellipse with a semi-major axis of $32\arcsec$, an ellipticity of 0.6, and a ${\rm PA}=92\fdg9$). The grey circles indicate the position of the fibres that were used for the dynamical analysis.}
\end{figure}

\begin{figure}
\begin{center}
  \includegraphics[scale=0.48]{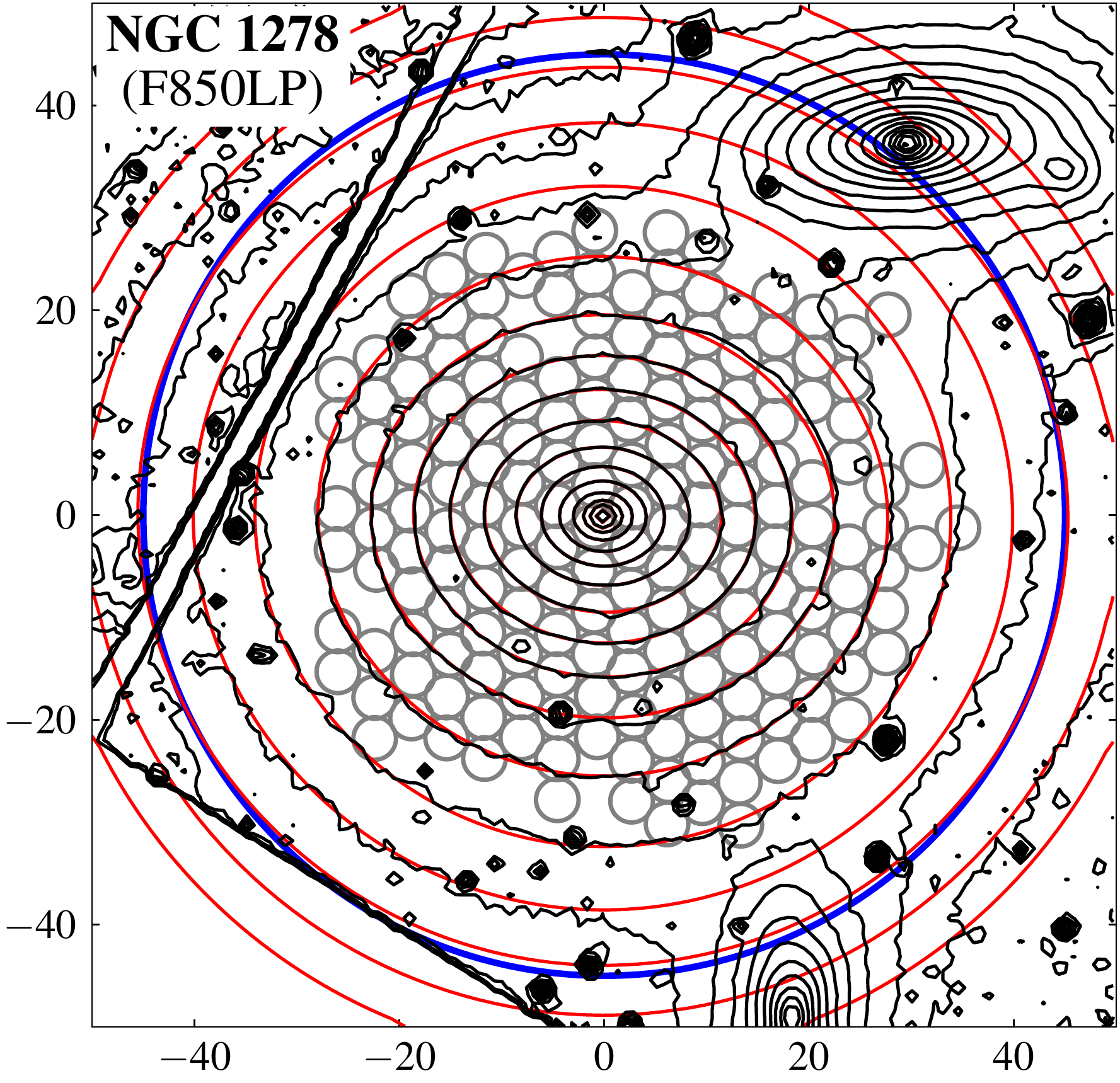}\\
  \end{center}
  \caption{\label{NGC1278_MGE} Same as Fig.~\ref{NGC1277_MGE}, but for NGC~1278 and the {\it HST} F850LP image. The fitted area (blue circle) is $45\arcsec$ in radius.}
\end{figure}

Tables~\ref{MGElum} and \ref{MGEbar} show the parameters of the \texttt{MGE} decompositions of the surface brightness and mass density maps of the two galaxies, respectively. The magnitude of the Sun in $r$ (F625W) and $z$ (F850LP) was assumed to be $M_{\odot r}=4.65\,{\rm mag}$ and $M_{\odot z}=4.50\,{\rm mag}$, respectively \citep{Willmer2018}. Figures~\ref{NGC1277_MGE} and \ref{NGC1278_MGE} allow a direct visual comparison of the isophotes of the original {\it HST} images (in black) and those from the model (in red).

\begin{figure*}
\begin{center}
  \includegraphics[scale=0.48]{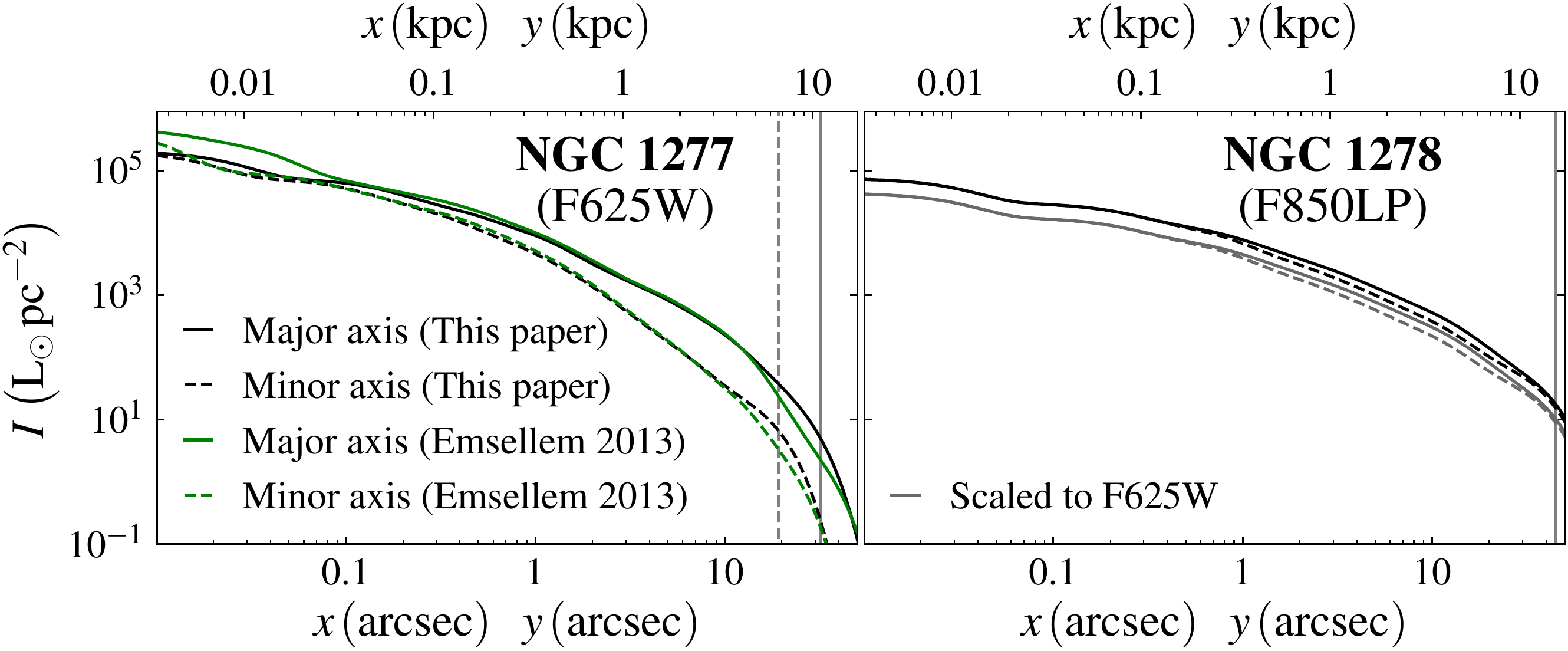}\\
  \end{center}
  \caption{\label{Surface_brightness_profiles} Major- (continuous line and coordinate $x$) and minor-axis (dashed line and coordinate $y$) surface brightness profiles of NGC~1277 ({\it left} panel; in F625W) and NGC~1278 ({\it right} panel; in F850LP). Black lines come from our \texttt{MGE} fit and green ones come from \citet{Emsellem2013} after a correction for extinction. The vertical grey lines indicate the maximum extent of the fitting area along the major (continuous) and the minor axis (dashed). For NGC~1278 the minor and major axis surface brightness profiles are nearly undistinguishable. The grey curves in the {\it right} panel correspond to the F625W surface brightness profiles calculated under the assumption that $r-z=0.74$ (see text).}
\end{figure*}

As a sanity check, we compared our \texttt{MGE} major- and minor-axis surface brightness profiles for NGC~1277 to those in \citet{Emsellem2013}. We corrected the latter for Galactic extinction to be directly comparable with ours. The profiles were computed following Eq.~1 in \citet{Cappellari2002}
\begin{equation}
\label{intensity}
 I(x,y)=\sum_jI_{0j}\,{\rm exp}\left[-\frac{1}{2\sigma_{{\star}j}^2}\left(x^2+\frac{y^2}{q^{\prime}_{{\star} j}}\right)\right],
\end{equation}
where the sum is made over all the Gaussian components $j$, $x$ and $y$ are coordinates in the plane of the sky (in the directions of the major and the minor axis of the galaxy, respectively), and $I_{0j}$, $\sigma_{{\star}j}$, and $q^{\prime}_{{\star}j}$ are the central surface brightness, the dispersion (size), and the projected axial ratio of the \texttt{MGE} components. The results are shown in Fig.~\ref{Surface_brightness_profiles}.

\begin{figure*}
\begin{center}
  \includegraphics[scale=0.48]{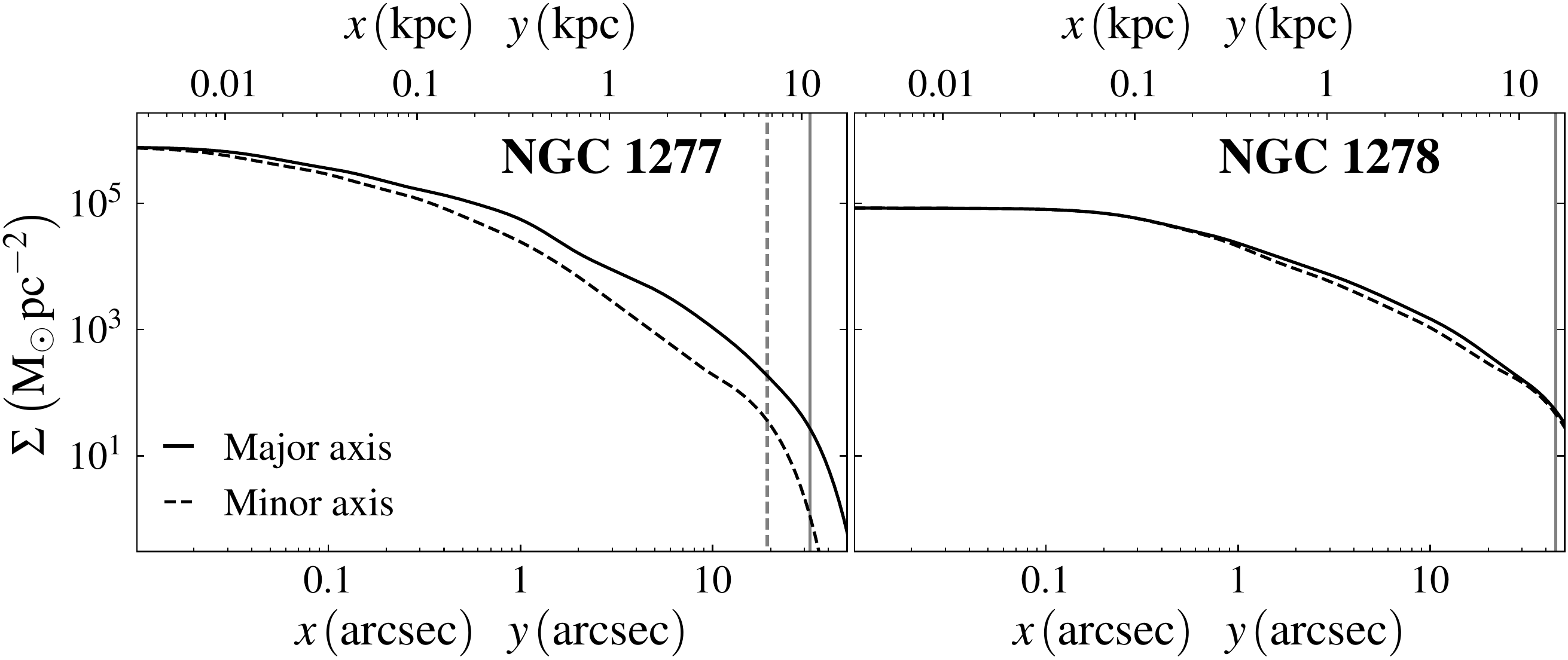}\\
  \end{center}
  \caption{\label{Surface_density_profiles} Major- (continuous line and coordinate $x$) and minor-axis (dashed line and coordinate $y$) stellar surface density profiles of NGC~1277 ({\it left} panel) and NGC~1278 ({\it right} panel). The vertical grey lines are as in Fig.~\ref{Surface_brightness_profiles}. The surface density estimates were calculated by deriving pixel-to-pixel mass-to-light ratios from colours following the prescriptions in \citet{Roediger2015}, and assuming a Salpeter IMF.}
\end{figure*}

Overall, the agreement between our parametrisation of the surface brightness map of NGC~1277 and that from \citet{Emsellem2013} is rather good. The apparently large difference within the central $0\farcs1$ only affects the central pixel. The other small discrepancies within  the central $1\arcsec$ can be attributed to differences in the handling of the small nuclear dust ring. Some large differences are seen at the edges of the fitting range, where the surface brightness profiles computed from our fit can be, at very specific radii, a factor of two or more above that obtained from \citet{Emsellem2013}. This occurs at a surface brightness of $\sim25\,{\rm mag\,arcsec^{-2}}$ and might be a consequence of small differences in sky subtraction, masking strategies, or the extent of the fitted area.

The total F625W luminosity of our NGC~1277 model can be calculated as \citep[][Eq.~10]{Cappellari2013}
\begin{equation}
\label{totallum}
 L=\sum_jL_j=\sum_j2\pi I_{0j}\sigma_{{\star}j}^2q_{{\star}j}^{\prime},
\end{equation}
where $L_j$ is the total luminosity of each of the individual Gaussian components, resulting in $L=2.6\times10^{10}\,{\rm L}_{\odot r}$. The luminosity estimated from the \citet{Emsellem2013} parametrisation (after the extinction correction) is $L=2.6\times10^{10}\,{\rm L}_{\odot r}$ too when using the same cosmology as us.

Our models show that NGC~1278, as expected, is more extended than NGC~1277. The central surface brightness of NGC~1278 is smaller than in NGC~1277, again indicating that the latter is a much more centrally concentrated object. The total F850LP luminosity of our NGC~1278 model is $L=8.7\times10^{10}\,{\rm L}_{\odot z}$. Given $r-z=0.74$ ($r-z=0.90$, when not accounting for differential extinction) as measured from a circular aperture $15^{\prime\prime}$ in radius centred in NGC~1278 in SDSS DR16 images \citep{Ahumada2020}, this corresponds to $L\approx5.8\times10^{10}\,{\rm L}_{\odot r}$, which is roughly two times the luminosity of NGC~1277 at the same band.

The circularised luminosity integrated within a cylinder of radius $R$ of an \texttt{MGE} model can be written as \citep[][Eq.~11]{Cappellari2013}
\begin{equation}
\label{cumulative}
L(R)=\sum_{j}L_j\left[1-{\rm exp}\,\left(-\frac{R^2}{2\sigma_{{\star}j}^2q_{{\star}j}^{\prime}}\right)\right].
\end{equation}
This expression can be used to find the effective radius of the galaxies by setting $L(R_{\rm e})=L/2$. For NGC~1277 we obtain $R_{\rm e}=3\farcs4$ (corresponding to 1.2\,kpc) and for NGC~1278 we find $R_{\rm e}=12\farcs3$ (4.2\,kpc). These values highlight again the extreme compactness of NGC~1277 when compared to a regular massive ETG. Our determination of the effective radius of NGC~1277 is comparable with the values of $R_{\rm e}=1.6\,{\rm kpc}$ obtained by \citet[][assumed distance $d=73\,{\rm Mpc}$]{Bosch2012}, $R_{\rm e}=\left(1.2\pm0.1\right)\,{\rm kpc}$ by \citet{FerreMateu2017}, and $R_{\rm e}=\left(1.3\pm0.1\right)\,{\rm kpc}$ by \citet[][$d=71\,{\rm Mpc}$]{Yildirim2017}. Because the two galaxies differ in luminosity by a factor of roughly two, it might be convenient to describe their compactness in terms of their effective radius relative to that of a given isophote. We chose $R_{23}$, the circularised radius of the $23\,{\rm mag\,arcsec^{-2}}$ extinction- and cosmological dimming-corrected F625W isophote. The circularised surface brightness profile was computed by setting $q^{\prime}_{\star j}\rightarrow1$ and $\sigma_{\star j}\rightarrow\sigma_{\star j}\sqrt{q^{\prime}_{\star j}}$ in Eq.~\ref{intensity} as suggested in \citet{Cappellari2009}. We obtained $R_{\rm 23}/R_{\rm e}=5.1$ and $R_{23}/R_{\rm e}=2.9$ for NGC~1277 and NGC~1278, respectively.

\setcitestyle{notesep={ }}

Using the data in Table~\ref{MGEbar} we derived the surface density profiles (Fig.~\ref{Surface_density_profiles}), which we found to follow rather closely the surface brightness profiles except for the inner arcsecond, where the colours differ from the rest of the galaxy due to the presence of dust rings. These profiles have to be taken with some care, since they are made under the assumption of a Salpeter IMF. Different IMFs might result in significant $\Upsilon_\star$ changes (see Sects.~\ref{jam} and \ref{imf} for a dynamical determination of $\Upsilon_\star$). Under this IMF assumption we use an expression analogous to Eq.~\ref{totallum},
\begin{equation}
 M_\star=\sum_jM_{\star j}=\sum_j2\pi \Sigma_{0j}\sigma_{{\star}j}^2q_{{\star}j}^{\prime},
\end{equation}
to find a total stellar mass of $M_\star=1.3\times10^{11}\,{\rm M}_\odot$ for NGC~1277 \citep[compatible with the value of $M_\star=\left(1.3\pm0.2\right)\times10^{11}\,{\rm M}_\odot$ in][for a distance $d=71\,{\rm Mpc}$]{Yildirim2017} and of $M_\star=2.9\times10^{11}\,{\rm M}_\odot$ for NGC~1278.

\setcitestyle{notesep={, }}

\section{\texttt{JAM} dynamical modelling}

\label{jam}

We used the Jeans Anisotropic Modelling (\texttt{JAM}\footnote{We used v6.3.2 of the \texttt{Python} package \texttt{JamPy}, available from \url{https://pypi.org/project/jampy/}.}) code to produce dynamical models of the galaxies. A basic assumption of \texttt{JAM} is that the galaxies are axisymmetric. Given \texttt{MGE} models of the luminous tracer and mass distributions, a galaxy inclination ($i$), and an anisotropy parameter ($\beta_z$ or $\beta_r$) that can be different for each \texttt{MGE} component, \texttt{JAM} computes the projected first and second moments of the velocity distribution of the dynamical tracers (stars in our case). \texttt{JAM} solves the Jeans equations to describe the motions of star in a state of hydrodynamical equilibrium \citep{Jeans1922} by assuming either a cylindrically \citep{Cappellari2008} or a spherically aligned velocity ellipsoid \citep{Cappellari2020}. The definition of the anisotropy parameter varies depending on the chosen alignment. In the cylindrically aligned case, with coordinates $(R,z,\theta)$, the anisotropy parameter is defined as $\beta_z\equiv1-\overline{V_z^2}/\overline{V_R^2}$, whereas for the spherically aligned case with coordinates $(r,\theta,\phi)$, $\beta_r\equiv1-\overline{V_\theta^2}/\overline{V_r^2}$ ($V_{z}$, $V_{R}$, $V_{\theta}$, and $V_{\rm r}$ stand for the velocities along the $z$, $R$, $\theta$, and $r$ coordinates, respectively). We fitted the \texttt{JAM} models to the $V_{\rm rms}$ values obtained from the GCMS and NIFS observations using \texttt{adamet}. Before running the fits, we rotated the $V_{\rm rms}$ maps so that the major axis of the galaxy determined in Sect.~\ref{kinematics} lied horizontally.

Tests of \texttt{JAM} using high-resolution $N$-body simulations \citep{Lablanche2012} and lower-resolution cosmological hydrodynamical simulations \citep{Li2016} have confirmed its good accuracy when using high-S/N data, and negligible bias in reproducing the total mass profiles. More recently, \texttt{JAM} was compared in detail against the \citet{Schwarzschild1979} method using samples of both observed galaxies, with circular velocities from interferometric observations of the CO gas, and numerical simulations respectively. These studies consistently found that the \texttt{JAM} method produces even more accurate (smaller scatter versus the true values) density profiles \citep[Figure~8 in][]{Leung2018} and enclosed masses \citep[Figure~4 in][]{Jin2019} than the more general Schwarzschild models. This may be due to the \texttt{JAM} assumption acting as an empirically motivated prior and reducing the degeneracies of the dynamical inversion. These extensive tests justify our choice of \texttt{JAM} for this study.

In general, our density models introduced into \texttt{adamet} include the dark matter halo, the stellar component, and a central mass to account for the supermassive black hole. The latter is introduced as a minute \texttt{MGE} component with a Gaussian sigma of $0\farcs01$.

We assumed spherical dark matter haloes described by a Navarro-Frenk-White profile \citep[NFW profile]{Navarro1996, Navarro1997}
\begin{equation}
\label{nfw}
 \rho_{\rm DM}=\frac{\rho_{\rm s}}{\frac{r}{r_{\rm s}}\left(1+\frac{r}{r_{\rm s}}\right)^2},
\end{equation}
where $r$ is the three-dimensional radius, $r_{\rm s}$ is the break radius, and $\rho_{\rm s}$ is a normalisation density. In order to reduce the number of free parameters, we set $r_{\rm s}=100\,{\rm kpc}$, which is a reasonable estimate for massive elliptical galaxies calculated from the stellar-to-halo mass relation from \citet{Girelli2020} and the halo mass-concentration relations from \citet{Klypin2011} and \citet[][see Sect.~\ref{dmhs} for further details]{Child2018}. As discussed in \citet{Cappellari2013}, the choice of a break radius is irrelevant because simulations systematically predict them to be found well outside the outermost regions sampled by the dynamical tracers. To include the  dark matter halo in our \texttt{JAM} models, we fitted the analytic form of the halo with the \texttt{mge\_fit\_1d} routine in the \texttt{MgeFit} package, which produces an accurate \texttt{MGE} approximation. The dark matter halo Gaussian components were characterised by their mass $M_{{\rm DM}\,j}$ and dispersion $\sigma_{{\rm DM}\,j}$.

The three-dimensional stellar mass distribution fed into \texttt{JAM} was that from the \texttt{MGE} stellar mass density models (Sect.~\ref{mges}). The intrinsic axial ratio of the Gaussian components was obtained by assuming an inclination angle $i$ \citep[][Eq.~14]{Cappellari2008}
\begin{equation}
\label{deproj}
 q_{{\star}j}=\frac{\sqrt{q_{{\star}j}^{\prime2}-{\rm cos}^2\,i}}{{\rm sin}\,i}.
\end{equation}
Once the intrinsic axial ratios of the components were computed, the volume density of stars was described as
\begin{equation}
 \rho_\star(R,z)=\alpha\sum_j\frac{M_{\star j}}{\left(\sqrt{2\pi}\sigma_{{\star}j}\right)^3q_{\star j}}\,{\rm exp}\,\left[-\frac{1}{2\sigma_{{\star}j}^2}\left(R^2+\frac{z^2}{q_{{\star}j}^2}\right)\right],
\end{equation}
where $R$ is the projection of the radius in the mid-plane of the galaxy, and where the formula from \citet{Cappellari2008} is scaled with the so-called mismatch parameter \citep{Cappellari2012}, which is defined as
\begin{equation}
 \alpha\equiv\frac{\Upsilon_\star}{\Upsilon_{\rm Salpeter}}=\frac{M_\star/L}{\left(M_\star/L\right)_{\rm Salpeter}}=\frac{M_\star}{\left(M_\star\right)_{\rm Salpeter}}.
\end{equation}
The mismatch parameter describes the stellar mass-to-light ratio with respect to that expected from a Salpeter IMF.

In practice, we fitted up to five parameters by feeding \texttt{JAM} models into \texttt{adamet}. These parameters are:
\begin{enumerate}[1)]
 \item The intrinsic axial ratio of the flattest \texttt{MGE} component of the stellar model, $q_{\star\,{\rm min}}$. It is related to the projected axial ratio $q_{\star\,{\rm min}}^{\prime}$ and the inclination $i$ through Eq.~\ref{deproj}. Fitting this parameter is equivalent to fit the inclination $i$.
 \item The shape of the velocity ellipsoid, parametrised by either $\beta_z$ or $\beta_r$.
 \item The mismatch parameter, that accounts for deviations from the Salpeter IMF that was assumed while constructing the stellar mass maps (Sect.~\ref{mges}). Finding this parameter is equivalent to  a dynamical determination of the stellar mass-to-light ratio.
 \item The black hole mass, $M_{\rm BH}$.
 \item The dark matter halo density normalisation. We introduced it as $f_{\rm DM}(6\,{\rm kpc})$, the fraction of matter that is dark within the inner 6\,kpc of the galaxy. The mass within 6\,kpc was calculated as $M(6\,{\rm kpc})=M_\star(6\,{\rm kpc})+M_{\rm DM}(6\,{\rm kpc})+M_{\rm BH}$. The seemingly arbitrary 6\,kpc radius corresponds to the radial coverage of the GCMS data of NGC~1277. When computing $f_{\rm DM}(6\,{\rm kpc})$ we have counted the central black hole mass as baryonic mass.
\end{enumerate}

We adopted constant priors on all parameters during the \texttt{adamet} Bayesian modelling. For most of the parameters, we refined the bounds of the fitting range after running the fit several times and examining the posterior probability distributions. In other cases, the boundaries come from physical limitations. For example, $M_{\rm BH}$ and $f_{\rm DM}(6\,{\rm kpc})$ cannot be smaller than zero. Also, the intrinsic axial ratio of the flattest \texttt{MGE} component cannot be larger than the projected one, setting an upper boundary $q_{\star\,{\rm min}}=q^{\prime}_{\star\,{\rm min}}$. Finally, to avoid unphysically thin components, we assume a minimum possible axial ratio for the flattest component at $q_{\star\,{\rm min}}=0.05$.

In order to calculate the mass of spherical dark matter halo \texttt{MGE}s within a specific radius we used the following expression from \citet[][Eq.~49]{Cappellari2008}
\begin{equation}
 M_{\rm DM}(r)=\sum_jM_{{\rm DM}\,j}\left[{\rm erf}\,\left(\frac{r}{\sqrt{2}\sigma_{{\rm DM}\,j}}\right)-\frac{\sqrt{\frac{2}{\pi}}r}{\sigma_{{\rm DM}\,j}}{\rm exp}\,\left(-\frac{r^2}{2\sigma_{{\rm DM}\,j}^2}\right)\right],
\end{equation}
where erf stands for the error function. For the flattened stellar \texttt{MGE} components, the mass contained within a radius $r$ is given by the following expression adapted from \citet[][Eq.~15]{Cappellari2013}
\begin{equation}
 M_\star(r)=\alpha\sum_jM_{\star j}\left[{\rm erf}\,\left(h_{\star j}\right)-2h_{\star j}{\rm exp}\,\left(-h_{\star j}^2\right)/\sqrt{\pi}\right],
\end{equation}
where we accounted for the mismatch parameter, and where
\begin{equation}
h_{\star j}=\frac{r}{\sqrt{2}\sigma_{\star j}q_{\star j}^{1/3}}.
\end{equation}

The instrumental pixel size and the seeing are accounted for by \texttt{JAM}. Since the pixels considered by \texttt{JAM} are square, for GCMS we assumed a pixel size of $3\farcs7$, equivalent in surface to our $4\farcs16$ in diameter fibres. We also assumed a PSF sigma of one arcsecond. For the NIFS data the situation is more complicated, because the data points correspond to the centre of Voronoi bins with variable size. We set the pixel size to the native value for NIFS, $0\farcs05$, because the data for the central region of NGC~1277 (the most sensitive to the black hole dynamics) has enough signal to remain unbinned. We used the same NIFS PSF parametrisation as \citet[][Appendix~A]{Krajnovic2018}, that is an inner component with a Gaussian sigma $0\farcs05$ and weight 0.65 and an outer component with sigma $0\farcs40$ and a weight 0.35.

\subsection{Dynamical modelling of NGC~1277}

\label{modelNGC1277}

\begin{figure*}
\begin{center}
  \includegraphics[scale=0.48]{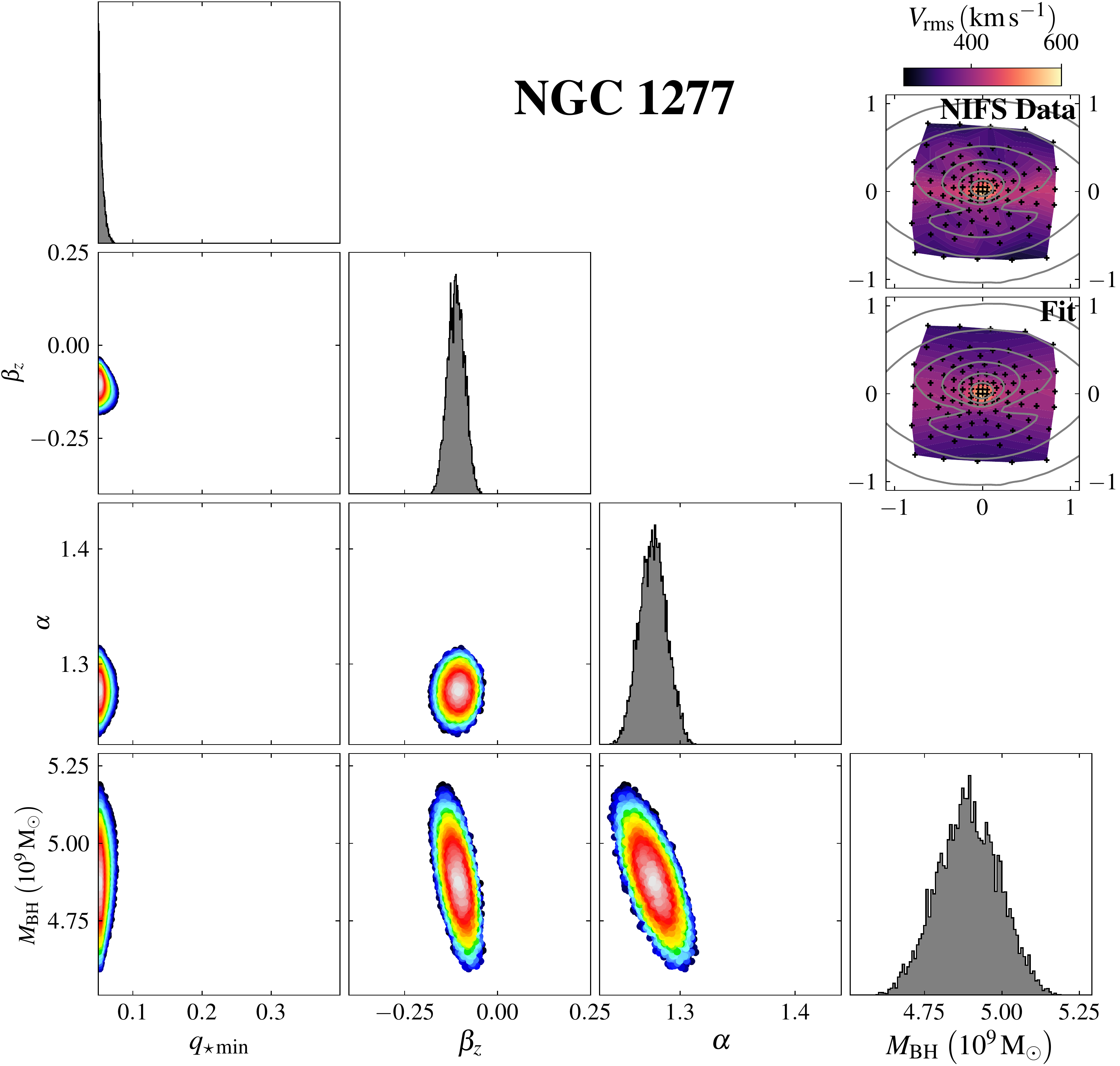}\\
  \end{center}
  \caption{\label{NGC1277_NIFS} Dynamical modelling of the NIFS $V_{\rm rms}$ map of NGC~1277 assuming a cylindrically aligned velocity ellipsoid (fit cA1). The corner plot shows the probability distribution of the fitted parameters marginalised over either two parameters for the contour plots, and over one for the histograms. In the contour plots blue regions correspond to the three-sigma confidence level and white regions denote maximum likelihood. The top-right corner shows the $V_{\rm rms}$ map and the fit, with grey isophotes indicating F625W contours separated by $0.5\,{\rm mag\,arcsec^{-2}}$ and the black crosses indicating the centre of the bins. The indentations in the isophotes are a clear sign of the presence of the dust ring, whose near side was masked while producing the \texttt{MGE} models. The fitted parameters were $q_{\star\,{\rm min}}$, $\beta_z$, $\alpha$, and $M_{\rm BH}$, whereas the dark matter fraction $f_{\rm DM}(6\,{\rm kpc})$ was fixed to zero.}
\end{figure*}

\begin{figure*}
\begin{center}
  \includegraphics[scale=0.48]{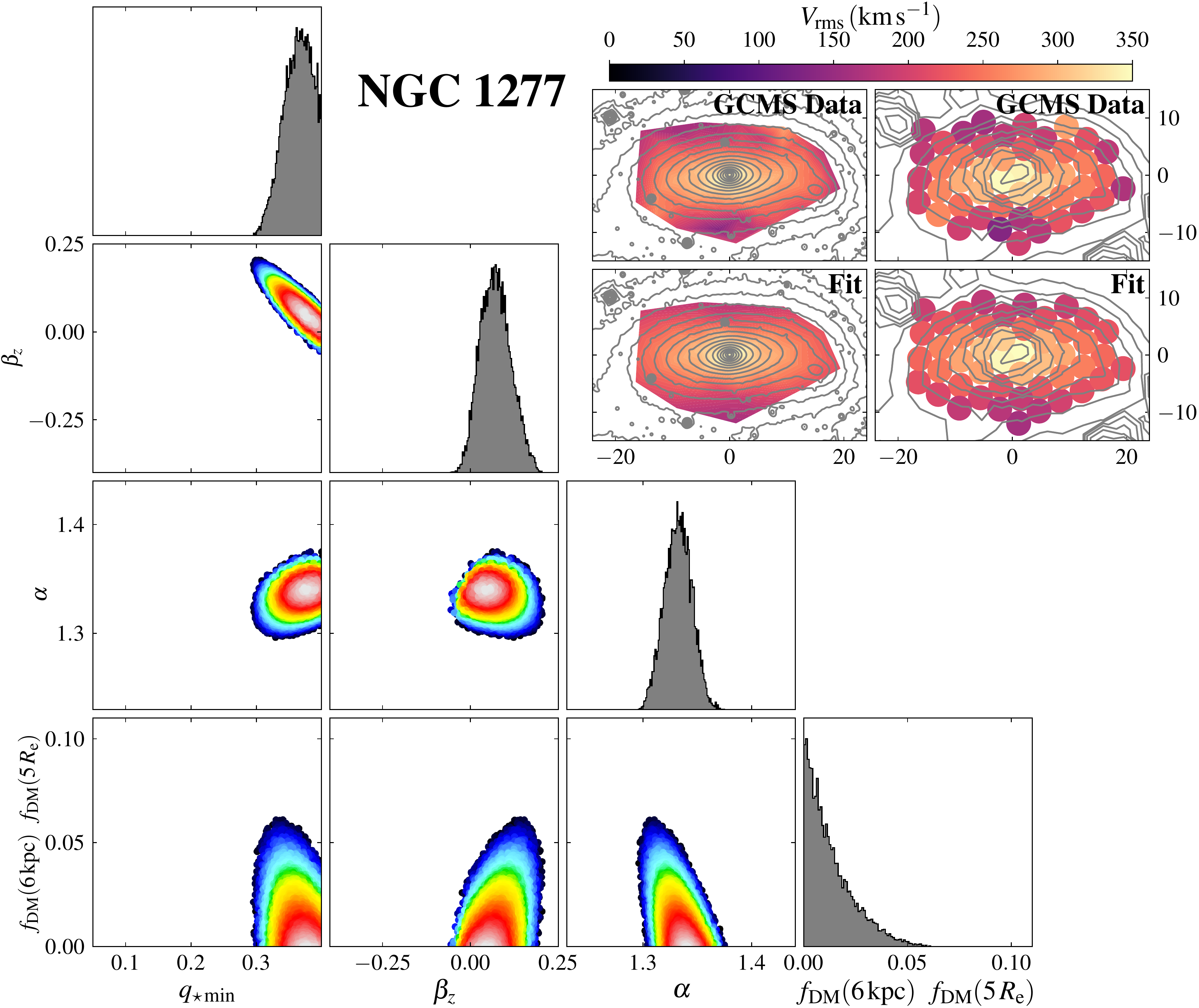}\\
  \end{center}
  \caption{\label{NGC1277_VIRUS} As Fig.~\ref{NGC1277_NIFS}, but this time the fitted $V_{\rm rms}$ map is that of the GCMS data, $f_{\rm DM}(6\,{\rm kpc})$ ($f_{\rm DM}(1.5\,R_{\rm e})$) is a free parameter and the central black hole mass is fixed to $M_{\rm BH}=4.70\times10^9\,{\rm M}_\odot$ (fit cA2). As in Fig.~\ref{NGC1277_kinmap}, the $V_{\rm rms}$ maps are displayed in their interpolated ({\it left}) and individual-fibre ({\it right}) forms. They surface brightness contours correspond to {\it HST} data for the interpolated maps and to photometry derived from the CGMS data for the non-interpolated ones.}
\end{figure*}

For NGC~1277 we have both GCMS and NIFS data. A particularity of the NIFS dataset is that the spaxel with the largest $\sigma$ does not correspond to the coordinate centre as provided by \citet{Walsh2016}. This small offset of $0\farcs02$ was corrected before running the fits.

\begin{figure}
\begin{center}
  \includegraphics[scale=0.48]{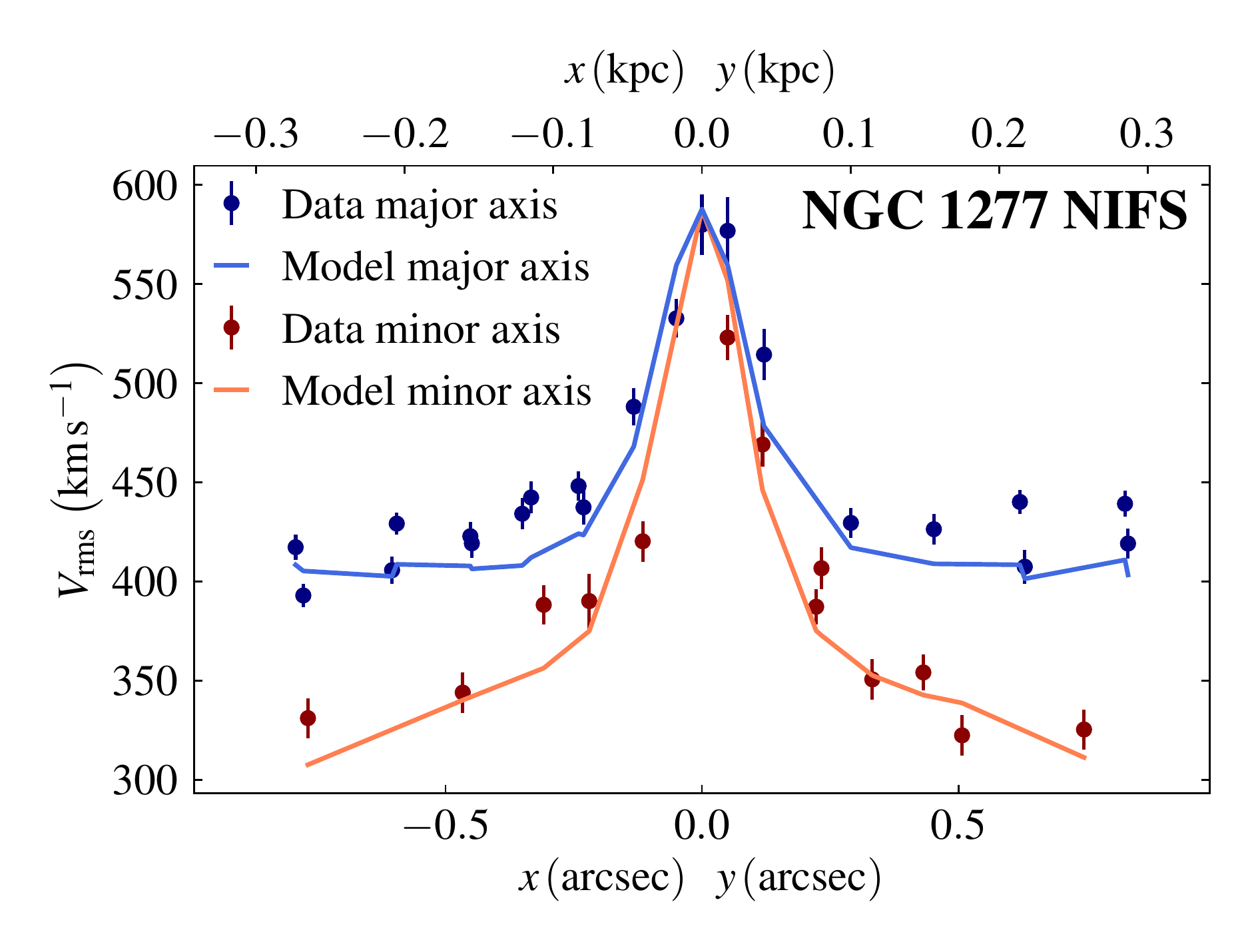}\\
  \end{center}
  \caption{\label{curve_NGC1277_NIFS} Major- (blue) and minor-axis (red) $V_{\rm rms}$ profiles for the NIFS data of NGC~1277 (symbols) and the cA1 fit (continuous line). The error bars represent one-sigma $V_{\rm rms}$ formal uncertainties as obtained by \texttt{pPXF} (see Sect.~\ref{kinematics}). We selected the bins whose centres are within a cone with an aperture of $30^{\circ}$ centred in the major (minor) axis.}
\end{figure}

\begin{figure}
\begin{center}
  \includegraphics[scale=0.48]{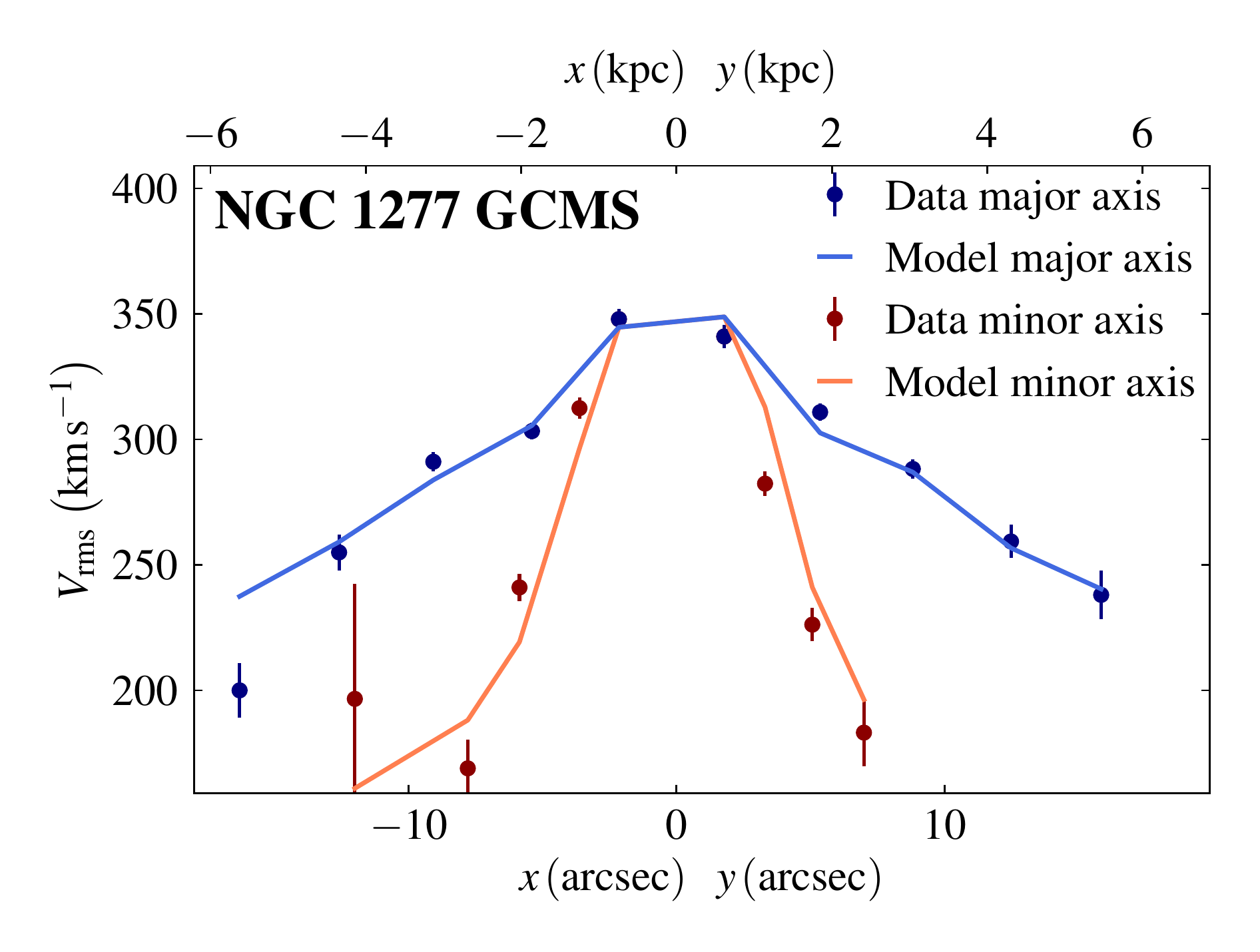}\\
  \end{center}
  \caption{\label{curve_NGC1277_VIRUS} Same as Fig.~\ref{curve_NGC1277_NIFS}, but this time for the GCMS data, the cA2 fit, and for fibres whose centre is within a fibre radius ($2.\!\!^{\prime\prime}08$) of the major and minor axes.}
\end{figure}

We tested three fitting approaches to check the internal consistency and gauge the uncertainties:
\setcitestyle{notesep={}}

\begin{enumerate}[A]
 \item \emph{Successively fit the NIFS and the GCMS data:} We first fitted the NIFS data (A1; Fig.~\ref{NGC1277_NIFS}). Since the field of view of the instrument only covers the inner 0.5\,kpc of NGC~1277, we assumed the dynamical effect of dark matter to be negligible and fixed $f_{\rm DM}(6\,{\rm kpc})=0$. Indeed, the dark matter fraction within 0.5\,kpc expected for a dark matter halo typical of a galaxy with the mass of NGC~1277 (see Sect.~\ref{dmhs}) is $\sim0.5\%$. Then, we fitted the GCMS data with the black hole mass $M_{\rm BH}$ fixed to the value obtained in A1 (A2; Fig.~\ref{NGC1277_VIRUS}). The rationale for this is that the black hole sphere of influence \citep{Peebles1972} is small compared to the field of view covered by the GCMS data ($\sim130\,{\rm pc}$ or $\sim0\farcs4$ for a $M_{\rm BH}=5\times10^9\,{\rm M}_\odot$ and for a central velocity dispersion $\sigma=400\,{\rm km\,s^{-1}}$) and we should not expect the latter to help constraining the black hole mass. When fitting for black hole masses with \texttt{JAM}, it is always recommended to restrict the field of view to the smallest region required to break the $M/L-$black hole degeneracy \citep[e.g.][]{Thater2022} not to risk biasing the black hole mass due to gradients in the velocity ellipsoid and the mismatch parameter, or due the choice of a parametrisation for the dark matter halo. In other words, a good global fit might be reached at the expense of a poor fit of the central region, resulting in a bad determination of the black hole mass. This approach is very different from the one usually employed with Schwarzschild models, which need large radii kinematics to constrain the orbital distribution and reduce degeneracies in the black hole mass.
 \item \emph{Simultaneous fit of the NIFS and the GCMS data:} Both sets of data were fitted simultaneously (plot not shown). We multiplied the uncertainties of $V_{\rm rms}$ in the NIFS dataset by a factor 2.5 so both datasets have a roughly equal weight in the chi-squared value \citep[the same procedure was followed in][ to merge data from SLUGGS and ATLAS$^{3{\rm D}}$]{Cappellari2015}.
 \item \emph{Fitting the GCMS data alone:} For this fit, we excluded the NIFS data (plot not shown). As detailed in the description of approach A, this procedure is not recommended because we run into the risk of biasing the black hole mass estimate. However, because we do not have NIFS data for NGC~1278, producing this fit allows for a direct comparison between the two galaxies.
\end{enumerate}
 \setcitestyle{notesep={,}}

We denote fits made assuming a cylindrically aligned velocity ellipsoid with a `c' prefix, and those with a spherically oriented one with an `s'. As discussed in \citet{Cappellari2020}, no single alignment holds for a full galaxy. A cylindrically oriented ellipsoid is only found in the case of a plane-parallel potential, whereas a spherically oriented one is found for a strictly spherical one. Nevertheless, a cylindrically oriented ellipsoid is a rather good approximation for a disc, especially close to the mid-plane \citep{Cappellari2007, Cappellari2008}. Since the \citet{Schwarzschild1979} dynamical modelling by \citet{Yildirim2015} does not show signs of a galaxy-wide pressure-supported component, we assumed a cylindrical alignment for our fiducial fit of NGC~1277 (cA).

The fitted parameters for all the approaches are listed in Table~\ref{parameters_NGC1277}. For parameters whose maximum likelihood values are not close to the boundary of the fitting interval, we present them as ${\rm value}\pm{\rm sigma}/2$, where sigma has been calculated as the interval between the 15.86 and 84.14 percentiles. In some cases, in particular for $f_{\rm DM}(6\,{\rm kpc})$ and $q_{\rm \star min}$, the probability distribution may peak close to the boundary of the fitting interval imposed by physics. We detected those cases are those where $\left|{\rm value}-{\rm boundary}\right|<{\rm sigma}/2$. In such cases, we determined one-sigma upper (lower) limits based on the 68.28 (31.72) percentile.

\begin{table*}
\caption{\texttt{JAM} model fit parameters for NGC~1277}
\label{parameters_NGC1277} 
\centering 
\setlength{\tabcolsep}{5.4pt}
\begin{tabular}{l l l c c c c c c}
\hline\hline 
&&Fitting approach& $q_{\star\,{\rm min}}$  & $\beta_z$ or $\beta_r$   & $\beta_{z,{\rm out}}$ or $\beta_{r,{\rm out}}$ & $\alpha$      & $M_{\rm BH}$                        & $f_{\rm DM}(6\,{\rm kpc})$\\
                &                   &             &    &              & & & $\left(10^9\,{\rm M}_\odot\right)$  &        $f_{\rm DM}(5\,{R_{\rm e}})$                    \\
\hline
\multirow{7}{*}{\rotatebox{90}{Cylindric}}&\multirow{4}{*}{\rotatebox{90}{Single $\beta_z$}}& cA1 NIFS only                           & $<0.056$ & $-0.11\pm0.03$ & $-$            &  $1.276\pm0.013$ & $4.88\pm0.11$ & \bf{0}           \\
&&\cellcolor{Gray}cA2 GCMS fixed $M_{\rm BH}$                                                  & \cellcolor{Gray}$0.37\pm0.02$   & \cellcolor{Gray}$0.05\pm0.05$  & \cellcolor{Gray}$-$    &  \cellcolor{Gray}$1.342\pm0.014$ & \cellcolor{Gray}\bf{4.88}     & \cellcolor{Gray}$<0.018$    \\
   &&cB NIFS+GCMS                                                                                                             & $>0.39$ & $-0.07\pm0.03$ & $-$            &  $1.317\pm0.014$ & $4.9\pm0.2$   & $<0.008$  \\
&&cC GCMS all free                                                                                                        & $0.33\pm0.03$   & $0.17\pm0.06$  & $-$            &  $1.40\pm0.03$   & $2.6\pm1.0$   & $<0.015$  \\\cline{2-9}
&\multirow{3}{*}{\rotatebox{90}{Two $\beta_z$}}&cA2$\beta_z$ GCMS fixed $M_{\rm BH}$                                          & $>0.35$   & $-0.10\pm0.11$  & $0.06\pm0.06$  &  $1.335\pm0.016$ & \bf{4.88}     & $<0.03$  \\
&&cB$\beta_z$ NIFS+GCMS                                                                                                           & $>0.35$   & $-0.31\pm0.04$ & $0.04\pm0.06$  &  $1.302\pm0.017$ & $5.49\pm0.16$   & $<0.05$    \\ 
&&cC$\beta_z$ GCMS all free                                                                                                 & $0.34\pm0.03$   & $0.09\pm0.13$ & $0.16\pm0.06$   &  $1.39\pm0.03$   & $2.5\pm1.0$   & $<0.02$    \\
\hline
\multirow{7}{*}{\rotatebox{90}{Spherical}}&\multirow{4}{*}{\rotatebox{90}{Single $\beta_r$}}&sA1 NIFS only                            & $<0.052$ & $0.26\pm0.06$  & $-$            &  $1.301\pm0.013$ & $3.5\pm0.3$   & \bf{0}           \\
&&sA2 GCMS fixed $M_{\rm BH}$                                                                                             & $>0.38$ & $0.16\pm0.06$  & $-$            &  $1.397\pm0.015$ & \bf{3.5}      & $<0.014$  \\
&&sB NIFS+GCMS                                                                                                                  & $0.338\pm0.016$ & $0.08\pm0.07$  & $-$            &  $1.330\pm0.012$   & $4.2\pm0.3$   & $<0.014$  \\ 
&&sC GCMS all free                                                                                                        & $0.386\pm0.013$ & $0.34\pm0.07$  & $-$            &  $1.51\pm0.03$   & $<3.9$   & $<0.019$    \\\cline{2-9}
&\multirow{3}{*}{\rotatebox{90}{Two $\beta_r$}}&sA2$\beta_r$ GCMS fixed $M_{\rm BH}$                                          & $>0.38$ & $0.0\pm0.2$  & $0.16\pm0.06$  &  $1.386\pm0.017$  & \bf{3.5}      & $<0.02$    \\
&&sB$\beta_r$ NIFS+GCMS                                                                                                           & $0.341\pm0.016$ & $0.13\pm0.15$  & $0.07\pm0.07$  &  $1.337\pm0.018$   & $4.0\pm0.6$   & $<0.018$    \\
&&sC$\beta_r$ GCMS all free                                                                                                 & $>0.38$ & $0.2\pm0.4$    & $0.30\pm0.13$  &  $1.48\pm0.08$   & $<3.6$   & $<0.02$    \\
\hline
\end{tabular}
\tablefoot{Fits cA1 and cA2 are shown in Figs.~\ref{NGC1277_NIFS} and \ref{NGC1277_VIRUS}, respectively. Values in boldface were fixed for the fit. The grey background indicates our fiducial fit. The errors and the upper and lower limits correspond to formal one sigma uncertainties (see text).}
\end{table*}

We first comment on the fiducial fit, cA. When fitting the NIFS data only we find a black hole mass $M_{\rm BH}=(4.88\pm0.11)\times10^9\,{\rm M}_\odot$ (Fig.~\ref{NGC1277_NIFS}), which is very similar to $M_{\rm BH}=(4.54\pm0.11)\times10^9\,{\rm M}_\odot$ obtained by \citet{Krajnovic2018} with the same data. The fitted values for $q_{\star\,{\rm min}}$ and $\beta$ are the same within the uncertainties. The subsequent GCMS data fit with a fixed $M_{\rm BH}$ (Fig.~\ref{NGC1277_VIRUS}) yields a negligible dark matter fraction with $f_{\rm DM}(6\,{\rm kpc})<0.018$. We find that the mismatch parameter is very similar for the NIFS and the GCMS fits (only a 5\% difference; $\alpha=1.342\pm0.014$ versus $\alpha=1.276\pm0.013$). This is relevant, since a large difference between the two would require considering an $\alpha$ varying with radius. The determination of the mismatch parameter in the cA2 fit allows us to re-evaluate the stellar mass of NGC~1277 to $M_\star\approx1.8\times10^{11}\,{\rm M}_\odot$.

In Figs.~\ref{curve_NGC1277_NIFS} and \ref{curve_NGC1277_VIRUS} we compare the major- and minor-axis $V_{\rm rms}$ profiles of the NIFS and GCMS data to those derived from the cA1 and cA2 fits, respectively. The agreement between the data and the fits is generally good, especially for the GCMS data (cA2), where the fitted $V_{\rm rms}$ passes close to most of the observed data points when accounting for three-sigma error bars. The central small $V_{\rm rms}$ peak is also well captured by the fit.

Both fitting approaches cB and cC yield very similar values for several of the parameters, with negligible dark matter fractions within 6\,kpc, a mismatch parameter $\alpha\approx1.3$, and a black hole mass in the range $2.6\times10^9\,{\rm M}_\odot\lesssim M_{\rm BH}\lesssim4.9\times10^9\,{\rm M}_\odot$. It is noteworthy that the high angular resolution NIFS data do not seem to be required to obtain a reasonable order of magnitude estimate of the black hole mass. This is probably due to the fact that we are introducing high angular resolution information through the \texttt{MGE} models of the luminosity and the surface brightness distributions (see Appendix~\ref{appendix}) and highlights the importance of high angular resolution information to properly constrain the mass of the central black hole. A further reason for the black hole mass agreement is that the mismatch parameter does not vary much with radius, as indicated by the small difference in the $\alpha$ values fitted in cA1 and cA2.

The intrinsic flattening of the flattest component $q_{\star\,{\rm min}}$ shows significant differences between fitting approaches. This parameter is found to be in the range between $q_{\star\,{\rm min}}=0.050$ and $q_{\star\,{\rm min}}=0.4$, which corresponds to an inclination between $i=67\degr$ and $i=90\degr$. The intermediate $q_{\star\,{\rm min}}$ values given by fit cA2 corresponds to $i\approx76\degr$, which is very close to the value deduced from the shape of the circumnuclear dust ring \citep[$i=75\degr$;][]{Bosch2012}. The value of $q_{\star\,{\rm min}}<0.06$ obtained from fitting the NIFS data only is considerably off the other values, probably because the NIFS FOV covers a region where the dominant \texttt{MGE} components are nearly round (Table~\ref{MGEbar}), hence offering little constraining power on the flatness of the galaxy.

The parameter $\beta_z$ is significantly smaller when fitting the NIFS data alone than when fitting GCMS data. This suggests a radial variation of the shape of the velocity ellipsoid. To explore this effect, we repeated all the fits where GCMS data are used, but this time we considered two $\beta_z$ parameters. The first one corresponded to stellar \texttt{MGE} components with a sigma smaller than the effective radius $R_{\rm e}$, whereas the second one ($\beta_{z,{\rm out}}$) described the more extended components. The fits with two $\beta_z$ values are denoted with the suffix `$\beta_z$'. The results of the fits (Table~\ref{parameters_NGC1277}) indicate that allowing for radial variations in $\beta_z$ does not have a significant effect in either $\alpha$, $M_{\rm BH}$, or $f_{\rm DM}(6\,{\rm kpc})$.

Following \citet{Cappellari2020}, we also fitted the data assuming a spherically aligned velocity ellipsoid (with an anisotropy parameter $\beta_r$) in order to have a better grasp of the uncertainties. The results remain broadly the same as with the cylindrically aligned models (Table~\ref{parameters_NGC1277}) and a negligible dark matter fraction is also recovered.

We have also made experiments with dark matter profiles other than NFW, such as the generalised NFW profile or the Einasto profile \citep{Einasto1965}. In all cases, the resulting dark matter fraction $f_{\rm DM}(6\,{\rm kpc})$ has been found to be compatible with zero. We also checked that uncertainties of the order of $0\farcs5$ in the determination of the centre of NGC~1277 within the field of view of the GCMS (see Sect.~\ref{kinematics}) do not affect our results.

\begin{figure*}
\begin{center}
  \includegraphics[scale=0.48]{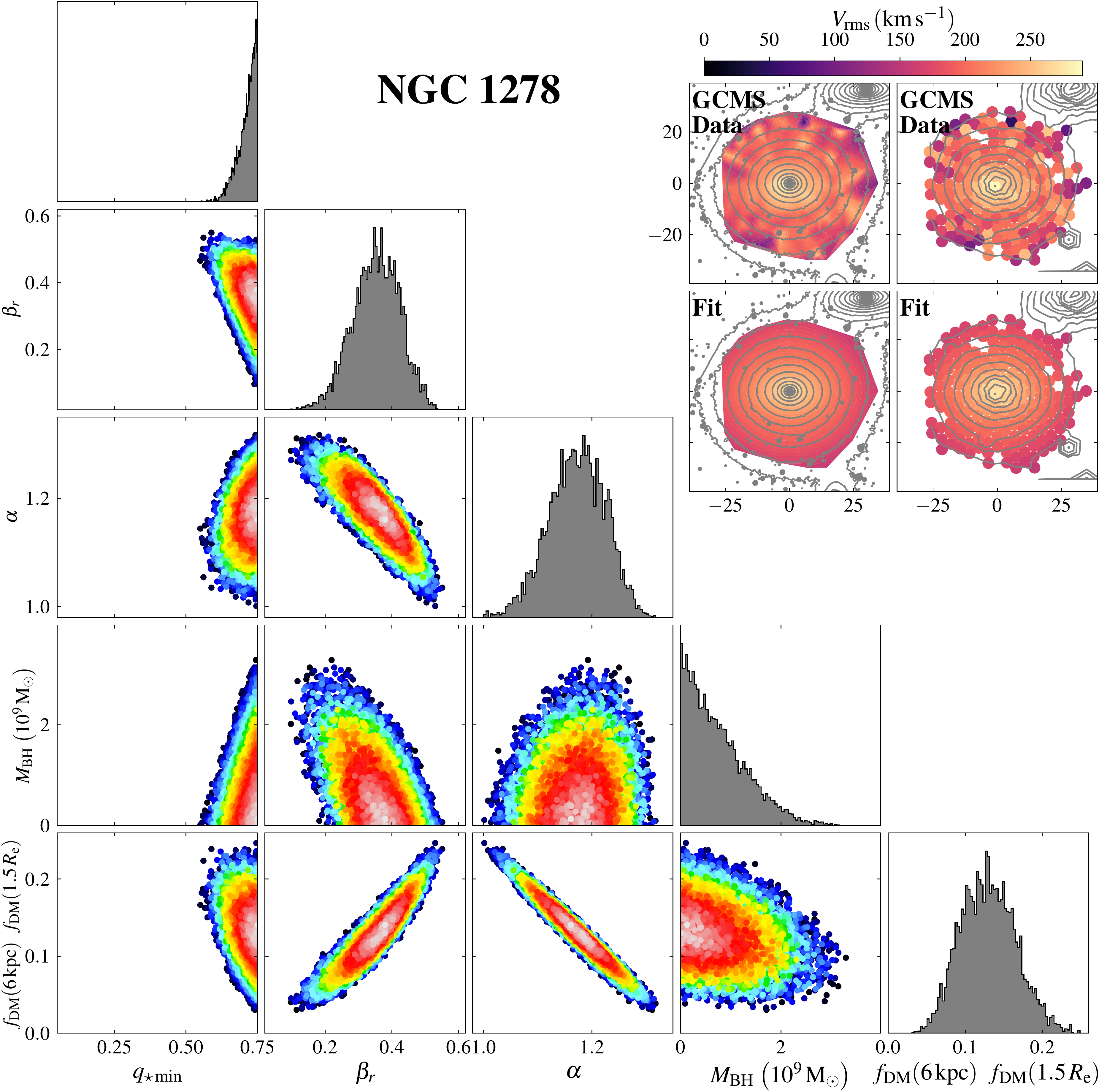}\\
  \end{center}
  \caption{\label{NGC1278_VIRUS_free} Same as Fig.~\ref{NGC1277_NIFS}, but this time for the GCMS $V_{\rm rms}$ map of NGC~1278 assuming a spherically aligned velocity ellipsoid (fit sB). The isophotes in the {\it left}-column maps come from an F850LP image. This is our fiducial fit for NGC~1278, where the fitted parameters are $q_{\star\,{\rm min}}$, $\beta_r$, $\alpha$, and $M_{\rm BH}$, and $f_{\rm DM}(6\,{\rm kpc})$ ($f_{\rm DM}(1.5\,R_{\rm e})$).}
\end{figure*}

\begin{figure}
\begin{center}
  \includegraphics[scale=0.48]{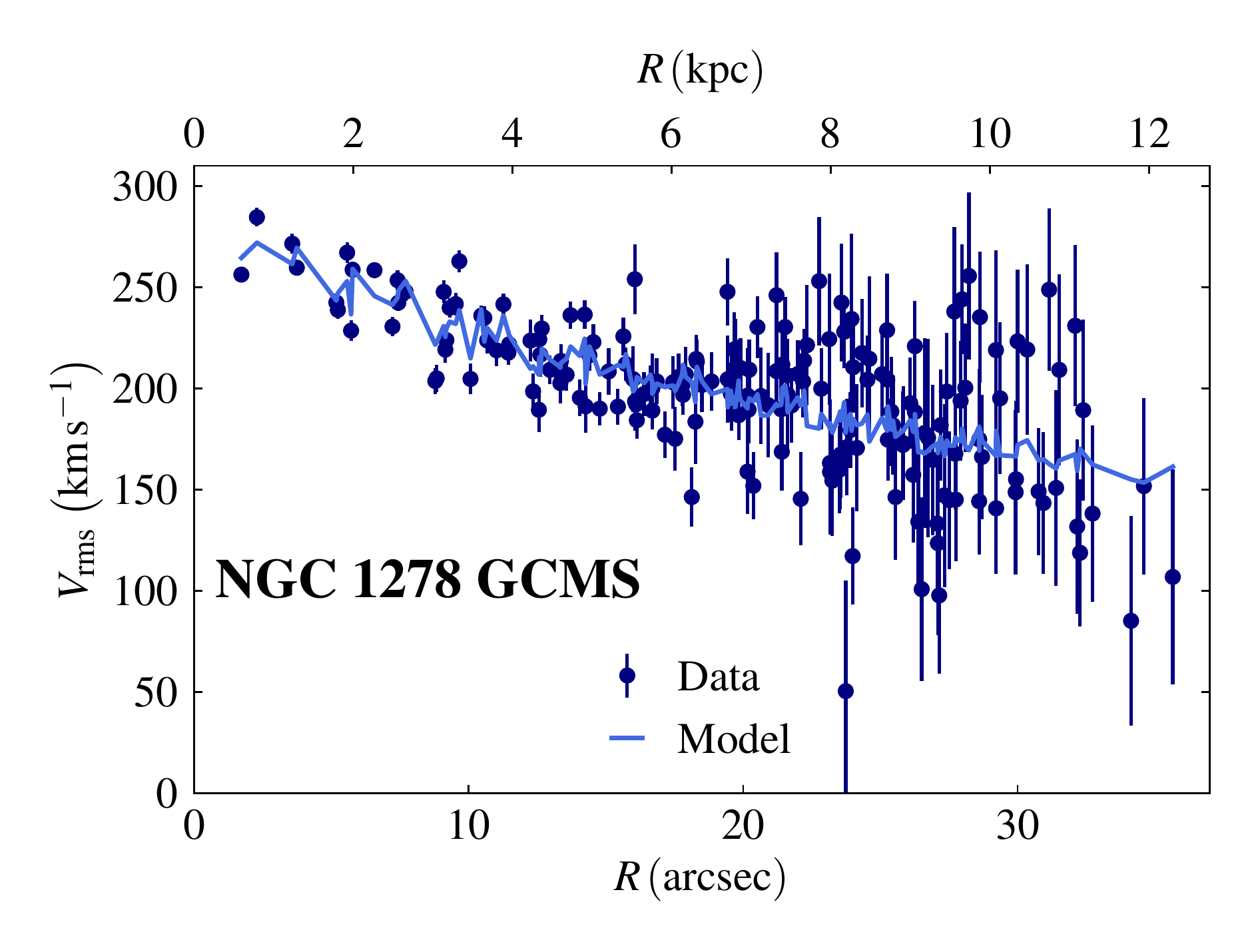}\\
  \end{center}
  \caption{\label{curve_NGC1278_VIRUS} Same as Fig.~\ref{curve_NGC1277_NIFS} but for NGC~1278 with GCMS data, and the SB fit. Because the galaxy is nearly spherical, we included data from all the fibres and sorted them according to their distance to the centre.}
\end{figure}

To sum up, the results of all our fitting approaches broadly coincide with those of the fiducial fit cA (with a cylindrically aligned velocity ellipsoid and a single $\beta_z$). This fit yields a mismatch parameter $\alpha=1.342\pm0.014$ (giving a probably too optimistically well-constrained stellar mass of $M_\star\approx(1.790\pm0.019)\times10^{11}\,{\rm M}_\odot$) and a black hole mass $M_{\rm BH}=\left(4.88\pm0.11\right)\times10^9\,{\rm M}_\odot$. We also find a negligible dark matter fraction within 6\,kpc ($5\,R_{\rm e}$), $f_{\rm DM}(6\,{\rm kpc})<0.018$. We stress that, irrespectively of the fitting approach, we find that dark matter has a negligible dynamical effect within the inner 6\,kpc of NGC~1277 under the assumption of a mismatch parameter that is constant with radius.

\subsection{Dynamical modelling of NGC~1278}

\label{modelNGC1278}

In order to fit the NGC~1278 kinematics we need to rely on the GCMS data only. This means that, contrary to the case of NGC~1277, we lack high angular resolution data able to provide tight constraints on the black hole mass. For these fits we followed two approaches:
\begin{enumerate}[A]
 \item \emph{Fit where the central black hole mass is fixed by scaling relations:} We use the $M_{\rm BH}-M_\star$ scaling relation in the Eq.~15 from \citet{Sahu2019} relating the black hole mass to the total stellar mass of elliptical galaxies (the plot for fit~A is not shown). By introducing the total stellar mass $M_\star=2.9\times10^{11}\,{\rm M_\odot}$ (Sect.~\ref{mges}) we obtained $M_{\rm BH}=1.5\times10^9\,{\rm M}_\odot$. However, because the residual mean square of the relation is 0.48\,dex, $M_{\rm BH}$ is expected to be in the rather large range between $M_{\rm BH}\approx5\times10^8\,{\rm M}_\odot$ and $M_{\rm BH}\approx5\times10^9\,{\rm M}_\odot$. Alternatively, we could have used the $M_{\rm BH}-\sigma$ relation. Assuming the expression in Eq.~7 from \citet{Kormendy2013} and a central velocity dispersion $\sigma=285\,{\rm km\,s^{-1}}$ we obtained a virtually identical black hole mass of ($M_{\rm BH}=1.4\times10^{9}\,{\rm M}_\odot$  with a 0.29\,dex scatter in the relation). 
 \item \emph{Fit with no constraints:} We left all the parameters free, including the black hole mass $M_{\rm BH}$. We chose it to be our fiducial fit (Fig.~\ref{NGC1278_VIRUS_free}) because the large scatter in the black hole scaling relations (see item A) could introduce a bias when fixing the black hole mass.
\end{enumerate}

\begin{table*}
\caption{\texttt{JAM} model fit parameters for NGC~1278}
\label{parameters_NGC1278} 
\centering 
\begin{tabular}{l l l c c c c c c}
\hline\hline 
&&Fitting approach& $q_{\star\,{\rm min}}$  & $\beta_z$ or $\beta_r$   & $\beta_{z,{\rm out}}$ or $\beta_{r,{\rm out}}$  & $\alpha$      & $M_{\rm BH}$                        & $f_{\rm DM}(6\,{\rm kpc})$\\
                &                   &             &    &              & & & $\left(10^9\,{\rm M}_\odot\right)$  &           $f_{\rm DM}(1.5\,{R_{\rm e}})$                         \\
\hline
\multirow{4}{*}{\rotatebox{90}{Cylindric}}&\multirow{2}{*}{\rotatebox{0}{Single $\beta_z$}}& cA $M_{\rm BH}$ fixed                                              & $>0.64$ & $0.04\pm0.03$ & $-$           & $1.325\pm0.018$ & \bf{1.5}     &$0.016\pm0.011$\\
&& cB  all free & $>0.66$ & $0.03\pm0.02$ & $-$ & $1.28\pm0.03$ & $3.2\pm0.8$ & $0.029\pm0.015$\\\cline{2-9}
&\multirow{2}{*}{\rotatebox{0}{Two $\beta_z$}}&cA$\beta_z$  $M_{\rm BH}$ fixed                                       & $0.61\pm0.09$ & $0.14\pm0.06$   & $-0.01\pm0.05$& $1.33\pm0.02$   & \bf{1.5}     & $0.040\pm0.013$   \\ 
&&cB$\beta_z$ all free       & $0.63\pm0.10$ & $0.12\pm0.07$   & $-0.01\pm0.05$& $1.32\pm0.04$   & $1.9\pm1.0$  & $0.035\pm0.015$  \\
\hline
\multirow{4}{*}{\rotatebox{90}{Spherical}}&\multirow{2}{*}{\rotatebox{0}{Single $\beta_r$}}& sA $M_{\rm BH}$ fixed                                              & $>0.71$ & $0.31\pm0.07$ & $-$            & $1.18\pm0.06$  &  \bf{1.5}    & $0.12\pm0.04$ \\
&&\cellcolor{Gray}sB all free &  \cellcolor{Gray}$>0.70$ &  \cellcolor{Gray}$0.38\pm0.08$ &   \cellcolor{Gray}$-$ &   \cellcolor{Gray}$1.16\pm0.06$ &   \cellcolor{Gray}$<0.7$ &   \cellcolor{Gray}$0.14\pm0.04$\\\cline{2-9}
&\multirow{2}{*}{\rotatebox{0}{Two $\beta_r$}}&sA$\beta_r$  $M_{\rm BH}$ fixed                                       & $>0.69$ & $0.31\pm0.09$ & $0.33\pm0.16$ & $1.16\pm0.12$  &  \bf{1.5}    & $0.13\pm0.07$ \\
&&sB$\beta_r$  all free      & $>0.69$ & $0.37\pm0.09$ & $0.40\pm0.17$ & $1.14\pm0.12$  &  $<0.9$ & $0.15\pm0.07$ \\
\hline
\end{tabular}
\tablefoot{Fit sB is shown in Fig.~\ref{NGC1278_VIRUS_free}. Values in boldface were fixed for the fit. The grey background indicates our fiducial fit.  The errors and the upper and lower limits correspond to formal one sigma uncertainties (see text in Sect.~\ref{modelNGC1277}).}
\end{table*}

The baryon distribution in NGC~1278 is found to be nearly circular in projection (Table~\ref{MGEbar}), which might be indicative of a far from flattened baryonic matter distribution. We tested this hypothesis by calculating the projected angular momentum within an effective radius to check whether NGC~1278 is a slow rotator, as follows. First, we estimated the projected angular momentum within $R_{\rm e}$ \citep[][ Eq.~6]{Emsellem2007}
\begin{equation}
 \lambda_{R_{\rm e}}=\frac{\sum_{i=1}^NF_iR_i\left|V_i\right|}{\sum_{i=1}^NF_iR_i\sqrt{V_i^2+\sigma_i^2}},
\end{equation}
where the sums are made over the fibres within $R_{\rm e}$, $R_i$ are the distances of the fibres to the centre of NGC~1278, and $F_i$ are the fluxes at those distances based on the \texttt{MGE} model calculated using the circularised version of Eq.~\ref{intensity} (see Sect.~\ref{mges}). We found $\lambda_{R_{\rm e}}=0.093$. Then, we measured the ellipticity of the \texttt{MGE} model using the routine \texttt{mge\_half\_light\_isophote} included in \texttt{JAM} and obtained $\varepsilon_{\rm e}=0.18$. Finally, we applied the test proposed in Eq.~19 in \citep{Cappellari2016}, who suggested that galaxies fulfilling
\begin{equation}
 \lambda_{R_{\rm e}}<0.080+\varepsilon_{\rm e}/4,
\end{equation}
as it is the case for NGC~1278, are slow rotators. Since this test proves that NGC~1278 is not likely to be a very flattened object, we chose a spherically aligned velocity ellipsoid for our fiducial fit. We present the results in Table~\ref{parameters_NGC1278}. As for NGC~1277, the fits are labelled with prefixes `c' and `s' so to denote the velocity alignments and a suffix `$\beta_z$' or `$\beta_r$', respectively, if they were produced assuming two different values for the shape of the velocity ellipsoid. Here we assumed that $\beta_{r,{\rm out}}$ and $\beta_{z,{\rm out}}$ affect the shape of the \texttt{MGE} components with $\sigma_\star$ larger than the effective radius, $R_{\rm e}=12\farcs3$.

Our fiducial fit sB (Fig.~\ref{NGC1278_VIRUS_free}) yields a significant fraction of dark matter within 6\,kpc, $f_{\rm DM}(6\,{\rm kpc})=0.14\pm0.04$. This fraction does not change if we keep the black hole mass fixed (fit sA). The fiducial fit yields a mismatch parameter $\alpha=1.16\pm0.06$, which implies that the stellar mass-to-light ratio is slightly larger than that expected from a Salpeter IMF. In principle, changes in $\alpha$ modify the $M_{\rm BH}$ obtained in the $M_{\rm BH}-M_\star$ scaling relation that is introduced as a constant in sA, but since $\alpha\approx1$, the shift in $M_{\rm BH}$ would be very small. The value of $M_{\rm BH}<0.7\times10^9\,{\rm M}_\odot$ obtained in the fiducial fit is compatible with what is expected from the scaling relations.

All the fits yield a $q_{\star\,{\rm min}}$ value peaking at or close to the boundary of the fitting range ($q_{\star\,{\rm min}}=0.75$). Since NGC~1278 is likely to be near-spherical, this is a poorly defined parameter and not much importance should be attached to this coincidence.

Contrarily to NGC~1277, where $f_{\rm DM}(6\,{\rm kpc})$ was consistently found to be negligible irrespective of the kind of fit, the fraction of dark matter within 6\,kpc in NGC~1278 is found to vary depending on whether a cylindrically or a spherically aligned velocity ellipsoid is assumed. The former orientation yields dark matter fractions $f_{\rm DM}(6\,{\rm kpc})=0.02-0.04$ and the latter have $f_{\rm DM}(6\,{\rm kpc})=0.10-0.15$. Fits with low $f_{\rm DM}(6\,{\rm kpc})$ compensate the lack of dark matter by increasing the baryonic mass through an increase in $\alpha$ from $\alpha\approx1.1-1.2$ (slightly larger than that expected for a Salpeter IMF) to up to $\alpha\approx1.3$ (indicating a bottom-heavy IMF comparable to that obtained for NGC~1277). For our fiducial fit, a zero dark matter fraction is disfavoured at a more than four-sigma level, but a strong covariance is seen between $\alpha$ and $f_{\rm DM}(6\,{\rm kpc})$ (Fig.~\ref{NGC1278_VIRUS_free}). Hence, an independent determination of $\alpha$ would be desirable in order to confirm the presence of a significant dark matter fraction. Such evidence was obtained from line index indicators, showing a much bottom-lighter IMF in NGC~1278 than in NGC~1277 (Ferré-Mateu et al.~in prep.).

Figure~\ref{curve_NGC1278_VIRUS} shows the radial $V_{\rm rms}$ profile. The agreement between the observation and the fit in the central region ($R\lesssim15^{\prime\prime}$) is remarkable. At larger radii, the error bars in $V_{\rm rms}$ for the individual fibres becomes as large as a few tens of ${\rm km\,s}^{-1}$, which causes a large scatter. Yet, the fit passes roughly through the middle of the data cloud. The reason why such data points with large error bars are not abundantly found for NGC~1277 is the abruptness of the surface brightness profile, which causes the radial range of fibres with barely acceptable kinematic fits to be very narrow.

As for NGC~1277, we checked whether registration errors of the GCMS data would have an impact on the results. We found that the uncertainties in $\alpha$ and $f_{\rm DM}(6\,{\rm kpc})$ introduced by a centring error of up to $1\arcsec$ are comparable to the formal one-sigma error bars of the fit.

In summary, for NGC~1278 we find a dark matter fraction $f_{\rm DM}(6\,{\rm kpc})=0.14\pm0.04$ and a black hole mass $M_{\rm BH}<0.7\times10^9\,{\rm M}_\odot$. The excess parameter is $\alpha=1.16\pm0.06$ (slightly larger than for a Salpeter IMF), which yields a total stellar mass $M_\star=(3.4\pm0.2)\times10^{11}\,{\rm M}_\odot$.

\section{Discussion}

\label{discussion}

\subsection{The mass-to-light ratio and the IMF of the stellar component of NGC~1277 and NGC~1278}

\label{imf}

\begin{figure}
\begin{center}
  \includegraphics[scale=0.48]{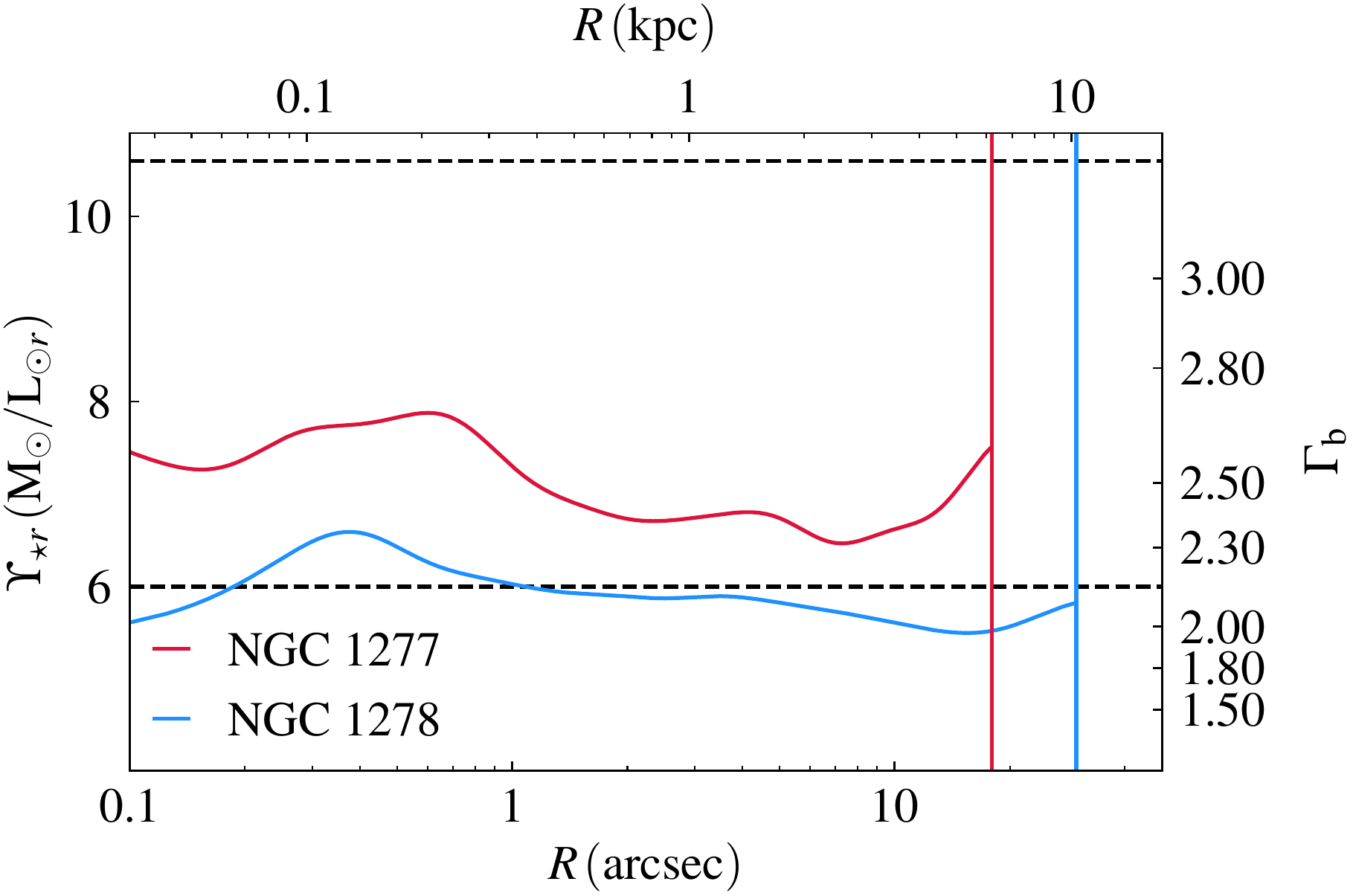}\\
  \end{center}
  \caption{\label{MLs} Circularised mass-to-light ratio $\Upsilon_{\star r}$ as a function of the radius for NGC~1277 and NGC~1278, in blue and red respectively. The vertical lines indicate the limit of the GCMS kinematic data, The right axis shows the correspondence with the absolute value of the outer exponent of a bimodal IMF, $\Gamma_{\rm b}$, for a stellar population with an age of 13.0\,Gyr and a metallicity $[{\rm Z}/{\rm H}]=0.26$. The horizontal black dashed lines denote the range of $\Gamma_{\rm b}$ values found by \citet{MartinNavarro2015} for NGC~1277, with the upper line ($\Gamma_{\rm b}=3.25$) corresponding to their upper limit for the inner parts of the galaxy and the lower line ($\Gamma_{\rm b}=2.15$) corresponding to their lower limit at a radius $R=1.4\,R_{\rm e}$. The surface brightness of NGC~1278 has been converted to F625W as in Sect.~\ref{mges}. The correspondence between $\Upsilon_{\star r}$ and $\Gamma_{\rm b}$ is only roughly valid for NGC~1277 because NGC~1278 is slightly younger and less metallic (Ferré-Mateu et al.~in prep).}
\end{figure}

Our determination of the stellar mass of NGC~1277 and NGC~1278 makes it possible to provide constraints on the IMF. Our fiducial fit of NGC~1277 yields a stellar mass-to-light ratio in F625W (similar to the $r$ band) $\Upsilon_{\star r}=7.0\,{\rm M}_\odot/{\rm L}_{\odot r}$. After converting the F850LP surface brightnesses to F625W through $r-z=0.74$ (measured from SDSS images, Sect.~\ref{mges}), we find $\Upsilon_{\star r}=5.8\,{\rm M}_\odot/{\rm L}_{\odot r}$ for NGC~1278.

If we know $\Upsilon_\star$, the age of the stellar population, and its metallicity, we can obtain information about the shape of the IMF from photometric predictions based on spectral energy distributions. Here, we used those based on E-MILES \citep{Vazdekis2012}\footnote{The E-MILES spectral energy distributions can be downloaded from \url{http://research.iac.es/proyecto/miles/pages/photometric-predictions-based-on-e-miles-seds.php}.} computed using BaSTI isochrones. These predictions are made assuming an IMF lower cutoff at $0.1\,{\rm M}_\odot$ and an upper cutoff at $100\,{\rm M}_\odot$. The $\Upsilon_\star$ inferences account for the stellar remnants (contrarily to the mass maps derived in Sect~\ref{mges}, with respect to which the mismatch parameter $\alpha$ from our fits was calculated). Based on \citet{MartinNavarro2015} and Ferré-Mateu et al.~(in prep.) we assume an age of 13.0\,Gyr and a metallicity of $[{\rm Z}/{\rm H}]=0.26$ for NGC~1277. 

Under the assumption of a bimodal IMF with a turning point at $0.4\,{\rm M}_\odot$ defined as in \citet{Vazdekis1996}, and where $-\Gamma_{\rm b}$ corresponds to the exponent of the power law in the high-end of the IMF, we obtain $\Gamma_{\rm b}=2.46$. This value is very similar with that found by \citet{MartinNavarro2015} and \citet{FerreMateu2017}, where the index varied between $\Gamma_{\rm b}\approx3.05\pm0.20$ at the very centre of NGC~1277 and $\Gamma_{\rm b}\approx2.55\pm0.40$ at a radius of about $1.4\,R_{\rm e}$. The fact that both line-strength indices and dynamical modelling yield similar results makes the inference of a bottom-heavy IMF for NGC~1277 very robust.

In Fig.~\ref{MLs} we plot the stellar mass-to-light ratio at F625W as a function of the radius  for NGC~1277 and NGC~1278. Here we define $\Upsilon_{\star r}$ as the ratio of the circularised surface density and the surface brightness \texttt{MGE} models corrected (multiplied) by the mismatch parameter. The right axis indicates the corresponding absolute value of the outer exponent of a bimodal IMF, $\Gamma_{\rm b}$, obtained from the E-MILES photometric predictions for a stellar population with age 13.0\,Gyr, metallicity $[{\rm Z}/{\rm H}]=0.26$ (this is roughly valid for NGC~1277 only, because NGC~1278 is slightly younger and less metallic; Ferré-Mateu et al.~in prep.). If the mismatch parameter, $\alpha$, were constant with radius as assumed in the \texttt{JAM} fits, the wiggles in $\Upsilon_\star$ would imply a genuine variation of the mass-to-light ratio. However, the observed changes in $\Gamma_{\rm b}$, are a combination of real variations driven by small radial changes in $\alpha$, and apparent variations due to small radial trends of the stellar population age and abundances. Nevertheless, it is reassuring to see that the estimated $\Gamma_{\rm b}$ index for NGC~1277 falls within the range of values obtained in \citet{MartinNavarro2015} and \citet{FerreMateu2017}. In NGC~1278, $\Upsilon_\star$ is systematically lower than in NGC~1277 at all radii. This is in line with the finding that relic galaxies have a bottom-heavier IMF than regular ETGs with similar properties \citep{MartinNavarro2023}.

\subsection{The dark matter haloes of NGC~1277 and NGC~1278}

\label{dmhs}

In this subsection we compare the fitted dark matter haloes to the expectations from models and simulations. To do so, two pieces of information are required:
\begin{enumerate}[1)]
 \item A stellar-to-halo mass relation (SHMR) providing a dark halo mass as a function of the measured stellar mass.
 \item A parametrisation of the three-dimensional radial profile of the dark matter halo as a function of its mass.
\end{enumerate}

We adopted the SHMR from \citet{Girelli2020}, who used abundance matching to relate the observed stellar mass of galaxies in the COSMOS field to dark matter halo masses extracted from {\sc dustgrain}-\emph{pathfinder} $\Lambda$CDM simulation \citep{Giocoli2018}. They offer SHMRs at different redshifts, which allows us to check $z=0$ galaxies, but also to test whether the relic galaxy NGC~1277 has properties compatible with those of galaxies at $z\sim2-3$. In this context, the dark matter halo mass $M_{\rm 200}$ is defined as the mass enclosed within a radius $r_{200}$ where the average density is a factor $\Delta=200$ times the critical density of the Universe
\begin{equation}
  \rho_{\rm c}=\frac{3H_0^2}{8\pi G}.
\end{equation}
Throughout this paper, the stellar masses in \citet{Girelli2020} have been consistently shifted by 0.23\,dex, so to convert their assumed \citet{Chabrier2003} IMF into a \citet{Salpeter1955} one.

Using the SMHR parametrisation in Eq.~6 of \citet{Girelli2020} and the parameters from their Table~2, we produced Montecarlo runs to estimate the dark matter halo mass accounting for the statistical uncertainties of the parameters and in the galaxy stellar mass (from our mismatch parameter uncertainties). The uncertainties were assumed to have a Gaussian distribution.

Once the masses of the dark matter haloes were estimated, we parametrised their three-dimensional radial profile using Eq.~18 in \citet{Child2018}. This expression gives the NFW dark matter halo concentration, defined as $c_{200}=r_{200}/r_{\rm s}$, as a function of the stellar-to-halo mass ratio $M_\star/M_{200}$. We adopted the free parameters from the `Stack, NFW' fit in their Table~1. We can then reformulate Eq.~\ref{nfw} as a function of the concentration
\begin{equation}
 \rho_{\rm DM}(r)=\frac{\rho_{\rm c}}{3A_{\rm NFW}}\frac{1}{\frac{r}{r_{200}}\left(\frac{1}{c_{200}}+\frac{r}{r_{200}}\right)^2},
\end{equation}
where $A_{\rm NFW}$ can be written as
\begin{equation}
 A_{\rm NFW}={\rm ln}\,\left(1+c_{200}\right)-\frac{c_{200}}{1+c_{200}}.
\end{equation}
Density profile uncertainties were computed from Montecarlo runs accounting for one-sigma error bars from our fit and those from the parameters in \citet{Girelli2020} and \citet{Child2018}.

\begin{figure*}
\begin{center}
  \includegraphics[scale=0.48]{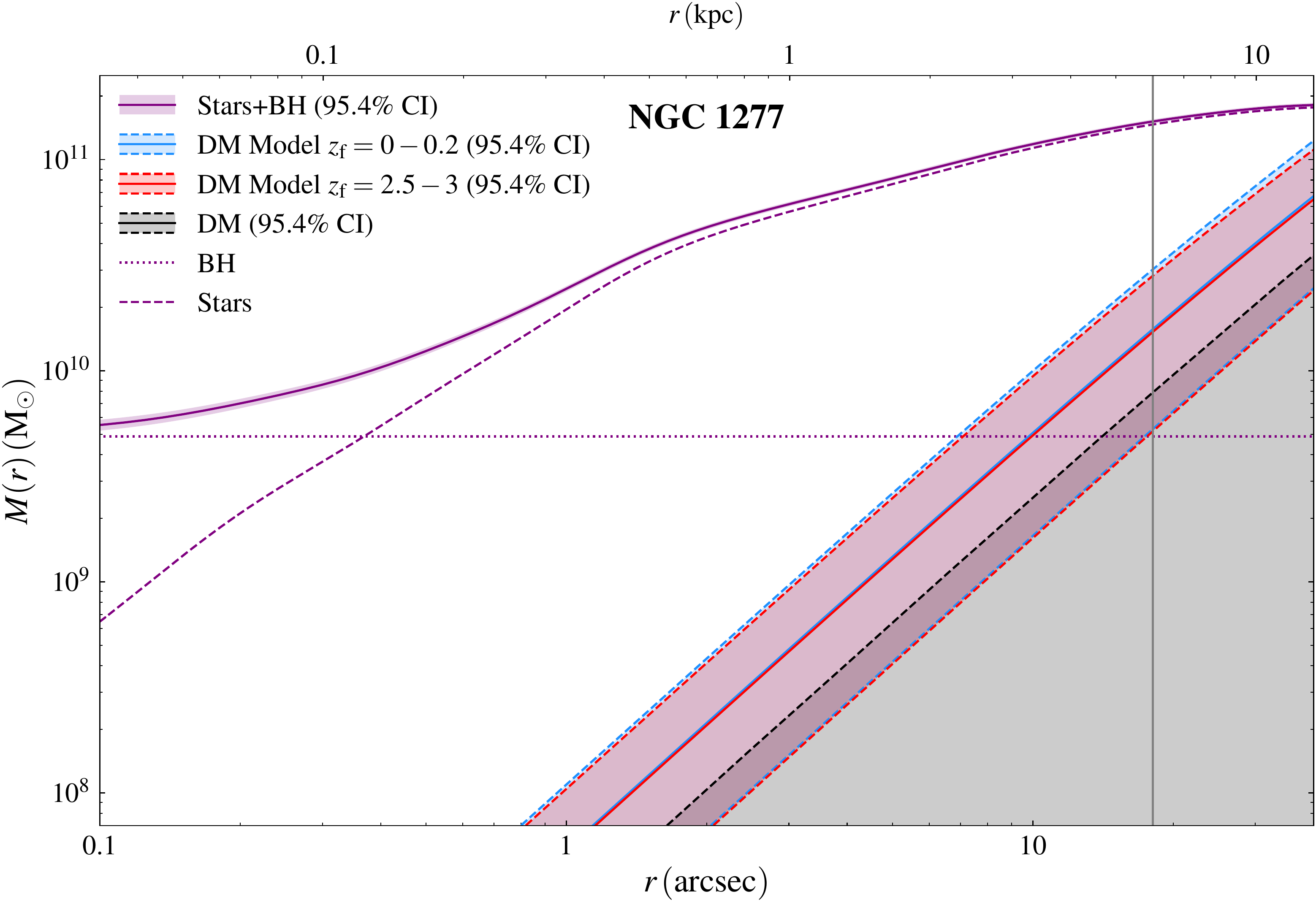}\\
  \end{center}
  \caption{\label{DM_NGC1277} Three-dimensional cumulative mass profiles of NGC~1277 for the baryonic component (stars plus black hole, in purple), the fitted dark matter halo (in black), and the predicted dark matter halo profiles for two different freezing redshifts (in blue and red for low and high $z_{\rm f}$, respectively) based on the SMHR parametrisation by \citet{Girelli2020} and the dark matter halo parametrisation by \citet{Child2018}. The bands correspond to 95.4\% confidence intervals. The baryonic component can be further subdivided into the black hole (dotted purple line) and the stellar component (dashed purple line). The grey vertical line indicates $r=6\,{\rm kpc}$ ($18^{\prime\prime}$), that is the radius within which we have kinematic data.}
\end{figure*}

\begin{figure*}
\begin{center}
  \includegraphics[scale=0.48]{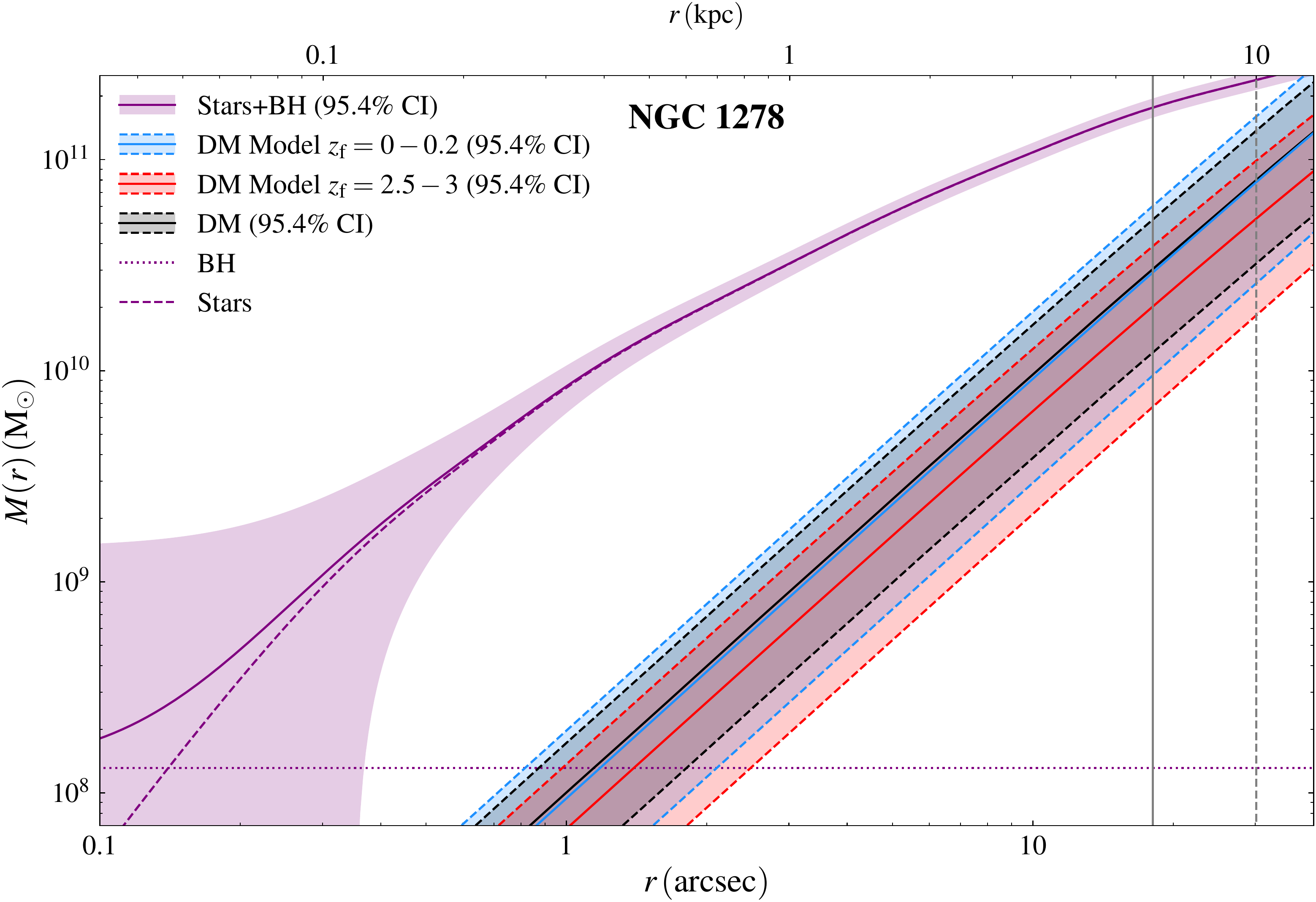}\\
  \end{center}
  \caption{\label{DM_NGC1278} Same as Fig.~\ref{DM_NGC1277}, but for NGC~1278. The dashed grey line indicates the radius out to which we have kinematic data, that is 10\,kpc or $30^{\prime\prime}$.}
\end{figure*}

In Fig.~\ref{DM_NGC1277} we display the cumulative baryonic (stellar plus black hole) and dark matter mass profiles from our fiducial fit to NGC~1277. We include 95.4\% confidence interval (two sigma) bands for both profiles. Since the $f_{\rm DM}(6\,{\rm kpc})$ probability distribution peaks at zero (the border of the fitting interval) we calculated the confidence intervals as the 95.4 percentile rather than the interval between the 2.3 and the 97.7 percentiles. We also display the predicted dark matter halo profiles for freezing redshifts $z_{\rm f}=0-0.2$ (meaning that the evolution of the galaxy has continued to the present, undergoing no freezing) and $z_{\rm f}=2.5-3$, as well as the 95.4\% confidence intervals based on the Montecarlo simulations. The predicted dark matter halo profiles for both freezing redshifts are very similar because the stellar mass of NGC~1277 roughly corresponds to that where the SHMR for all the redshifts intersect \citep[Fig.~8 in][]{Girelli2020}.

The cumulative mass  profiles in Fig.~\ref{DM_NGC1277} indicate that within 6\,kpc (equivalent to $5R_{\rm e}$, which is the radius traced by our kinematic data), NGC~1277 has a mass-follows-light behaviour, with a maximum allowed dark matter fraction (at a two-sigma level) of $f_{\rm DM}(6\,{\rm kpc})=0.05$. This is in tension with $f_{\rm DM}(6\,{\rm kpc})=0.09$ predicted for both freezing redshifts $z_{\rm f}=0$ and $z_{\rm f}=3$. However, it is compatible with the minimum two-sigma mass fractions allowed by the models according to our Montecarlo runs based on the \citet{Girelli2020} and \citet{Child2018} works ($f_{\rm DM}(6\,{\rm kpc})=0.03$). The negligible amount of dark matter within 6\,kpc, and hence the tension between the fit and the expectations, is a common feature among all the experiments listed in Table~\ref{parameters_NGC1277}. Our results are compatible with those from \citet{Yildirim2015}, who found a fraction of dark matter within the effective radius, $f_{\rm DM}(R_{\rm e})$ compatible with zero.

The discrepancy between the fitted and the expected dark matter fraction for NGC~1277 arises from a combination of several factors. The extreme compactness of NGC~1277 might make it baryon-dominated over most of the disc, so it may become hard to characterise the dynamical effects of the underlying dark matter distribution. Also, the reported formal error bars ignore the systematics that might be introduced by the hypotheses used in the Jeans modelling (such as the use of a single mismatch parameter across the whole galaxy). However, those systematics are probably small because if the fitted parameters are robust against changes in the assumed velocity ellipsoid (as it occurs for $f_{\rm DM}(6\,{\rm kpc})$), they can be assumed to be reliable \citep{Cappellari2020}. The result is also robust against changes in the positioning of the galaxy centre and the estimated distance.

Contrary to the case of NGC~1277, Fig.~\ref{DM_NGC1278} shows that the dark matter profile obtained in our fiducial fit of NGC~1278 falls right on top of the $z_{\rm f}=0$ prediction. The fitted dark matter profile is above, but still compatible, with what would be expected for a $z_{\rm f}=3$ galaxy. For NGC~1278, $f_{\rm DM}(6\,{\rm kpc})$ varies with the varying fitting approaches. Nevertheless, regardless of the fitting approach and other input parameters (such as the position of the centre of the galaxy and the estimated distance), we always find that NGC~1278 has a non-negligible dark matter fraction.

In summary, we find that the fact that the extreme relic galaxy NGC~1277 has no detectable dark matter within five effective radii is in tension with the expectations from models. On the other hand, we find that the regular massive ETG NGC~1278 has a detectable dark matter halo which is compatible with the models.

\subsection{The dark matter fraction within one and five effective radii and within 6\,kpc}

\label{sectfdm}

\begin{figure*}
\begin{center}
  \includegraphics[scale=0.48]{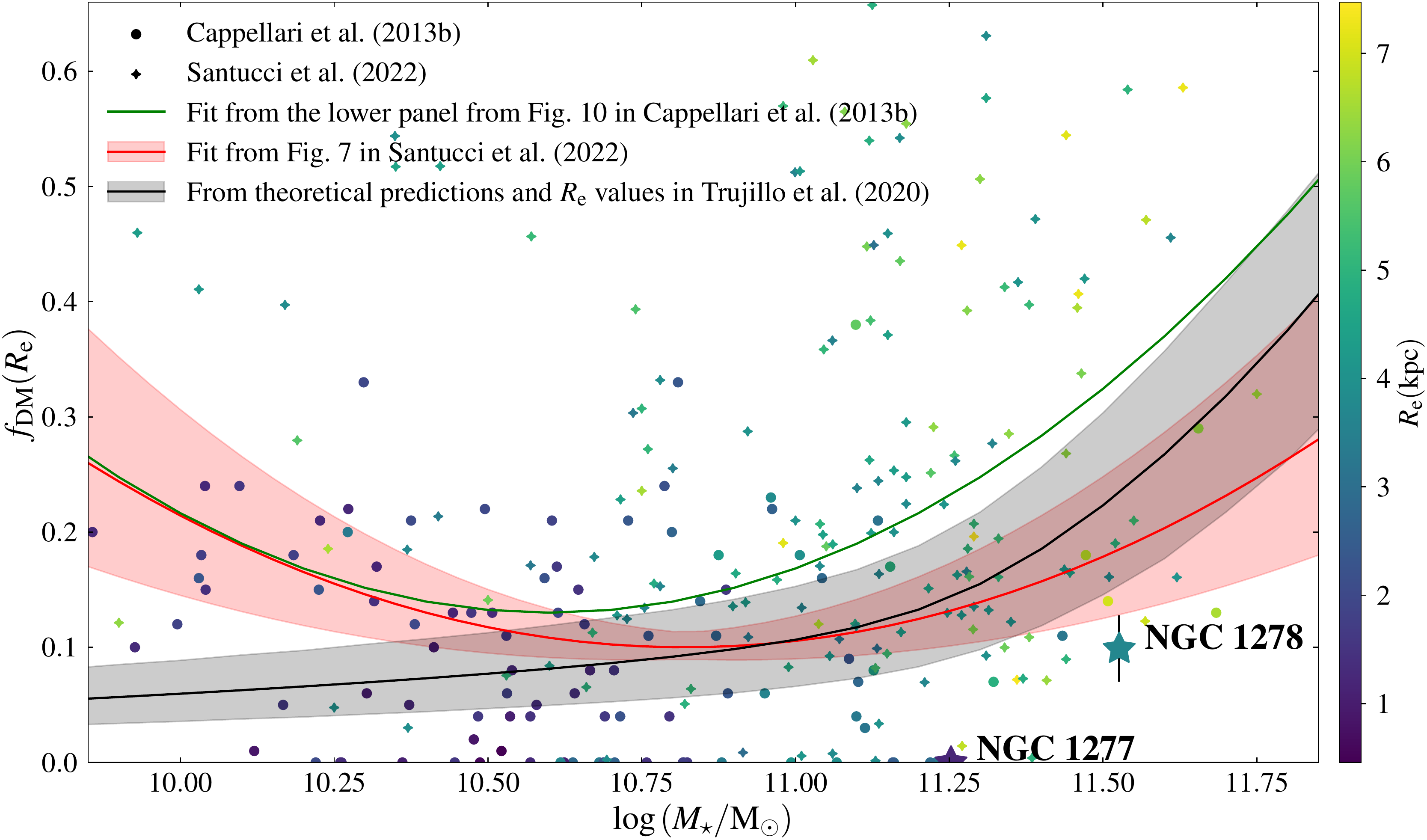}\\
  \end{center}
  \caption{\label{fdm} Dark matter fraction within the effective radius, $f_{\rm DM}(R_{\rm e})$, as a function of the stellar mass for NGC~1277 and NGC~1278 (star symbols). The error bars represent one sigma confidence intervals (that corresponding to NGC~1277 is smaller than the symbol size). Circles correspond to the 103 ATLAS$^{\rm 3D}$ galaxies with good fits in \citet{Cappellari2013} and the diamond symbols correspond to the data for a sample of 161 passive galaxies from the SAMI Galaxy Survey from \citet{Santucci2022}. The colour coding of the symbols indicates the effective radius of the galaxies as shown by the side-bar. The ATLAS$^{\rm 3D}$ data points were derived by merging the galaxy distances from \citet{Cappellari2011}, the effective radii and luminosities from \citep{Cappellari2013}, and the $f_{\rm DM}(R_{\rm e})$ and $\Upsilon_\star$ values from \citet{Cappellari2013a}. The effective radii for the SAMI galaxies were kindly provided by G.~Santucci. The green line indicates the $M_\star-f_{\rm DM}(R_{\rm e})$ relation derived from \texttt{JAM} models with a fixed cosmologically motivated NFW dark matter halo from \citet{Cappellari2013}. The red line and confidence band correspond to the fit represented in Fig.~7 from \citet{Santucci2022}. The black line shows a prediction of the $M_\star-f_{\rm DM}(R_{\rm e})$ relation made by combining the dark matter haloes derived from the SHMR from \citet{Girelli2020} and the dark matter halo parametrisation from \citet{Child2018} with the $R_{\rm e}$ values derived from a linear fit (Eq.~\ref{res_trujillo}) to the size measurements for galaxies with a morphological type $T\leq-1$ in \citet{Trujillo2020}. The latter relation was calculated assuming a \citet{Vaucouleurs1948} profile for the stellar  component and it is complemented by a one-sigma uncertainty band calculated using a Montecarlo simulation. For the \citet{Trujillo2020} and the \citet{Santucci2022} data and the \citet{Girelli2020} SHMR, the stellar masses were corrected by a factor 0.23\,dex to convert the assumed \citet{Chabrier2003} IMF into a \citet{Salpeter1955} one (see Eq.~\ref{mass2} in Sect.~\ref{mges}).}
\end{figure*}

\begin{figure}
\begin{center}
  \includegraphics[scale=0.48]{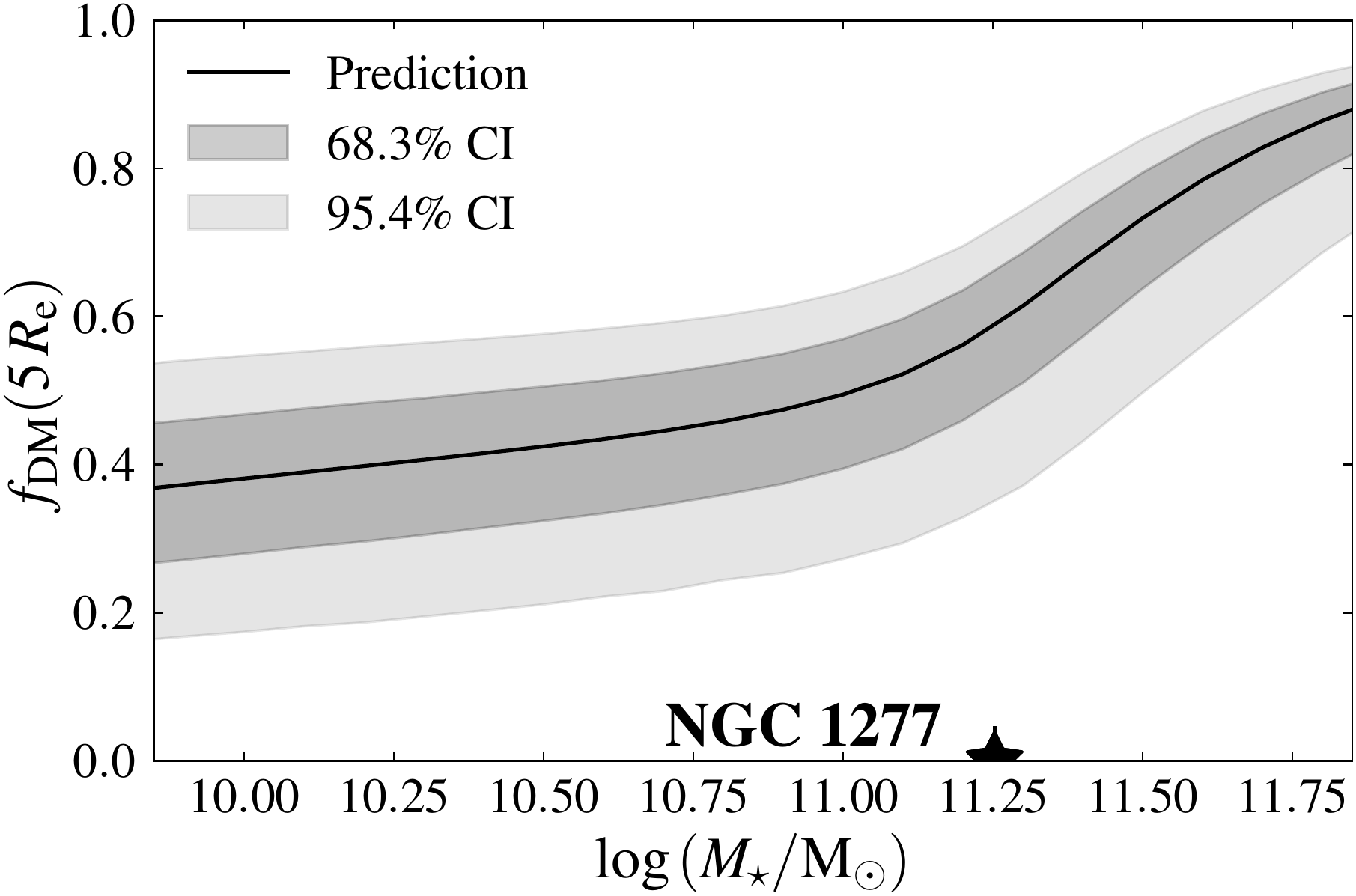}\\
  \end{center}
  \caption{\label{fdm5Re} Dark matter fraction within five effective radii $f_{\rm DM}(5\,R_{\rm e})$ for NGC~1277 (star) with the 95.4\% confidence interval indicated by a vertical line. The black line indicates the predicted $M_\star-f_{\rm DM}(5\,R_{\rm e})$ relation based on the dark matter haloes profiles and $R_{\rm e}$ values derived as in Fig.~\ref{fdm}. The grey shades indicate the 68.3\% (darker) and 95.4\% (lighter) confidence intervals of the relation calculated using a Montecarlo simulation. A \citet{Vaucouleurs1948} profile was assumed for the stellar density profile of the galaxies.}
\end{figure}

\begin{figure}
\begin{center}
  \includegraphics[scale=0.48]{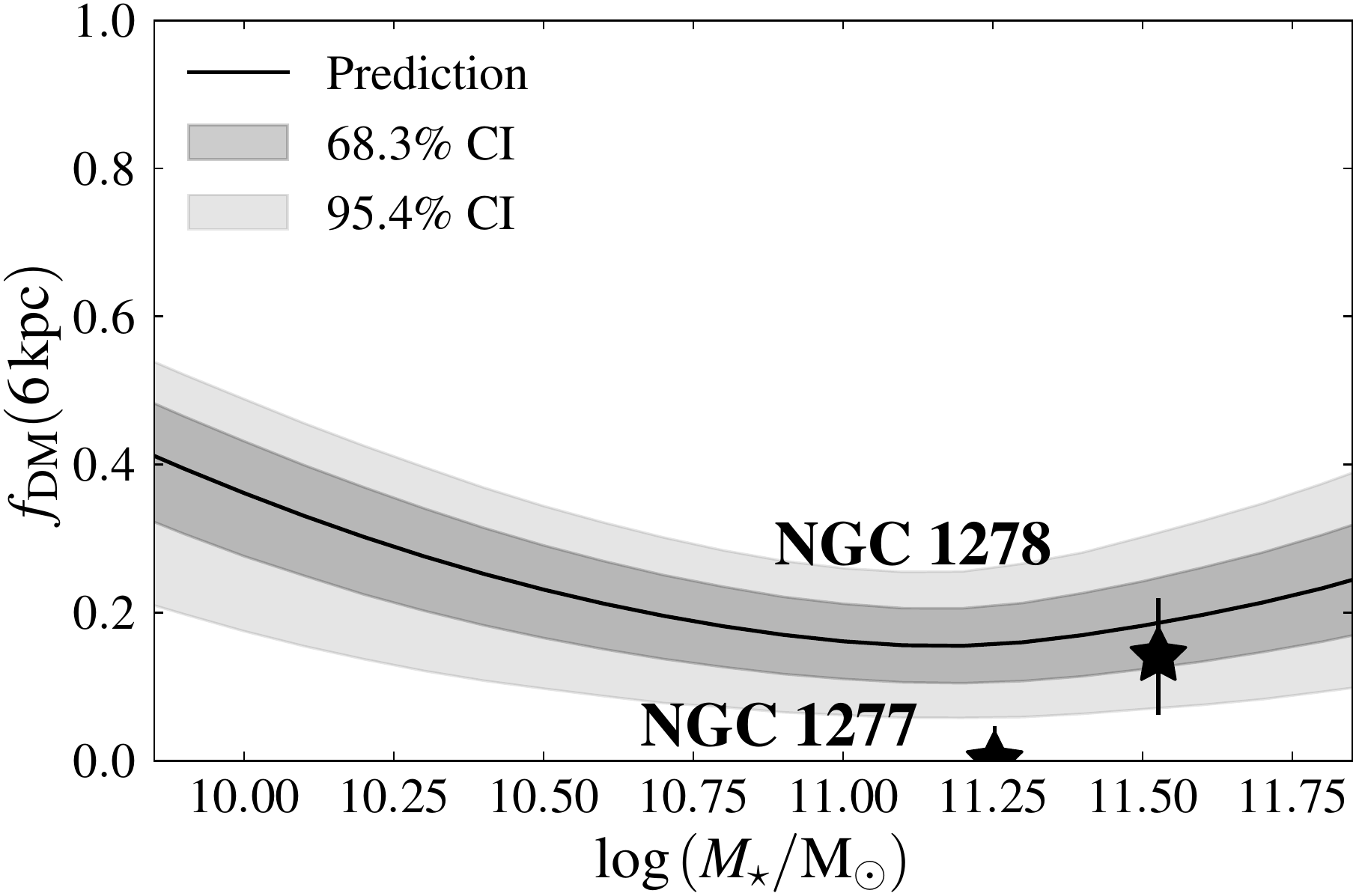}\\
  \end{center}
  \caption{\label{fdm6kpc} Same as Fig.~\ref{fdm5Re}, but this time for the dark matter fraction enclosed within 6\,kpc, $f_{\rm DM}(6\,{\rm kpc})$, and including the data point corresponding to NGC~1278.}
\end{figure}

\setcitestyle{notesep={}}

In this section, we compare the fitted dark matter fractions of NGC~1277 and NGC~1278 with those of other massive ETGs from the literature, as well as with expectations from combining theoretical models with measurements of the sizes of massive ETGs. Our deep observations allow us to explore as far as $5\,R_{\rm e}$ for NGC~1277 and $2.5\,R_{\rm e}$ for NGC~1278. However, studies of large samples \citep[e.g.][]{Cappellari2013, Santucci2022}, typically only cover the central regions of the galaxies. Thus, here we compare the dark matter fractions within $R_{\rm e}$, $f_{\rm DM}(R_{\rm e})$. Given the small radius probed by these studies and the increased fraction of baryons as we approach the central regions of galaxies, this dark matter fraction comparison is of limited interest because these previous studies sampled radii where dark matter usually plays a minor role.

In Fig.~\ref{fdm}, we compare the dark matter fractions $f_{\rm DM}(R_{\rm e})$ for NGC~1277 and NGC~1278 to the values obtained for a sub-sample of 103 galaxies from the ETG survey ATLAS$^{\rm 3D}$ \citep{Cappellari2011} shown in the top panel of Fig.~10 from \citet{Cappellari2013} and a sample of 161 passive galaxies from the SAMI Galaxy Survey \citep{Croom2012, Bryant2015} shown in Fig.~7 from \citet{Santucci2022}. The former data points correspond to the fits deemed to be good by the authors and come from \texttt{JAM} models similar to ours. The stellar masses of the galaxies from \citet{Santucci2022} were positively shifted by 0.23\,dex to convert their assumed \citet{Chabrier2003} IMF into a \citet{Salpeter1955} one as we do in Sect.~\ref{mges}. Fig.~\ref{fdm} also shows two $M_\star-f_{\rm DM}(R_{\rm e})$ relations derived from the above-mentioned works. The green line comes from the lower panel in Fig.~10 in \citet{Cappellari2013}, and was derived from \texttt{JAM} fits where the NFW halo parameters were fixed to cosmologically motivated values obtained from \citet{Moster2010}. The red line indicates the fit from Fig.~7 in \citet{Santucci2022} and is again shifted by 0.23\,dex.

Figure~\ref{fdm} also displays a third $M_\star-f_{\rm DM}(R_{\rm e})$ relation (black line) that was obtained from the dark matter haloes predicted by combining the SHMR from \citet[][, which is corrected to a Salpeter IMF too]{Girelli2020} and the dark matter halo parametrisation from \citet[][, see Sect.~\ref{dmhs}]{Child2018}. In order to assign an effective radius to each galaxy stellar mass, we produced a linear fit between ${\rm log}\,R_{\rm e}$ and ${\rm log}\,M_\star$ using the observational data for ETGs (Hubble-Reynolds type $T\leq1$) from \citet{Trujillo2020}. After correcting $M_\star$ by 0.23\,dex to shift the IMF to a Salpeter one, we obtained
\begin{equation}
\label{res_trujillo}
{\rm log}\,R_{\rm e}=-5.0+0.51\,{\rm log}\,M_\star,
\end{equation}
where $R_{\rm e}$ is in kpc and $M_\star$ in ${\rm M}_\odot$. To estimate the fraction $X_\star(r=R_{\rm e})$ of the stellar mass that is found within a three-dimensional radius $r=R_{\rm e}$ we assumed a \citet{Sersic1963} profile with an index $n=4$ \citep[a][ profile]{Vaucouleurs1948}. The fraction $X_\star$ for the $n=4$ specific case  was calculated by combining Eqs.~30 and 41 and from \citet{Baes2011}
\begin{equation}
 X_\star(r)=1.523\left(\frac{r}{R_{\rm e}}\right)^2G^{8,1}_{1,9}\left[\begin{array}{cc}0 & \\ \vec{b},&-1\end{array}\left|0.7132\left(\frac{r}{R_{\rm e}}\right)^2\right.\right],
\end{equation}
where $G$ is the Meijer~G function \citep{Meijer1936} and
\begin{equation}
 \vec{b}=\left(\frac{1}{8}, \frac{2}{8}, \frac{3}{8}, \frac{4}{8}, \frac{5}{8}, \frac{6}{8}, \frac{7}{8}, \frac{1}{2}\right).
\end{equation}
We found that for a \citet{Vaucouleurs1948} profile $X_\star(R_{\rm e})=0.415$. The $M_\star-f_{\rm DM}(R_{\rm e})$ relation is complemented by a one-sigma uncertainty band that was calculated with a Montecarlo simulation using the uncertainties quoted by \citet{Girelli2020} and \citet{Child2018}, as well as an intrinsic sigma of 0.057 in ${\rm log}\,R_{\rm e}$ from Table~3 in \citet{Trujillo2020}.

We found that the $f_{\rm DM}(R_{\rm e})$ value for NGC~1278 is compatible with the $M_\star-f_{\rm DM}(R_{\rm e})$ relation from \citet{Santucci2022}. NGC~1278 is below  the $M_\star-f_{\rm DM}(R_{\rm e})$ relations from \citet{Cappellari2013}, but is within the scatter of the galaxies in the lower panel of their Fig.~10. NGC~1278 also falls below the two other $M_\star-f_{\rm DM}(R_{\rm e})$ relations examined in Fig.~\ref{fdm}. The fact that NGC~1278 falls below all the $M_\star-f_{\rm DM}(R_{\rm e})$ relations while it has a dark matter halo compatible with the expectation for its mass (Fig.~\ref{DM_NGC1278}) is because its effective radius ($R_{\rm e}=4.2\,{\rm kpc}$) is smaller than the average for ETGs of its mass ($R_{\rm e}=7.6\,{\rm kpc}$ from Eq.~\ref{res_trujillo}).

NGC~1277 is well below the three $M_\star-f_{\rm DM}(R_{\rm e})$ relations shown in Fig.~\ref{fdm} with $f_{\rm DM}(R_{\rm e})<0.005$ (95.4\% confidence interval). Based on the $M_\star-f_{\rm DM}(R_{\rm e})$ relation that we established using models and the effective radii in \citet{Trujillo2020}, the expected dark matter fraction of a galaxy with the stellar mass of NGC~1277 is $f_{\rm DM}(R_{\rm e})=0.14^{+0.06}_{-0.05}$ at a one-sigma confidence level and $f_{\rm DM}(R_{\rm e})=0.14^{+0.13}_{-0.10}$ at a two-sigma confidence level. Calculating $f_{\rm DM}(R_{\rm e})$ is another way of demonstrating the dark matter deficient nature of this object in comparison with other galaxies with a similar stellar mass. However, as NGC~1277 is very compact, we emphasise that its extreme lack of dark matter is only fully demonstrated by finding little to no dark matter within 6\,kpc ($5\,R_{\rm e}$).

If two galaxies have the same dark matter halo and the same stellar mass, the most compact one has the smallest $f_{\rm DM}(R_{\rm e})$. We colour-coded the symbols in Fig.~\ref{fdm} according to their $R_{\rm e}$ to check whether galaxies with a small $f_{\rm DM}(R_{\rm e})$ are exclusively compact ($R_{\rm e}\approx1\,{\rm kpc}$). We find that galaxies with a small $f_{\rm DM}(R_{\rm e})$ can be relatively extended (with effective radii of a few kpc), and that some compact galaxies have a non-negligible central dark matter fraction. Therefore, the reason why \citet{Cappellari2013} and \citet{Santucci2022} find galaxies with $f_{\rm DM}(R_{\rm e})\approx0$ cannot be solely the compactness of the objects. Instead, these small $f_{\rm DM}(R_{\rm e})$ values are likely because the fraction of dark matter within $1\,R_{\rm e}$ is so small, that it becomes hard to constrain it. Indeed, ATLAS$^{\rm 3D}$ was designed to observe galaxies out to $1\,R_{\rm e}$ \citep{Cappellari2011} and most of the \citet{Santucci2022} galaxies are covered out to less than $2\,R_{\rm e}$ (see their Fig.~1). Figure~4 in \citet{Cappellari2015} illustrates this point by showing that, whereas the stellar and the total mass profiles of massive ETGs do not differ much in the inner $1\,R_{\rm e}$, the stellar density profile falls much faster than the total one in the outskirts. Hence, our large radial coverage ($5\,R_{\rm e}$ for NGC~1277 and $2.5\,R_{\rm e}$ for NGC~1278) ensures a precision in the characterisation of the dark matter component that cannot be achieved if we are limited to the inner $1-2\,R_{\rm e}$.

To stress the importance of the fact that we find little to no dark matter within $5\,R_{\rm e}$ for NGC~1277, in Fig.~\ref{fdm5Re} we compare the properties of NGC~1277 with a $M_\star-f_{\rm DM}(5\,R_{\rm e})$ relation computed by combining the works by \citet{Child2018}, \citet{Girelli2020}, and \citet{Trujillo2020} as done for the $M_\star-f_{\rm DM}(R_{\rm e})$ relation in black in Fig.~\ref{fdm}. This time the fraction of the stellar mass lying within the radius of interest is $X_\star(5\,R_{\rm e})=0.842$. Additionally, we used a Montecarlo simulation to estimate confidence intervals, yielding $f_{\rm DM}(5\,R_{\rm e})=0.59^{+0.07}_{-0.10}$ (at a 68.3\% confidence interval) and $f_{\rm DM}(5\,R_{\rm e})=0.59^{+0.13}_{-0.24}$ (at a 95.4\% confidence interval) for an ETG with the stellar mass of NGC~1277. This is clearly incompatible with the two-sigma limit $f_{\rm DM}(5\,R_{\rm e})<0.05$ obtained from our dynamical fit (Sect.~\ref{dmhs}).

As NGC~1277 is very compact compared to a typical galaxy of its mass \citep[1.2\,kpc versus 5.5\,kpc from the $M_\star-R_{\rm e}$ relation derived from][]{Trujillo2020}, the lack of dark matter within $5\,R_{\rm e}$ could be interpreted as a consequence of the high density of the stellar component. To emphasise that NGC~1277 is genuinely dark matter deficient, the dark matter fraction within a given physical radius could be checked instead. In Fig.~\ref{fdm6kpc} we compare $f_{\rm DM}(6\,{\rm kpc})$ for NGC~1277 and NGC~1278 with that of a prediction calculated as for Figs.~\ref{fdm} and \ref{fdm5Re}. Here, $X_{\star}(6\,{\rm kpc})$ had to be computed in a stellar mass by stellar mass basis, as this fraction varies as a function of $6\,{\rm kpc}/R_{\rm e}$. We again show that, whereas NGC~1278 falls within the expected trend, NGC~1277 is below it. Although the discrepancy between the fit and the prediction is not as large as for the $M_\star-f_{\rm DM}(5\,R_{\rm e})$ relation, a tension is still found ($f_{\rm DM}(6\,{\rm kpc})<0.05$ two-sigma limit versus $f_{\rm DM}(6\,{\rm kpc})=0.16\pm0.05$ and $f_{\rm DM}(6\,{\rm kpc})=0.16\pm0.10$ at 68.3\% and 95.4\% confidence intervals, respectively).

The tension between the measured dark matter fraction and the predictions would pose a challenge for $\Lambda$CDM if NGC~1277 were found to be incompatible with being a statistical outlier. According to \citet{Quilis2013}, relic galaxies (defined as those that have grown less than 10\% in mass through mergers) are expected to be today one in one thousand among massive galaxies with $M_\star>8\times10^{10}\,{\rm M}_\odot$. Assuming a normal distribution, relic galaxies would be 3.3-$\sigma$ outliers of the distribution. If we use this number to gauge the level of tension between the measured dark matter fraction in NGC~1277 and that predicted from the simulations, we find that the answer depends on whether the dark matter fraction is measured within a given physical scale or within a given number of effective radii. Whereas in the former case NGC~1277 might be compatible with being a statistical outlier (the upper two-sigma limit of the dark matter fraction in NGC~1277 is close to the lower two-sigma limit of the expectation), in the latter one there is a strong (more than 5\,$\sigma$) tension. Given that we lack a clear-cut answer about the relevant radius within which the dark matter fraction should be measured, we are refraining from speculating whether the tension observed for NGC~1277 challenges the current cosmological paradigm.

The decaying rotation curve of NGC~1277 (Fig.~\ref{rotcurve}) bears a qualitative resemblance with those of the high-redshift massive star-forming galaxies studied by \citet{Genzel2017}. The latter six galaxies are also found to be dark matter-poor, with $f_{\rm DM}(R_{\rm e})\lesssim0.2$ and compatible with zero in four cases. However, these galaxies cannot be straightforwardly compared to NGC~1277. The reason is that the typical effective radius in the \citet{Genzel2017} sample is $R_{\rm e}\approx8\,{\rm kpc}$, so these objects are not nearly as compact as NGC~1277. Also, in five out of six galaxies, the data coverage stops at $R\approx2\,R_{\rm e}$, so it misses the galactic outskirts where dark matter is presumed to be playing a major role and might be able to provide tight constraints on $f_{\rm DM}(R_{\rm e})$.

In conclusion, we find that NGC~1277 lacks dark matter within $1\,R_{\rm e}$, $5\,R{\rm e}$, and $6\,{\rm kpc}$ when compared to the expectations from models. On the other hand, we find that NGC~1278 behaves according to what is expected for a massive ETG.

\setcitestyle{notesep={, }}

\subsection{How could a compact dark matter-deficient galaxy form?}

We have found evidence that, in addition to being the best massive relic candidate in the local Universe, NGC~1277 is dark matter deficient (it has a fraction of dark matter within $5\,R_{\rm e}$ of less than 5\% at a two sigma confidence level, whereas the expected value is $\sim60\%$; Sect.~\ref{sectfdm} and Fig.~\ref{fdm5Re}). The principle of parsimony suggests that the two oddities have a common cause (or that one is the consequence of the other). However, another relic, Mrk~1216 \citep{FerreMateu2017}, has been found to contain significant fractions of dark matter within $R_{\rm e}$ \citep{Yildirim2015, Buote2019}. The situation poses a conundrum: either 1) relic galaxies can form with largely different dark matter contents, or 2) NGC~1277 and Mrk~1216 are different kinds of objects born and/or evolved under different circumstances. The latter possibility is supported by the work by \citet{FerreMateu2017}, who found Mrk~1216 to be the galaxy with the smallest `degree of relic' (that is, the one that is the most similar to a regular ETG) among their three-galaxy sample (which includes NGC~1277). They also found that the burst of star formation that generated the bulk of the baryonic mass in Mrk~1216 was more extended in time than for the other galaxies in the sample, as well as traces of recent accretion. As a consequence, they posit that Mrk~1216 is a relic that is on its way to become a `regular' massive ETG \citep[see also][]{Yildirim2015}.

How did NGC~1277, one of the densest known galaxies, form within a not so dense dark matter halo? This seems counterintuitive and, according to a study by \citet{FloresFreitas2022} based on the TNG50 cosmological simulation \citep{Pillepich2019, Nelson2019}, the progenitors of present-day massive relic galaxies inhabited at $z\sim4$ dark matter haloes that were more massive than those of an average quiescent galaxy progenitor with a similar stellar mass. Also, Figure~7 in \citet{Lapiner2023} shows that simulations predict that even the most compact blue nuggets have non-negligible dark matter fractions ($f_{\rm DM}(R_{\rm e})\gtrsim0.1$).

If it is possible for a few rare galaxies not to have dark matter, it becomes easy to explain why they do not undergo the envelope accretion phase. In the absence of a halo, nearby galaxies undergo parabolic flybys instead of spiralling-in due to dynamical friction \citep{Chandrasekhar1943}. This hypothesis would hold even if we had underestimated $f_{\rm DM}(5\,R_{\rm e})$, because the stripping of the outskirts of the halo (see discussion below) would still be a plausible mechanism to explain the lack of accretion. On the other hand, regular massive ETGs like NGC~1278 are effective at cannibalising small dark matter rich-galaxies, which impacts both $R_{\rm e}$ and $f_{\rm DM}$ \citep[e.g.][]{Hilz2012, Tortora2022}. If we assume that the lack of an envelope in NGC~1277 is due to a dark matter deficiency, it must have either lost its dark matter early (before significant accretion occurred) or was dark matter deficient ab initio. Hence, the big question about NGC~1277 shifts from `how did it avoid to accrete an envelope?' to `why is it dark matter deficient?'.

If dynamical friction were suppressed in NGC~1277, it would be extremely unlikely that its central black hole were accreted as suggested by \citet{Shields2013}. Its relatively large mass should be then attributed to the fact that the black hole-related scaling relations are established based on galaxies that have accreted an envelope. Indeed, present-day regular massive ETGs are significantly more massive than they were in their blue nugget phase.

A possible way to generate a dark matter-poor galaxy that resembles a massive relic was proposed by \citet{FloresFreitas2022}. They used the TNG50 cosmological simulation to identify five relic analogues, of which the one residing in subhalo~6 is the purest and the most similar to NGC~1277. The simulated galaxy is born in isolation within a massive $\sim10^{12}\,{\rm M}_\odot$ dark matter halo that is significantly stripped as it enters a cluster. As the galaxy loses most ($\sim90\%$) of its dark matter, it also leaves behind most of its gas, which prevents further star formation from happening. During this process, the stellar body of the galaxy remains mostly intact, as it loses only about $20\%$ of its mass. The resulting galaxy has 1.4 times more mass in dark matter than in stars  \citep{FloresFreitas2022}, and virtually all the dark matter outside a radius of 30\,kpc has been stripped away (Flore-Freitas, private communication). This phenomenon is similar to the stripping of dark matter of satellites as they interact with their main galaxy \citep{Engler2021}. \citet{Moreno2022} also explored simulated dwarf satellite galaxies that lose most of their dark matter as they have a close interaction with their parent galaxy. They proposed that in such events the stellar component is harder to remove than the dark matter one because stellar orbits are more circular than those of dark matter particles. Hence, repeated interactions would be efficient at stripping dark matter particles when they are close to their apocentre.

The main caveat of the \citet{FloresFreitas2022} scenario is given by the authors themselves. Indeed, as subhalo~6 falls within the cluster, the apparent stripping might in part be caused by some of the material being assigned to the parent halo. Thus, the mass loss could be an artefact introduced by the algorithm assigning particles to the different dark matter haloes. Another problem is that, whereas the outer parts of the halo are efficiently stripped, the dark matter mass within the inner 6\,kpc diminishes by a factor of less than two (Flores-Freitas; private communication), so the mechanism does not seem sufficient to remove most of the dark matter from the central parts of the galaxy. On the other hand, the stripping of the outer parts of the halo would probably be sufficient to greatly reduce dynamical friction and perhaps prevent the accretion of an envelope. A possible way around all these caveats is that subhalo~6 contains the best relic analogue within the TNG50 volume, but this volume of $1.4\times10^5\,{\rm Mpc}^3$ is relatively small compared to the observed local number density of relics \citep[three in $5\times10^6\,{\rm Mpc}^3$;][]{FerreMateu2017} and might be insufficient to cover the most rare and extreme objects.

A scenario similar to that put forward by \citet{FloresFreitas2022} was given by \citet{Yu2018}. These authors studied the Illustris simulation \citep{Vogelsberger2014} searching for dark matter-deficient galaxies. They found a galaxy akin to NGC~1277 that is, compact ($R_{\rm e}\approx2\,{\rm kpc}$), massive ($M_\star=9.42\times10^{10}\,{\rm M}_\odot$), and with a total dark matter mass fraction below 0.1. According to \citet{Yu2018}, a strong interaction with the environment at redshift $z<1$ would have stripped about half of the stars from the galaxy and most of its dark matter (their Fig.~2). The result would be a dense compact galaxy with little dark matter. Only one such galaxy is found within the $(106.5\,{\rm Mpc})^3\approx1.2\times10^6\,{\rm Mpc}^{3}$ Illustris box, which is tantalisingly close to the observational density constraints by \citet{FerreMateu2017} and of the same order of magnitude as the number density of relics estimated by \citet{Quilis2013}. It is also compatible with the estimated density of massive ultra-compact galaxies for the $z<0.3$ Universe of $(1.0\pm0.4)\times10^{-6}\,{\rm Mpc}^{-3}$ from \citet{Buitrago2008} and the lower limit for the relic density of one relic within $10^7\,{\rm Mpc}^3$ from \citet{Spiniello2021a}.

A possible caveat of the work by \citet{Yu2018} is that the Illustris simulation yields lower dark matter fractions for ETGs than those measured through gravitational lensing \citep{Mukherjee2022}. Another caveat is that, since the mass loss in their scenario occurred quite late ($z<1$), the interaction should have stripped both the envelope and the population of blue globular clusters \citep[discovered to be absent in][]{Beasley2018}. This possibility is discussed in \citet{Beasley2018} and found to be unlikely because, whereas tidal stripping could explain the lack of an envelope, it could not explain the extreme density of the centre of the galaxy. As a consequence, they conclude that NGC~1277 was already dense and compact at birth. Another sign that NGC~1277 never had a significant envelope can be found by comparing this galaxy with the regular ETG NGC~1278. Indeed, we find that NGC~1278 has signs of a relatively young ($<10\,{\rm Gyr}$) accreted stellar population all the way to its centre (Ferré-Mateu et al.~in prep.) which should be observed in NGC~1277 too if it ever had an envelope that was later lost. Furthermore, NGC~1277 lacks shells and streams down to $\mu_r=26.8\,{\rm mag\,arcsec^{-2}}$ \citep[implying that no more than 2\% of the stellar mass of the galaxy is missing from the deepest {\it HST} imaging;][]{Trujillo2014}. This indicates a paucity of interactions affecting the stellar component in the last few Gyr \citep{Mancillas2019}. Within the framework of the two-phase formation scenario the simplest way to reconcile the stripping hypothesis with the observations is by assuming that the dark matter loss occurred very early \citep[as seen in the work by][]{FloresFreitas2022}, before NGC~1277 had time to accrete a significant envelope.

In a bottom-up hierarchical paradigm, galaxies grow by mergers. In the early Universe, galaxies were richer in gas than nowadays and mergers and cold streams caused a wet compaction that led to the formation of the compact high-redshift galaxies known as blue nuggets \citep[e.g.][]{Lapiner2023}. It is possible that for a few major mergers the geometry of the encounter (impact parameter and relative velocity) were such that the collisionless dark matter would escape the system. The escape velocity from a $10^{11}\,{\rm M}_\odot$ body at a distance of $1\,{\rm kpc}$ is $\approx1000\,{\rm km\,s^{-1}}$, so under this hypothesis the relative velocity of the two bodies should be of the order of $2000\,{\rm km\,s^{-1}}$ in order to remove the dark matter from the system. Whereas these relative velocities can hardly be found in the field, they are consistent with the high-velocity tail within rich clusters. For example the Perseus cluster where NGC~1277 resides, is a rich cluster \citep{Voelk2000} with a velocity dispersion of $\sigma=1288^{+90}_{-77}\,{\rm km\,s^{-1}}$ \citep{Girardi1993}. Hence, we speculate that infrequent high-velocity collisions during the proto-cluster phase are possible. Indeed, assuming a Hubble-Lemaître parameter at $z=2$, $H(z=2)\approx200\,{\rm km\,s^{-1}}$, we obtain from the relations in \citet{Ferragamo2022} that to reach a velocity dispersion of $1000\,{\rm km\,s^{-1}}$ (so a few outliers have velocities of $2000\,{\rm km\,s^{-1}}$), a cluster mass of $3\times10^{14}\,{\rm M}_\odot$ is sufficient. This mass is less than half the current mass of the Perseus cluster, $M=8.5\times10^{14}\,{\rm M}_\odot$ \citep{Mathews2006}.

In this high-velocity collision scenario, the situation would be somewhat similar to that of the Bullet Cluster \citep{Clowe2004, Markevitch2004}, where dissipative matter (gas) is left behind. However, the analogy cannot be pushed too far because the gas in the blue nugget progenitor fragments would be orders of magnitude denser and cooler than that in the intra-cluster medium. 

In the high-velocity encounter framework, the difference between the progenitor of NGC~1277 and that of a blue nugget on its way to become a regular ETG would be that the latter keeps most of its dark matter. This might be because the mergers that generated the progenitors of regular ETGs were mostly minor \citep[as suggested by][]{Zolotov2015} or because the relative velocity of the merging fragments was not large enough to prevent the dark matter from falling back.
\setcitestyle{notesep={ }}

The high-velocity merger hypothesis has of course several caveats. The first one is that NGC~1277 looks like the passively evolved product of the average blue nugget which is presumed to have a regular fraction of dark matter. As a consequence, if our suggestion is valid, young massive ETGs should look similar irrespective of their dark matter content. This is not completely unreasonable because the dark matter fraction in the central regions of blue nuggets and massive ETGs is often small \citep[][and Fig.~\ref{fdm}, respectively]{Lapiner2023}. However, it remains to be seen whether the lack of a dense extended dark matter halo would not result in the loss of many of the baryons at the onset of intense star formation. Another problem of this scenario is related to the time-scales involved in the process. On one hand, intense star formation would likely be triggered by the collision. On the other hand, the star formation time-scale should be longer than the dynamical time-scale ($\tau_{\rm dyn}\sim1/\sqrt{G\overline{\rho}}\sim40\,{\rm Myr}$, where $\overline{\rho}$ is the average density of NGC~1277) for the gas to have time to settle in a disc in order to produce a mostly rotation-supported stellar body. Also, the density of the gas should have been high enough to ensure a fast cooling to prevent the escape of a large fraction of the gas shock-heated by the collision.

\setcitestyle{notesep={, }}

A third possibility to explain the apparent lack of dark matter in NGC~1277 is a strong departure from a NFW dark matter profile. Our data do not reach the depth required to probe the matter distribution beyond 6\,kpc. This leaves the door open for a halo with a core much flatter than that predicted by a NFW profile. It is known that density fluctuations introduced by AGN and stellar feedback at the centre of galaxies can erase cusps \citep[e.g.][]{Martizzi2013}. The extremely massive central black hole could have powered the feedback, but such processes are also efficient at expelling baryons, so they should have occurred after most of the gas was transformed into stars. Therefore it is unclear how an extremely concentrated stellar component could sit within a halo that is more cored than average unless a large amount of fresh gas were accreted after the AGN had modified the density profile. Furthermore, a cored halo would not prevent dynamical friction from happening, so it would fail to naturally explain the absence of an envelope. In this case, the lack of mergers would be the consequence of a fortuitous lack of nearby galaxies to interact with.

Finally, a mass-follows-light scenario could be considered. In this case the dark matter distribution would be similar to that of the baryons. The presence of dark matter would be observed as an increase in the mass-to-light ratio, which would manifest as a large mismatch parameter $\alpha$ and mimicking a bottom-heavy IMF. At first glance the high value of $\alpha$ for NGC~1277 makes this explanation plausible. However, the dynamically estimated mass-to-light ratio broadly agrees with that that can be extracted from the line-strength index-based study by \citet{MartinNavarro2015}. The fact that a bottom-heavy IMF is found using two completely independent methods leaves little room for a significant mass-follows-light dark matter component.

One might be tempted to use the fact that NGC~1277 lacks detectable dark matter to speculate about the possibility of the (in)existence of Milgromian dynamics \citep[also known as MOND;][]{Milgrom1983} or other alternatives to the $\Lambda$CDM paradigm. Given a centrally concentrated baryonic mass of $M_\star\approx1.6\times10^{11}\,{\rm M}_\odot$ and an acceleration constant $a_0=1.24\times10^{-10}\,{\rm m\,s^{-2}}$ \citep{McGaugh2011}, a radius $R=13\,{\rm kpc}$ should be explored to be able to probe the fully Milgromian regime. This is about twice the radius that we cover and therefore our data do not permit studying the Milgromian regime unless the Newtonian-Milgromian transition were not abrupt. In a Milgromiam framework the lack of dark matter could not be invoked to explain the absence of dynamical friction and the missing envelope and, as in the case of the cored halo discussed above, the lack of an envelope would have been caused by the paucity of mergers for this particular galaxy.

NGC~1277 is similar to a gigantic rotating globular cluster, where the stars are the main contributor to the mass. Our favoured hypothesis to explain the formation of such a peculiar object is that NGC~1277 was born with little dark matter or that it lost most of it very early at or soon after its blue nugget phase. A plausible mechanism for that is that most of the dark matter of the galaxy was lost through interactions as it entered and moved through the Perseus cluster. Alternatively, NGC~1277 could have formed in a high-velocity collision of gas-rich fragments, where baryons coalesced and formed stars as dark matter left the system because of a velocity exceeding the escape one.

\section{Summary and conclusions}

\label{conclusion}

Relic galaxies are fascinating objects thought to be the frozen remains of the high redshift massive ETGs ($z\gtrsim2$). Whereas according to the two-phase formation scenario for massive ETGs the vast majority of these objects have suffered a complex merger history and have accreted an envelope, relic galaxies are thought to be the relatively untouched and passively evolved descendants of the blue nuggets observed in the early Universe. Very few relic galaxies are known, the most pristine of which is NGC~1277 \citep{Trujillo2014}. NGC~1277 is a massive (stellar mass $M_\star\approx1.8\times10^{11}\,{\rm M}_\odot$) and centrally concentrated (effective radius $R_{\rm e}\approx1.2\,{\rm kpc}$) galaxy in the Perseus cluster.

We have revisited NGC~1277 using data from the fibre spectrograph GCMS at the 2.7\,m Harlan J.~Smith telescope (Fig.~\ref{map_snr}). The data are of unprecedented interest because the long 2.5\,hour exposure allows us to study the outskirts of the galaxy (going from a radius of $13^{\prime\prime}$ in literature studies up to $18^{\prime\prime}$ here). The field of view also covers the regular massive ETG NGC~1278, a galaxy that constitutes an excellent benchmark against which to compare NGC~1277. In order to complement the GCMS data of NGC~1277, we use kinematics of the innermost $1\farcs6\times1\farcs6$ of the galaxy obtained by \citet{Walsh2016} using the AO-assisted NIFS integral field spectrograph.
 
Our aim was to produce dynamical models of both NGC~1277 and NGC~1278 and recover their mass distribution. We used {\it HST} images to compute colour-derived mass density maps to be able to disentangle the dark and the baryonic matter. Both the surface brightness and the surface density maps were parametrised using multi-Gaussian expansions (\texttt{MGE}s; Tables~\ref{MGElum} and \ref{MGEbar} and Figs.~\ref{Surface_brightness_profiles} and \ref{Surface_density_profiles}) produced with the code by \citet{Cappellari2002}.

We fitted the $V_{\rm rms}\equiv\sqrt{V^2+\sigma^2}$ maps of the two galaxies (Figs.~\ref{NGC1277_kinmap} and \ref{NGC1278_kinmap}) by feeding Jeans anisotropic models provided by \texttt{JAM} \citep{Cappellari2008, Cappellari2020} into the \texttt{adamet} Bayesian analysis code \citep{Cappellari2013}. For each galaxy, we fitted (among other parameters) the mismatch parameter indicating how the mass-to-light ratio compares to that expected from a \citet{Salpeter1955} IMF, $\alpha$, and the dark matter fraction within 6\,kpc, $f_{\rm DM}(6\,{\rm kpc})$. The 6\,kpc radius matches the maximum coverage of the good-quality GCMS data of NGC~1277 and corresponds to five effective radii ($5\,R_{\rm e}$) for this particular galaxy. During the fitting process we assumed that the galaxies are axisymmetric and that they have a single mismatch parameter throughout their stellar body.

Our fits (Figs.~\ref{NGC1277_NIFS} and \ref{NGC1277_VIRUS}) show that NGC~1277 has a large mismatch parameter $\alpha\approx1.34$, which is a consequence of a bottom-heavy IMF. Our most surprising result is that there is little to no dark matter within the explored radius and that the maximum permissible dark matter fraction within 6\,kpc (corresponding to $5\,R_{\rm e}$) is $f_{\rm DM}(6\,{\rm kpc})=0.05$ (95.4\% confidence interval; Fig.~\ref{DM_NGC1277}), which is in tension with the $f_{\rm DM}(6\,{\rm kpc})=0.09$ expectation based on the $\Lambda$CDM cosmology as obtained from stellar-to-halo mass relations by \citet{Girelli2020} and the dark matter halo radial profile parametrisation in \citet{Child2018}. This low dark matter fraction is on an even stronger tension with the prediction of $f_{\rm DM}(5\,R_{\rm e})\sim0.6$ (Fig.~\ref{fdm5Re}) obtained from combining the above two works with a $M_\star-R_{\rm e}$ relation for massive ETGs extracted from the data in \citet{Trujillo2020}.

NGC~1278 has a mismatch parameter $\alpha\approx1.16$, which indicates a shallower IMF slope than for NGC~1277 (Fig.~\ref{NGC1278_VIRUS_free}). According to our fiducial fit, the dark matter fraction of NGC~1278 is $f_{\rm DM}(6\,{\rm kpc})=0.14\pm0.08$ (95.4\% confidence interval), which which agrees with the expectations from the $\Lambda$CDM cosmology (Fig.~\ref{DM_NGC1278}).

The large mismatch parameter of NGC~1277 indicates a high value of the stellar mass-to-light ratio, $\Upsilon_{\star r}=7.0\,{\rm M_\odot}/L_{\odot r}$. This provides a dynamically derived constraint to the IMF that is compatible with the bottom-heavy IMFs determined by \citet{MartinNavarro2015} and \citet{FerreMateu2017} using line-strength indices (Fig.~\ref{MLs}).

The lack of a dynamically significant dark matter component in NGC~1277 confirms the findings by \citet{Yildirim2015}. The fact that one of the galaxies with the densest stellar component lies within an underdense dark matter halo is extremely puzzling. On the other hand it explains why it has not accreted an envelope, since the lack of an extended dark matter component would shut down dynamical friction and hence the spiralling down of nearby dwarf galaxies. Hence, we propose that NGC~1277 either lost its dark matter early, before it accreted a significant envelope, or was dark matter deficient ab initio.

A possible way to create a dark matter-poor object such as NGC~1277 has been proposed by \citet{FloresFreitas2022} who, based on the TNG50 simulation, posited that relic galaxy analogues may have lost a large fraction of their dark matter through interactions with the environment as they entered a cluster. A similar scenario has been proposed by \citet{Yu2018} based on a search for dark matter deficient galaxies in the Illustris simulation. It is suggestive that the density of compact, massive and dark matter deficient galaxies in the Illustris simulation is very close to that found in the local Universe by \citet[][one in $\approx1.2\times10^{6}\,{\rm Mpc}^3$ versus three in $\approx5\times10^6\,{\rm Mpc}^3$]{FerreMateu2017} and compatible with other density estimates by \citet{Buitrago2018} and \citet{Spiniello2021a}. Simulations have also shown a similar mechanism at work for dwarf galaxies through interactions with their parent galaxies \citep{Engler2021, Moreno2022}. Alternatively, an NGC~1277-like object could have been born in a high-velocity collision of gas-rich objects where dark matter would have escaped the system leaving behind a rotating stellar body (blue nugget). The high relative velocities required for this process to work could only be found in a cluster-like environment and, based on \citet{Ferragamo2022}, we speculate that they might have been possible in the progenitors of the rich galaxy clusters at $z\sim2$.

We have found that NGC~1277 lacks any detectable dark matter within 6\,kpc (corresponding to five effective radii, $5\,R_{\rm e}$) of its centre. Massive dark matter-deficient galaxies are unexpected within the framework of the consensus cosmological model $\Lambda$CDM. We propose that the galaxy was dark matter deficient ab initio or soon after birth and that the relic nature of NGC~1277 stems from the lack of an extended halo. In order to confirm this hypothesis more deep integral field studies of pure relic galaxies are required. It would also be interesting to check whether even deeper observations of NGC~1277 can find traces of a tenuous dark matter halo by finding a flattening of the rotation curve at a radius $R\gtrsim10\,{\rm kpc}$.

\begin{acknowledgements}We thank the referee, Dr.~Eric Emsellem, for his very helpful comments that have helped strengthen this article.

We thank Giulia Santucci for kindly sharing information about Fig.~7 in \citet{Santucci2022}. We thank Rodrigo Flores-Freitas and Ana L.~Chies-Santos for sharing valuable insights about the intricacies of their work in \citet{FloresFreitas2022}. We thank Michalina Maksymowicz-Maciata for her careful reading of the paper that has led to the discovery of a couple of typos.

SC is indebted to all the participants at the `Cosmic nuggets -- A feast of compact and massive Galaxies across the Universe', in particular Dr.~Claudia Pulsoni, for the useful idea exchanges and for the references that were essential for the discussion in this paper. SC also thanks Drs.~Jorge S\'anchez Almeida, Alexandre Vazdekis, Claudio dalla Vecchia, and Tod Lauer for useful comments and enlightening discussions.

SC and MALL acknowledge support from the Ram\'on y Cajal programme funded by the Spanish Government (references RYC2020-030480-I and RYC2020-029354-I). SC acknowledges funding from the State Research Agency (AEI-MCINN) of the Spanish Ministry of Science and Innovation under the grants `The structure and evolution of galaxies and their central regions' with reference PID2019-105602GB-I00/10.13039/501100011033, and `Thick discs, relics of the infancy of galaxies' with reference PID2020-113213GA-I00. IT acknowledges support from the ACIISI, Consejer\'{i}a de Econom\'{i}a, Conocimiento y Empleo del Gobierno de Canarias and the European Regional Development Fund (ERDF) under grant with reference PROID2021010044 and from the State Research Agency (AEI-MCINN) of the Spanish Ministry of Science and Innovation under the grant PID2019-107427GB-C32 and  IAC project P/300624, financed by the Ministry of Science and Innovation, through the State Budget and by the Canary Islands Department of Economy, Knowledge and Employment, through the Regional Budget of the Autonomous Community. FB also acknowledges support from the grants PID2020-116188GA-I00 and PID2019-107427GB-C32 by the Spanish Ministry of Science and Innovation. MALL acknowledges support from the Spanish grant PID2021-123417OB-I00. AFM has received support from the Spanish Ministry grants CEX2019-000920-S, PID2021-123313NA-I00 and RYC2021-031099-I. This research was supported by the International Space Science Institute (ISSI) in Bern, through ISSI International Team project No.~505 (The Metal-THINGS Survey of Nearby Galaxies).

Based on observations made with the NASA/ESA {\it Hubble Space Telescope}, and obtained from the Hubble Legacy Archive, which is a collaboration between the Space Telescope Science Institute (STScI/NASA), the Space Telescope European Coordinating Facility (ST-ECF/ESA) and the Canadian Astronomy Data Centre (CADC/NRC/CSA). This research has made use of the NASA/IPAC Extragalactic Database (NED), which is funded by the National Aeronautics and Space Administration and operated by the California Institute of Technology. We acknowledge the usage of the HyperLeda database (\url{http://leda.univ-lyon1.fr}).

This research made use of \texttt{NumPy} \citep{Harris2020}, \texttt{SciPy} \citep{Virtanen2020}, \texttt{SymPy} \citep{Meurer2017}, and \texttt{Astropy},\footnote{\url{http://www.astropy.org}} a community-developed core Python package for Astronomy \citep{Collaboration2013, Collaboration2018}. This research made use of \texttt{Photutils}, an \texttt{Astropy} package for detection and photometry of astronomical sources \citep{Bradley2022}

This research was partially based on data from the MILES project.

This research has made use of the integral-field spectroscopy data-reduction tool \texttt{P3D}, which is provided by the Leibniz-Institut f\"ur Astrophysik Potsdam (AIP).

\end{acknowledgements}

\bibliographystyle{aa}
\bibliography{NGC1277}

\begin{appendix}

\section{Could the {\it JWST} be used to find the black hole masses of relic progenitors at $z\gtrsim1$?}

\label{appendix}

We have found that the mass of the black hole sitting at the centre of NGC~1277 is $M_{\rm BH}=\left(4.88\pm0.11\right)\times10^{9}\,{\rm M}_\odot$. This very precise determination was made using the NIFS data from \citet[][our cA1 fit for NGC~1277]{Walsh2016}. Surprisingly, if we use the GCMS data only we obtain a black hole mass of the right order of magnitude ($M_{\rm BH}=\left(2.6\pm1.0\right)\times10^{9}\,{\rm M}_\odot$; cC fit). Black hole masses of a few times $10^9\,{\rm M}_\odot$ are consistently found for both a cylindrically and a spherically aligned velocity ellipsoid, irrespective of whether we use the data from NIFS, from GCMS, or both datasets.

One might wonder whether these roughly concordant black hole mass values between fits resorting to datasets with a very different angular resolution is merely a coincidence. The fact that by using adaptive optics or the {\it JWST} we can now obtain integral field data of the progenitor of a massive ETG at $z\gtrsim1$ with a physical resolution comparable to that given by the GCMS for NGC~1277 makes it worth to study this issue a bit further.

A possible explanation for the GCMS data alone being able to constrain the black hole mass is that some high angular resolution information is introduced through \texttt{MGE} parametrisations of {\it HST} images. In order to simulate the \texttt{MGE}s of a galaxy at high redshift seen through the {\it JWST}, we first need to degrade the {\it HST} images.

\begin{table}[b]
\caption{Parameters of the \texttt{MGE} decompositions of the degraded surface brightness map of NGC~1277}
\label{MGElum_degraded} 
\centering                                      % used for centering table
\begin{tabular}{c c c}          % centered columns (4 columns)
\hline\hline                        % inserts double horizontal lines
${\rm log}\,I_0\,({\rm log}\,({\rm L}_{\odot}\,{\rm pc^{-2}}))$ & $\sigma_{\star}\,({\rm arcsec})$ & $q^{\prime}_{\star}$ \\    % table heading
\hline
4.3&0.346&0.650\\
3.6&1.139&0.650\\
3.3&1.835&0.650\\
2.4&3.506&0.650\\
2.8&3.660&0.450\\
2.4&6.915&0.450\\
2.1&8.775&0.546\\
1.7&15.633&0.650\\
\hline
\end{tabular}
\tablefoot{$I_0$ stands for the central surface brightness of the projected Gaussian components, $\sigma_{\star}$ denotes their dispersion, and $q^{\prime}_{\star}$ their projected axial ratio.}
\end{table}

\begin{table}[b]
\caption{Parameters of the \texttt{MGE} decompositions of the degraded surface mass density map of NGC~1277}
\label{MGEbar_degraded} 
\centering                                      % used for centering table
\begin{tabular}{c c c}          % centered columns (4 columns)
\hline\hline                        % inserts double horizontal lines
${\rm log}\,\Sigma_0\,({\rm log}\,({\rm M}_\odot\,{\rm pc^{-2}}))$ & $\sigma_{\star}\,({\rm arcsec})$ & $q^{\prime}_{\star}$ \\    % table heading
\hline
4.8&0.683&0.700\\
4.1&1.928&0.700\\
3.3&3.380&0.450\\
3.4&4.867&0.450\\
3.2&7.511&0.498\\
2.1&15.620&0.633\\
2.1&15.620&0.700\\
\hline
\end{tabular}
\tablefoot{Same as Table~\ref{MGElum_degraded}, but with $\Sigma_0$ indicating the projected central surface mass density of the Gaussian components.}
\end{table}

\begin{figure}[b]
\begin{center}
  \includegraphics[scale=0.48]{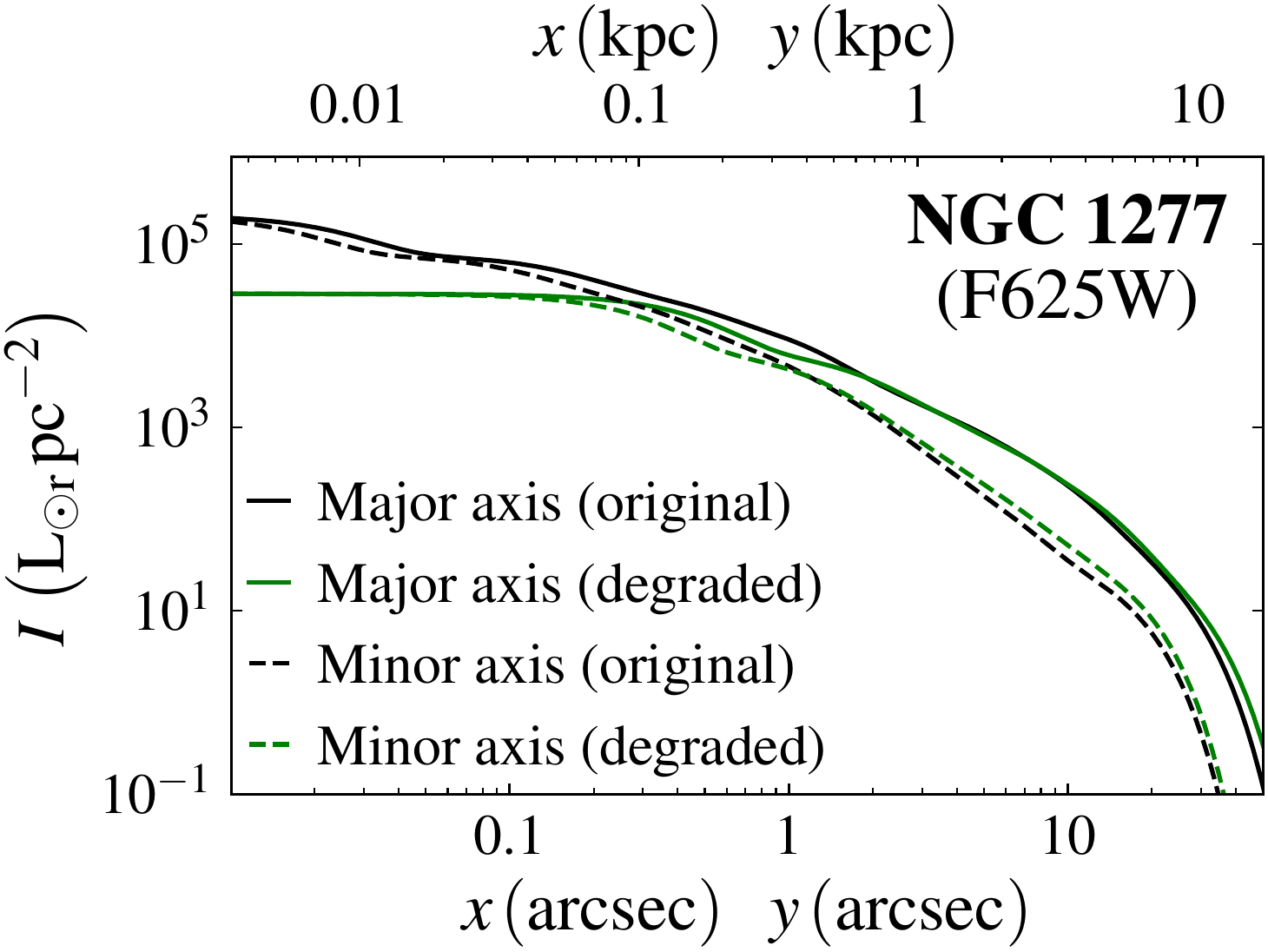}\\
  \end{center}
  \caption{\label{surface_brightness_degraded} Major- (continuous line and coordinate $x$) and minor-axis (dashed line and coordinate $y$) surface brightness profiles of NGC~1277. Black lines come from our original \texttt{MGE} parametrisation (Fig.~\ref{Surface_brightness_profiles} and Sect.~\ref{mges}) and green ones were obtained from {\it HST} data that were degraded to a resolution corresponding to that of a galaxy at $z=1.2$ as seen by the {\it JWST}.}
\end{figure}

\begin{figure}[!b]
\begin{center}
  \includegraphics[scale=0.48]{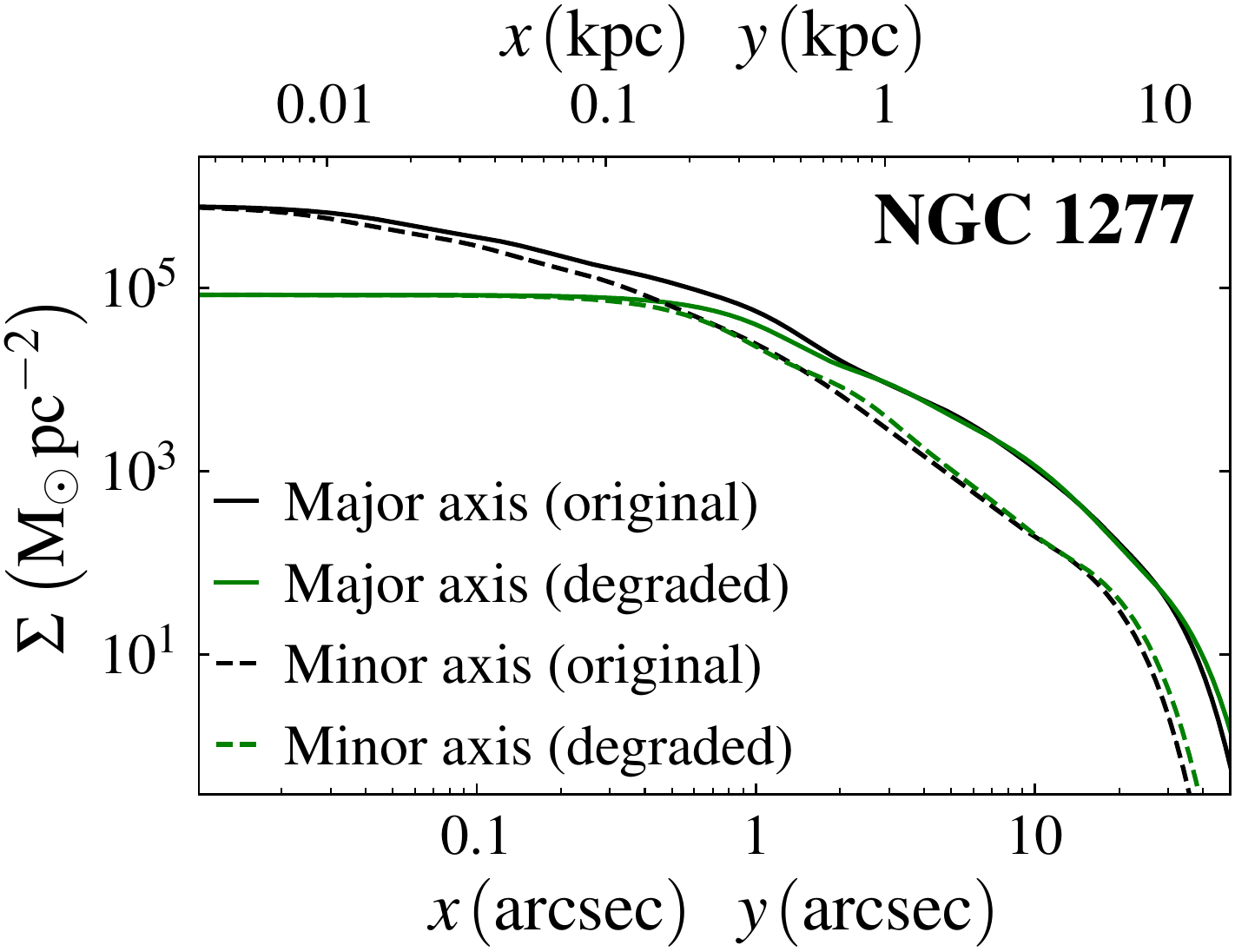}\\
  \end{center}
  \caption{\label{surface_density_degraded} As Fig.~\ref{surface_brightness_degraded}, but for the surface density profiles. As in Fig.~\ref{Surface_density_profiles}, a Salpeter IMF was assumed.}
\end{figure}

In the cosmology assumed in this paper, at $z=1.2$ one arcsecond corresponds to roughly 8.55\,kpc (compared to 345\,pc at the distance of NGC~1277). To reproduce how NGC~1277 would look at $z=1.2$ seen through the {\it JWST}, we degraded the {\it HST} F475W, F625W, and F850LP images by reducing their side by a factor 16 (that is, by coadding the flux from a box of 16\,pixels by 16\,pixels to generate a single pixel). The factor 16 comes from the difference in angular size ($8550\,{\rm pc/345}\,{\rm pc}\approx25$) and the $0\farcs05$ versus $0\farcs031$ difference in pixel size between the {\it HST} ACS and the {\it JWST} Near Infrared Camera \citep[NIRCam;][]{Rieke2005}. Then, we convolved the images with a {\it JWST} PSF simulated by WebbPSF \citep{Perrin2014}. Given the wavelength shift for a $z=1.2$ object, we convolved the F625W image with a F140M PSF. The mass map is built from the F475W and the F850LP filters (Sect.~\ref{mges}) that, given the redshift, would roughly correspond to the F115M and F182M filters, respectively. However, the PSF mismatch between these two filters is rather large and introduces artefacts in the colour map used to compute the mass map. This is why, we convolved the two filters with the F182M PSF, so we simulated a PSF matching. Effectively, the resampling and convolution corresponds to a degradation of the width of the core of the PSF of a factor of $\sim15$.

We produced a mask based on the degraded images, hence missing small stars and the circumnuclear dust ring that are obvious in the original {\it HST} images. The degraded F625W image and the mass map were parametrised with the \texttt{MGE}s listed in Tables~\ref{MGElum_degraded} and \ref{MGEbar_degraded}, respectively.

\begin{figure*}
\begin{center}
  \includegraphics[scale=0.48]{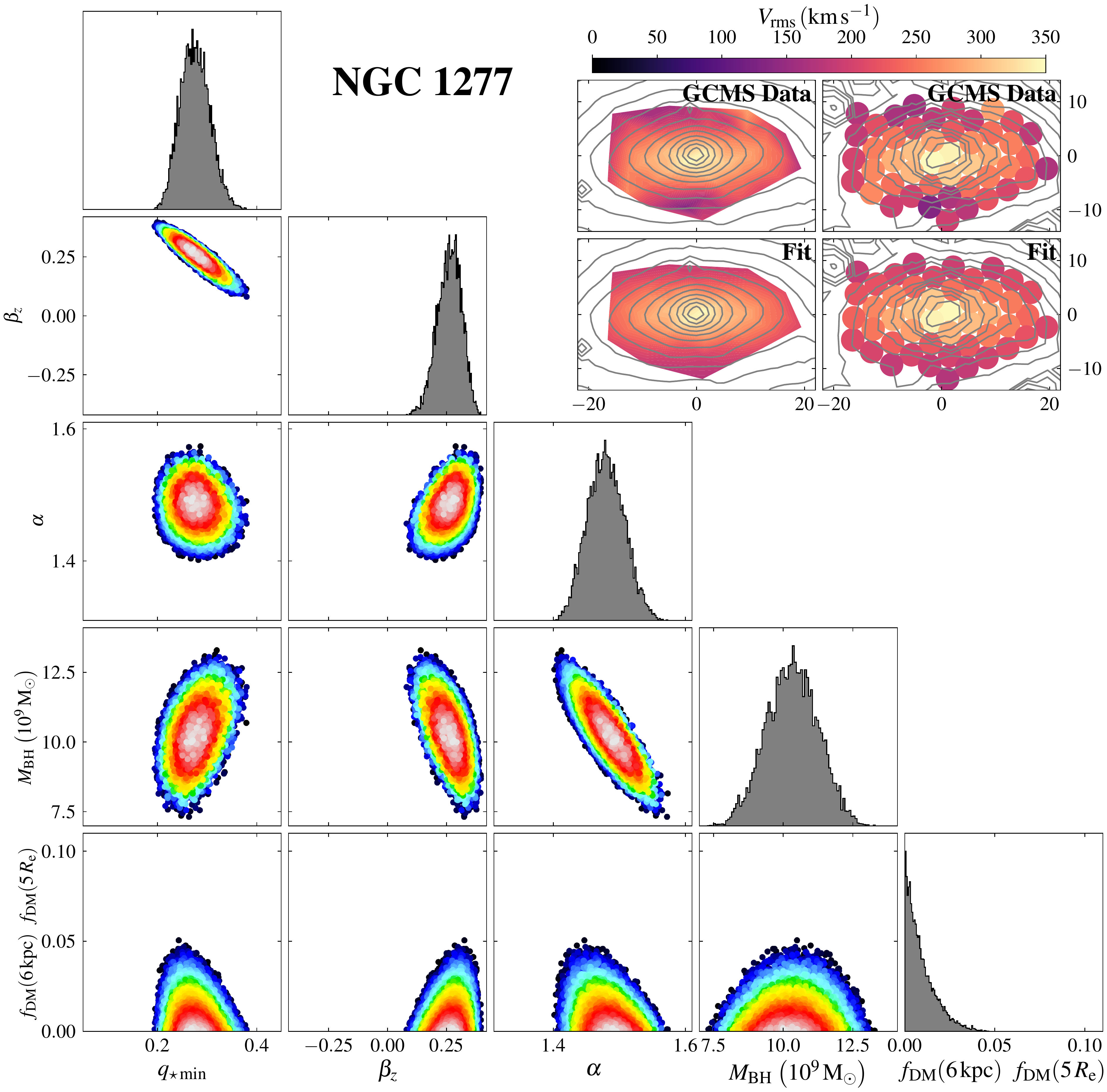}\\
  \end{center}
  \caption{\label{NGC1277_VIRUS_degraded} Same as Fig.~\ref{NGC1277_VIRUS}, but this time we also fitted $M_{\rm BH}$ and we used \texttt{MGE} parametrisations obtained from degraded {\it HST} images so to simulate how NGC~1277 would look like at $z=1.2$ (fit cCd).}
\end{figure*}

The total luminosity of NGC~1277 as obtained from the \texttt{MGE} decomposition of the degraded image is $L=2.7\times10^{10}\,L_{\odot r}$ (similar to what was found for the original image). The total mass obtained from the degraded mass map is $M=1.3\times10^{11}\,{\rm M}_\odot$ (similar to what was found for the original map before applying the correction for the mismatch parameter).

In Figs.~\ref{surface_brightness_degraded} and \ref{surface_density_degraded} we show the surface brightness and surface density profiles obtained from the degraded data and we compare them to those obtained from the original {\it HST} data (Figs.~\ref{Surface_brightness_profiles} and \ref{Surface_density_profiles}). Unsurprisingly, the density peaks within the central arcsecond are washed out by the image degradation.

\begin{table*}
\caption{\texttt{JAM} model fit parameters for NGC~1277 with the \texttt{MGE} parametrisation obtained from degraded {\it HST} data simulating {\it JWST} observations of a galaxy at a redshift $z=1.2$}
\label{parameters_NGC1277_degraded} 
\centering 
\begin{tabular}{l l l c c c c c c}
\hline\hline 
&&Fitting approach& $q_{\star\,{\rm min}}$  & $\beta_z$    & $\beta_{z,{\rm out}}$ & $\alpha$      & $M_{\rm BH}$                        & $f_{\rm DM}(6\,{\rm kpc})$\\
                &                   &             &    &              & & & $\left(10^9\,{\rm M}_\odot\right)$  &               $f_{\rm DM}(5\,{R_{\rm e}})$                    \\
\hline
Cylindric&Single $\beta_z$&cCd GCMS all free                                                                                                        & $0.27\pm0.03$   & $0.28\pm0.06$  & $-$            &  $1.50\pm0.03$   & $10.1\pm1.0$   & $<0.013$  \\\cline{2-9}
\hline
\end{tabular}
\tablefoot{Fit cCd is shown in Fig.~\ref{NGC1277_VIRUS_degraded}. The errors and upper limits correspond to formal one sigma uncertainties.}
\end{table*}

We produced Jeans models of NGC~1277 as in Sect.~\ref{jam}, but this time using the \texttt{MGE}s obtained from degraded data. Since we want to determine the black hole mass, we did a fit akin cC in Table~\ref{parameters_NGC1277}, that is with a free $M_{\rm BH}$, a cylindrical velocity ellipsoid, and no variation in $\beta_z$ with radius. We call this model cCd (`d' stands for degraded). The fit is shown in Fig.~\ref{NGC1277_VIRUS_degraded} and the fitted parameters are listed in Table~\ref{parameters_NGC1277_degraded}.

Once again, we find a negligible dark matter fraction within 6\,kpc. Perhaps the most surprising result is the detection of an extremely massive black hole, with $M_{\rm BH}=(1.01\pm0.10)\times10^{10}\,{\rm M}_\odot$. This mass is about two times larger than that found in Sect.~\ref{modelNGC1277}. The mismatch parameter that we find here, $\alpha\approx1.5$, is also significantly larger than in Sect.~\ref{modelNGC1277} ($\alpha\approx1.3$).

Why is the black hole mass so large when using low angular resolution data? Accounting for the mismatch parameter, the stellar mass within the central kpc of NGC~1277 is $M_\star(1\,{\rm kpc})=5.5\times10^{10}\,{\rm M}_\odot$ for the original \texttt{MGE} parametrisation. But this goes down to only $M_\star(1\,{\rm kpc})=4.6\times10^{10}\,{\rm M}_\odot$ when using the degraded data. The central $V_{\rm rms}$ peak is the consequence of both a black hole and a peaked stellar distribution. If the peak in the stellar distribution is smoothened or erased, the algorithm assigns a larger mass to the black hole to compensate for the loss of the stellar cusp. Hence, our simple test tentatively indicates that black hole mass estimates obtained from insufficiently sharp data might be biased towards high values. It is noteworthy that in spite of the likely bias, we recover a black hole mass that is only a factor of roughly two off the one obtained using high-angular-resolution data.

Our test underlines the importance of properly characterising the central stellar distribution to be able to give an order-of-magnitude estimate of the mass of the central black hole. Using images with the {\it JWST} resolution concomitantly with a deconvolution method such as the Lucy-Richardson algorithm \citep{Richardson1972, Lucy1974} it should be possible to recover information about the central stellar peak. Hence, such techniques should allow obtaining reasonable black hole mass estimates for galaxies with a central velocity dispersion cusp. Before being applied, this possibility hinted at by our relatively simple experiments should be confirmed with thorough tests made using mock simulated data.

\end{appendix}

\end{document}